\documentclass[onecolumn,secnumarabic,amssymb, nobibnotes, aps, prd]{revtex4-2}
\usepackage{version}

\usepackage{natbib}
\usepackage{amsmath,amsfonts}
\usepackage[table,usenames,dvipsnames]{xcolor}
\usepackage{color}
\usepackage{multirow}
\usepackage{caption}
\usepackage{graphicx}
\usepackage{subcaption}
\usepackage{tikz}
\usepackage{etoolbox}
\usepackage{ulem}
\usepackage{comment}
\usepackage[ruled,vlined]{algorithm2e}
\DeclareRobustCommand\full  {\tikz[baseline=-0.6ex]\draw[blue,thick] (0,0)--(0.5,0);}
\DeclareRobustCommand\fullblack  {\tikz[baseline=-0.6ex]\draw[black,thick] (0,0)--(0.5,0);}
\DeclareRobustCommand\dotted{\tikz[baseline=-0.6ex]\draw[thick,dotted] (0,0)--(0.54,0);}
\DeclareRobustCommand\dashed{\tikz[baseline=-0.6ex]\draw[thick,dashed] (0,0)--(0.54,0);}
\DeclareRobustCommand\chain {\tikz[baseline=-0.6ex]\draw[blue, thick,dash dot ] (0,0)--(0.5,0);}
\DeclareRobustCommand\chainblack {\tikz[baseline=-0.6ex]\draw[black, thick,dash dot ] (0,0)--(0.5,0);}
\newrobustcmd*{\mycircle}[1]{\tikz{\filldraw[draw=#1,fill=#1] (0,0) circle [radius=0.05cm];}}
\newrobustcmd*{\mytriangle}[1]{\tikz{\filldraw[draw=#1,fill=#1] (0,0)--(0.2cm,0) -- (0.1cm,0.2cm);}}

\DeclareRobustCommand\fullgrey  {\tikz[baseline=-0.6ex]\draw[Gray,thick] (0,0)--(0.5,0);}

\begin{document}

\title{Ensemble-variational assimilation of statistical data \\ in large eddy simulation}

\author{Vincent Mons}
\affiliation{DAAA, ONERA, Universit\'e Paris Saclay, F-92190 Meudon, France \\
            Department of Mechanical Engineering, Johns Hopkins University, Baltimore, MD 21218, USA}
\author{Yifan Du}%
\affiliation{Department of Mechanical Engineering, Johns Hopkins University, Baltimore, MD 21218, USA}
\author{Tamer A.~Zaki}%
\thanks{corresponding author, email: \url{t.zaki@jhu.edu}}
\affiliation{Department of Mechanical Engineering, Johns Hopkins University, Baltimore, MD 21218, USA}

\begin{abstract}
    A non-intrusive data assimilation methodology is developed to improve the statistical predictions of large-eddy simulations (LES). The ensemble-variational (EnVar) approach aims to minimize a cost function that is defined as the discrepancy between LES predictions and reference statistics from experiments or, in the present demonstration, independent direct numerical simulations (DNS). This methodology is applied to adjust the Smagorinsky subgrid model and obtain data assimilated LES (DA-LES) which accurately estimate the statistics of turbulent channel flow. To separately control the mean and fluctuations of the modeled subgrid tensor, and ultimately the first- and second-order flow statistics, two types of model corrections are considered. The first one optimizes the wall-normal profile of the Smagorinsky coefficient, while the second one introduces an adjustable steady forcing in the momentum equations to independently act on the mean flow. Using these two elements, the data assimilation procedure can satisfactorily modify the subgrid model and accurately recover reference flow statistics. The retrieved subgrid model significantly outperforms more elaborate baseline models such as dynamic and mixed models, in a posteriori testing. The robustness of the present data assimilation methodology is assessed by changing the Reynolds number and considering grid resolutions that are away from usual recommendations. Taking advantage of the stochastic formulation of EnVar, the developed framework also provides the uncertainty of the retrieved model.  
\end{abstract}

\maketitle

\section{Introduction}
Numerical predictions of turbulent flows at high Reynolds numbers generally require modeling of a least a fraction of the turbulent scales to be computationally tractable. Reynolds-averaged Navier-Stokes (RANS) and large eddy simulation (LES) are among the most popular approaches to address this modeling challenge. In contrast to  direct numerical simulation (DNS) where all flow scales down to the Kolmogorov eddies are resolved, both RANS and LES invoke modeling assumptions: RANS computes the mean flow only, and models all the turbulent scales. LES \citep{Pope2000_cambridge,Meneveau2000_afm,Sagaut2006_springer} on the other hand resolves the largest turbulent scales, and requires a representation of the effect of the unresolved, or subgrid scales, on the resolved ones. While LES requires a large computational cost compared to RANS modeling, its predictions include the unsteady evolution of the large turbulent scales, which enables a better evaluation of large-scale coherent phenomena and intermittent flows, for example strong flow separation, recirculation, vortex shedding and transition, compared to RANS \citep{Rumsey2011_ijhff,Lee2015_ja,Blocken2018_bs,Lardeau2012_ftc}. LES is thus a valuable tool for the numerical investigation of flows with such characteristics, although the accuracy of its predictions may still be impacted by the modeling of the subgrid scales. 
In the present work, we will introduce a data-assimilation approach to infuse LES with available statistical observations in order to improve the fidelity of LES, and will assess performance in canonical turbulent channel flow.

Whether RANS or LES modeling is considered, the accuracy of predictions are partly determined by the quality of modeling the unresolved turbulent scales. RANS requires a closure relationship for the Reynolds stress tensor, while LES models the subgrid tensor. This tensor arises from spatial filtering of the Navier-Stokes equations and accounts for the interactions between resolved and subgrid scales. In parallel with more traditional approaches to improve turbulence modeling \citep{Piomelli2014_ptrcl,Durbin2018_afm}, there is a growing interest in rigorously employing reference data from DNS or experiments to enhance the fidelity of RANS or LES predictions through data assimilation techniques \citep{Lewis2006_cambridge}. Generally speaking, data assimilation enables the merging of experimental and numerical approaches and to overcome their inherent limitations, namely the scarcity of experimental data and the uncertainties in modeling and in the inputs of simulations \citep{zaki2021_prf}. Incidentally, as an alternative to DNS \citep{Colburn2011_jfm,Suzuki2012_jfm,Gronskis2013_jcp,Mons2016_jcp,Wang2019_jcp,Li2020_jfm,wang2021_jfm}, recent studies have explored the possibilities of relying on LES for flow reconstruction/state estimation through data assimilation \citep{Meldi2017_jcp,Chandramouli2020_jcp,Labahn2020_ftc,Bauweraerts2021_jfm}. However, contrary to RANS where a number of studies exploited data assimilation to adjust turbulence models \citep{Kato2013_cf,Kato2015_jcp,Singh2016_pof,Xiao2016_jcp,Li2017_jcp,Xiao2019_pas,franceschini2020_prf}, the potential of data assimilation to enhance LES modeling has not been investigated to a similar extent, with very few exceptions \citep{Chandramouli2020_jcp,Sirignano2020_jcp}.

In \citep{Chandramouli2020_jcp}, adjoint-based data assimilation \citep{LeDimet1986_tellus} was adopted to optimize a spatially-varying coefficient in the location uncertainty model \cite{Memin2014_gafd} based on synthetic particle image velocimetry (PIV)-like measurements of the flow past a cylinder at $Re=3900$. In a more usual framework in terms of LES formulation \citep{Sirignano2020_jcp}, adjoint-based optimization was used to adjust a deep neural network that represents the divergence of the subgrid tensor relying on filtered velocity fields obtained by DNS of freely decaying isotropic turbulence.

In data assimilation, a cost function is defined as the sum of the squared discrepancies between observations and a numerical estimation. While the adjoint technique, which was employed in the above studies, is efficient at evaluating the gradient of the cost function with respect to high-dimensional control vectors, i.e.\,the quantities being optimized, it generally requires significant resources to implement and to use. This is particularly true for unsteady problems, because the storage of the forward flow solution is required for the backward-in-time integration of the adjoint model over the whole considered time window \cite{wang2019_jfm,wang2021_jfm}. The computational burden is thus very large for 3D flows, which is perhaps at the origin of the short time-window considered in \cite{Chandramouli2020_jcp} and of the restriction of the the assimilation to pairs of consecutive instants in \cite{Sirignano2020_jcp}. As an alternative to adjoint-based optimization, we propose use of stochastic data assimilation to infer corrections to subgrid models for LES; specifically, we adopt ensemble-based variational data assimilation (EnVar) \citep{Liu2008_mwr,Yang2015_cf,Mons2016_jcp,Mons2017_jweia,Jahanbakhshi2019_jfm,Mons2019_jcp,Jahanbakhshi2021_jfm,buchta2021_jfm}. In the EnVar approach, the data assimilation problem is still formulated as the minimization of a cost function that measures the discrepancies between available observations and a numerical approach. Instead of relying on an adjoint model, however, the minimization is performed based on an ensemble representation of the considered control vector. The EnVar approach may thus be considered as merging the robustness of variational techniques with the ease of implementation of more usual ensemble-based approaches such as the ensemble Kalman filter \citep{Evensen2009_springer}.

An EnVar-based methodology for the optimization of subgrid models will be introduced and assessed in canonical turbulent channel flow \citep{Kim1987_jfm,Moser1999_pof}. Building on the popular Smagorinsky model \citep{Smagorinsky1963_mwr}, the developed data assimilation procedure will be used to infer corrections to this baseline model, and to obtain what we term data-assimilated LES (DA-LES) predictions. Contrary to previous studies \citep{Chandramouli2020_jcp,Sirignano2020_jcp} where detailed instantaneous velocity fields were employed, we here investigate the possibility of relying on more limited data, namely statistical quantities.  In other words, instead of targeting the reproduction of a specific spatio-temporal flow realization through LES, we here aim to optimize the Smagorinsky model in order to accurately recover reference statistics for turbulent channel flows.

Two types of data assimilation experiments will be performed in this study. We will first consider filtered statistical quantities, such as the mean of the subgrid tensor, as observations to adjust the Smagorinsky model. In this case, we attempt to improve the performance of LES in a manner similar to a priori testing \citep{Sagaut2006_springer,ClarkDiLeoni2021_jfm}. This approach does not fully take into account the impact of discretization on LES predictions \citep{Vreman1994_cnme,Ghosal1996_jcp,Majander2002_ijnmf}. The importance of numerical methods in LES is well established, and was reaffirmed in recent studies that applied deep learning to subgrid modeling \citep{Beck2019_jcp,Sirignano2020_jcp}: calibrating a subgrid model without taking into account the specific discretization likely leads to poor performance of the actual LES solver. In a second set of experiments, more commonly available statistics such as the mean flow and/or the Reynolds stresses will be employed as observations. Such time-averaged quantities are more representative of actual measurements that can be generated from experiments and that are of direct interest for industrial applications. These data assimilation experiments take into account the implementation of the LES fully, including discretization, and enhance LES predictions according to a posteriori testing. Our observation data are obtained from independent DNS, but can equally be from experimental measurements.

In addition the data assimilation technique itself and the type of considered observations, another important aspect of the present methodology is the choice of the correction to the subgrid model. This choice has to be related to the nature of the observations in order to ensure the well-posedness of the data assimilation problem. The most ideal and detailed correction could consist of adding a spatial- and time-dependent term to the divergence of the subgrid tensor, which would offer maximal flexibility to address the functional deficiencies of the subgrid model. However, as discussed in \citep{Chandramouli2020_jcp}, even if instantaneous field data are available, the very large dimension of such a correction would make its identification through data assimilation computationally expensive. In addition, translating the result to a predictive engineering approach would remain challenging.

Considering more constrained corrections, a natural choice in the case of the Smagorinsky model is to optimize the so-called Smagorinsky coefficient $C_{\mathrm{s}}$. Since in the case of turbulent channel flow statistical inhomogeneity occurs in the wall-normal direction $y$ only, data assimilation may be formulated as identifying an optimal profile $C_{\mathrm{s}}(y)$ given reference statistics. However, this choice can not overcome all limitations of the Smagorinsky model. In particular, as suggested by \cite{hartel1994_pof,Hartel1998_jfm} and confirmed in the present study, it may be of interest to have the ability to independently adjust the dissipation of resolved mean and turbulent kinetic energies to improve the estimation of both first- and second-order statistics. Partly inspired by recent studies on the application of data assimilation to RANS modeling \citep{franceschini2020_prf}, we consider the introduction of a steady forcing term $\sigma(y)$ in the momentum equations which allows a separate adjustment of the mean of the subgrid tensor, and thus of the predicted mean flow. It will be confirmed in the present study that the simultaneous consideration of the coefficient $C_{\mathrm{s}}(y)$ and of the forcing $\sigma(y)$ as control vectors in the data assimilation procedure indeed provides an efficient correction to the Smagorinsky model in order to accurately reproduce statistical quantities of interest.

In \S\ref{sec:numerics_LES_models}, the characteristics of the present LES solver are first provided and the different subgrid models that are adopted in this study are specified. The Smagorinsky model forms the baseline model in the data assimilation procedure. In addition, the dynamic model \citep{Germano1991_pof,Lilly1992_pof} and a mixed model based on the scale-similarity hypothesis \citep{Liu1994_jfm} will be used to further benchmark the proposed methodology. The design of an appropriate correction form to the Smagorinsky model is also discussed in this section. The proposed EnVar data assimilation approach for optimizing subgrid models is detailed in \S\ref{sec:DA_methodology}, with a particular emphasis on ensemble generation. The DA-LES method is then first applied in \S\ref{sec:results_DA_filtered} considering filtered statistics as observations. Through the corresponding data assimilation experiments, the sensitivity of the LES predictions to changes in the coefficient $C_{\mathrm{s}}(y)$ is evaluated, along with the impact of discretization errors and the limitations of adjusting $C_{\mathrm{s}}(y)$ alone. In \S\ref{sec:results_DA_statistics}, the DA-LES procedure is further assessed using statistical observations that are directly accessible from experiments, i.e.\,unfiltered data. Variations in the grid resolution and Reynolds number are carried out to confirm the efficacy and robustness of the proposed procedure. Finally, conclusions are provided in \S\ref{sec:conclusions}.

\section{LES modeling and limitations}\label{sec:numerics_LES_models}

\subsection{Governing equations and numerical method}\label{sec:numerical_method}

We consider the turbulent flow of an incompressible Newtonian fluid in a channel \citep{Kim1987_jfm,Moser1999_pof}. The governing equations for large eddy simulations (LES) are derived from the Navier-Stokes equations through the application of a low-pass filter in order to retain large-scale contributions only \citep{Pope2000_cambridge,Sagaut2006_springer}. The velocity component $u_i$ in Cartesian coordinates is thus decomposed as
\begin{equation}
    u_i=\overline{u}_{i}+u'_{i},
\end{equation}
where the overbar refers to resolved scales, while the remaining term $u'_{i}$ includes small-scale, or subgrid effects. 
When statistical quantities are evaluated, the Reynolds decomposition is adopted, 
\begin{equation}
    u_i=\left\langle u_i \right\rangle+u_i'',
\end{equation}
where $\left \langle \cdot \right\rangle$ denotes ensemble averaging and $u_i''$ is the associated fluctuation.

The governing equations for the resolved part $\overline{u}_{i}$ of the velocity field may be written as
\begin{equation}\label{eq:filtered_NS_1}
    \frac{\partial \overline{u}_{i}}{\partial t}+\frac{\partial }{\partial x_{j}}(\overline{u}_{i}\overline{u}_{j})=-\frac{\partial \overline{p}}{\partial x_{i}}+2\nu\frac{\partial \overline{S}_{ij}}{\partial x_{j}}-\frac{\partial \tau_{ij}}{\partial x_{j}}, \qquad 
    \frac{\partial \overline{u}_{j}}{\partial x_{j}}=0,
\end{equation}
where the fluid density has been normalized by its constant reference value, $\overline{p}$ is the filtered pressure field, $\nu$ is the constant fluid kinematic viscosity, $\overline{S}_{ij}$ refers to the resolved strain-rate tensor.  The subgrid tensor, $\tau_{ij}=\overline{u_i u_j}-\overline{u}_i\overline{u}_j$, includes the interactions between resolved and subgrid scales and has to be closed in order to solve (\ref{eq:filtered_NS_1}). The different subgrid models for $\tau_{ij}$ that will be considered in this study are discussed in \S\ref{sec:subgrid_models}. 

The computational domain to solve (\ref{eq:filtered_NS_1}) is a three-dimensional box with streamwise, wall-normal and spanwise extents equal to $2\pi \delta$, $2\delta$ and $\pi \delta$, respectively, where $\delta$ is the channel half height. Periodic boundary conditions are imposed in the streamwise and spanwise directions, while the bottom and top walls correspond to no-slip surfaces. The friction Reynolds number $Re_{\tau}=u_{\tau}\delta/\nu$, where $u_{\tau}$ is the friction velocity, will be set to either $Re_{\tau}=590$ or $Re_{\tau}=1{,}000$ in this study.  

Equation (\ref{eq:filtered_NS_1}) is solved in a finite-volume framework using a fractional step algorithm \citep{Rosenfeld1991_jcp}; advection terms are discretized using a second-order Adams-Bashforth scheme and the viscous terms, including the eddy-viscosity part of the subgrid models discussed in \S\ref{sec:subgrid_models}, are treated implicitly using Crank-Nicolson. This numerical method was extensively validated in previous studies \citep{Jelly2014_pof,Lee2015_ftc,Wang2019_jcp}.

The grid is uniform in the streamwise ($x$) and spanwise ($z$) directions, while a hyperbolic stretching is used for the wall-normal coordinate ($y$). Various resolutions will be considered and are detailed in table \ref{tab:grid_resolutions}. Quantities are reported in non-dimensional form using wall scaling. A superscript $*$ indicates that a reference friction velocity $u_{\tau}$ from DNS is adopted, while a superscript $+$ is used when $u_{\tau}$ is obtained from the individual LES calculations.

The first grid DNS590 will be used to perform direct numerical simulation (DNS) at $Re_{\tau}=590$ from which reference filtered quantities such as the subgrid tensor will be evaluated. Reference statistics for $Re_{\tau}=1{,}000$ are obtained from the Johns Hopkins Turbulence Databases \citep{Graham2016_jot}. For the LES, two grids will be used when targeting $Re_{\tau}=590$. First, a relatively fine grid (LES590f) will be employed, which corresponds to standard recommendations for wall-resolved LES \citep{Sagaut2006_springer,Meyers2007_pof}, namely $\Delta x^{*}=50$ and $\Delta z^{*}=20$, to satisfactorily capture energetic large-scale phenomena in the buffer layer. In addition, this grid is appropriately refined close to the walls with the first off-wall grid point located at $y^{*}=0.5$. The consideration of a coarser grid (LES590c) in the streamwise and spanwise directions will provide further assessment of the data assimilation procedure of \S\ref{sec:DA_methodology}. While the minimum requirements for LES grids suggested in the literature are not unique, the spanwise resolution of grid LES590c in particular is below most recommendations \citep{Sagaut2006_springer,Piomelli2002_arfm}. The calculations at $Re_{\tau}=1{,}000$ are performed on grid LES1000 which corresponds to the same spatial resolution in wall units as the coarse grid LES590c.

\begin{table}
\begin{center}
\begin{tabular}{ccccccccc} 
 \hline
Grid & $Re_{\tau}$ & $N_x$ & $N_y$ & $N_z$ & $\Delta x^{*}$ & $\Delta z^{*}$ &$\Delta y^{*}_{\mathrm{min}}$&$\Delta y^{*}_{\mathrm{max}}$ \\ 
 \hline
DNS590 & ~~590 &$444$ & $256$ & $554$ & $~~8$ & $~4$ &$0.4$ & $12$\\ 
LES590f & ~~590 &$~72$ & $192$ & $~92$ & $50$ & $20$ & $0.5$ & $15$\\ 
LES590c & ~~590 & $~54$ & $192$ & $~54$ & $70$ & $35$ & $0.5$ & $15$\\
LES1000 & 1{,}000 &$~90$ & $320$ & $~90$ & $70$ & $35$ & $0.5$ & $15$\\
\hline
\end{tabular}
\end{center}
\caption{\label{tab:grid_resolutions}Summary of the different grids adopted in this study, including the target $Re_{\tau}$, number of grid points and spatial resolution in every direction.}
\end{table}

\subsection{Subgrid models}\label{sec:subgrid_models}

\subsubsection{Smagorinsky model}\label{sec:Smagorinsky model}

In this study, we will primarily rely on the most classical approach to close the subgrid tensor $\tau_{ij}$ in (\ref{eq:filtered_NS_1}), namely the Smagorinsky model \citep{Smagorinsky1963_mwr} which relies on the Boussinesq turbulent viscosity hypothesis to emulate the essentially forward energy cascade from resolved to subgrid scales. This approach models the deviatoric part $\tau^{\mathrm{d}}_{ij}$ of the subgrid tensor only according to
\begin{equation}\label{eq:smagorinsky_model}
    \tau^{\mathrm{d}}_{ij}=-2 \nu_{\mathrm{sgs}} \overline{S}_{ij}, \qquad    
    \nu_{\mathrm{sgs}}=\left(C_{\mathrm{s}}\overline{\Delta}\right)^{2}\left(2\overline{S}_{ij}\,\overline{S}_{ij}\right)^{\frac{1}{2}},
\end{equation}
where $C_{\mathrm{s}}$ is a scalar that may vary in space and time, and $\overline{\Delta}$ refers to the cutoff scale that is associated with the filtering operation to obtain the LES equations (\ref{eq:filtered_NS_1}). In this framework, the quantity $\frac{1}{3}\tau_{kk}$ is implicitly integrated into the filtered pressure $\overline{p}$ in (\ref{eq:filtered_NS_1}). In the following, note that by definition $\tau^{\mathrm{d}}_{ij}=\tau_{ij}$ for off-diagonal components. In order to take into account anisotropy and inhomogeneity effects in turbulent channel flows, standard adjustments include choosing $C_{\mathrm{s}}$ and the cutoff scale $\overline{\Delta}$ as
\begin{equation}\label{eq:smagorinsky_constant}
    C_{\mathrm{s}}(y^{+})=C_{\mathrm{I}}(1-\exp(-(y^{+}/25)^{3}))^{\frac{1}{2}},\qquad 
    \overline{\Delta}(y)=\left( \Delta x\Delta y (y)\Delta z\right)^{\frac{1}{3}}.
\end{equation} 
The form of $C_{\mathrm{s}}$ in (\ref{eq:smagorinsky_constant}), which is here a function of the wall-normal coordinate only, enforces an asymptotic cubic behavior for the eddy viscosity at the walls \citep{Pope2000_cambridge}. It involves the constant $C_{\mathrm{I}}$, which may be chosen as $\sim 0.2$ from the consideration of isotropic turbulence. The cutoff scale $\overline{\Delta}$ corresponds to the cubic root of the volume of a mesh cell at the corresponding wall-normal location \citep{Deardorff1970_jfm}. The Smagorinsky model (\ref{eq:smagorinsky_model})-(\ref{eq:smagorinsky_constant}) will be the baseline subgrid model in the following data assimilation experiments.

\subsubsection{Dynamic model}

Another well-known approach is the dynamic Germano-Lilly model \citep{Germano1991_pof,Lilly1992_pof}, which is still based on the Smagorinsky form (\ref{eq:smagorinsky_model}) but allows an automatic determination of $C_{\mathrm{s}}$ through the application of a second level of filtering and the use of the Germano identity \citep{Germano1991_pof}. In the case of turbulent channel flows \citep{Piomelli1993_pof}, $C_{\mathrm{s}}$ may be determined through
\begin{equation}\label{eq:dynamic_model}
C_{\mathrm{s}}^{2}(y,t)=  \frac{\left\langle m_{ij}A_{ij}^{\mathrm{d}}\right\rangle_{xz}}{\left\langle m_{kl}m_{kl}\right\rangle_{xz}},
\end{equation}
with
\begin{equation}\label{eq:dynamic_model_2}
  A_{ij}=\widetilde{\overline{u}_{i}\overline{u}_{j}}-\widetilde{\overline{u}_{i}}\widetilde{\overline{u}_{j}}, \quad
  m_{ij}=\alpha_{ij}-\widetilde{\beta}_{ij}, \quad  
  \alpha_{ij}=-2\widetilde{\overline{\Delta}}^{2}\left(2\widetilde{\overline{S}}_{kl}\,\widetilde{\overline{S}}_{kl}\right)^{\frac{1}{2}}\widetilde{\overline{S}}_{ij}, \quad \beta_{ij}=-2\overline{\Delta}^{2}\left(2\overline{S}_{kl}\,\overline{S}_{kl}\right)^{\frac{1}{2}}\overline{S}_{ij},
\end{equation}
where the notation $\widetilde{\cdot}$ refers to the application of the second filter, and $\left\langle \cdot \right\rangle_{xz}$ denotes averaging along the homogeneous directions. The cutoff scale $\widetilde{\overline{\Delta}}(y)$ which is associated to the second level of filtering is chosen as $\widetilde{\overline{\Delta}}(y)=\left(2 \Delta x\Delta y(y)2\Delta z\right)^{\frac{1}{3}}$. This second filtering is performed by applying a top-hat filter that is approximated by 
Simpson's quadrature rule.

\subsubsection{Mixed model}

Even though the dynamic model avoids the need to prescribe the coefficient $C_{\mathrm{s}}$, it is still restricted by the Boussinesq hypothesis. While the purely dissipative character of the Smagorinsky model form  is desirable for numerical considerations, it is not necessarily appropriate for wall turbulence where strong backward energy transfers from subgrid to resolved scales occur \citep{Horiuti1989_pof,Piomelli1991_pof,hartel1994_pof}. Moreover, the subgrid tensor that is predicted by the Smagorinsky model is known to be poorly correlated with the true one \citep{Majander2002_ijnmf}. Therefore, in addition to the above models, we also consider in this study a specific mixed model that combines the scale-similarity approach proposed in \cite{Liu1994_jfm} with a Smagorinsky contribution in order to ensure a sufficiently dissipative character of the complete model. It is expressed as
\begin{equation}\label{eq:mixed_model}
\tau_{ij}=C_{\mathrm{sim}}\left(\widetilde{\overline{u}_{i}\overline{u}_{j}}-\widetilde{\overline{u}_{i}}\widetilde{\overline{u}_{j}}\right)-2\nu_{\mathrm{sgs}} \overline{S}_{ij},
\end{equation}
where $C_{\mathrm{sim}}$ is an adjustable constant that weights the scale-similarity contribution in the mixed model (\ref{eq:mixed_model}) and may be evaluated as $\sim 1$. The eddy-viscosity part in (\ref{eq:mixed_model}) corresponds to the Smagorinsky model (\ref{eq:smagorinsky_model})-(\ref{eq:smagorinsky_constant}). The constant $C_{\mathrm{I}}$ in (\ref{eq:smagorinsky_constant}) is decreased to $0.141$ for the mixed model in order to emphasize the importance of the scale-similarity contribution relatively to the eddy-viscosity part. The second level of filtering that is denoted by $\widetilde{\cdot}$ and used to evaluate the scale-similarity contribution is chosen as the same one that is considered for the dynamic model in (\ref{eq:dynamic_model})-(\ref{eq:dynamic_model_2}).

The dynamic and mixed models will help in the assessment of the performance of the calibrated Smagorinsky model (\ref{eq:smagorinsky_model})-(\ref{eq:smagorinsky_constant}) through the methodology described in \S\ref{sec:DA_methodology}.  
In such comparisons, the dynamic model will be regarded as the best choice of $C_{\mathrm{s}}$ according to the Germano identity, and the mixed model is supposed to be better correlated with the true subgrid scale tensor and allows backward energy transfers \citep{Liu1994_jfm,Meneveau2000_afm}.

\subsection{Minimal correction to adjust LES statistics}\label{sec:minimal_correction_LES} 

We aim to improve the predictions of LES through adjustments of the Smagorinsky model (\ref{eq:smagorinsky_model})-(\ref{eq:smagorinsky_constant}) that rely on the data assimilation methodology that will be detailed in \S\ref{sec:DA_methodology}. Specifically, we target the correct estimation of usual statistical quantities such as the mean flow and the Reynolds stresses. The scalar $C_{\mathrm{s}}$ appears as a leading candidate for adjustment of the Smagorinsky model. However, an important consideration is whether adjusting $C_{\mathrm{s}}$ alone is sufficient to accurately predict both mean and second-order statistical quantities. This may be investigated through the consideration of the subgrid dissipation, namely the energy drain from resolved to subgrid scales, whose correct estimation is generally the main goal in functional modeling approaches for LES \citep{Sagaut2006_springer}. In the framework of wall turbulence, it was proposed in \cite{hartel1994_pof,Hartel1998_jfm} to split the mean subgrid dissipation $\varepsilon_{\mathrm{sgs}}$ according to 
\begin{equation}\label{eq:esgs_decomposition}
\varepsilon_{\mathrm{sgs}}=-\left\langle \tau_{ij}\overline{S}_{ij} \right\rangle=\varepsilon^{\mathrm{ms}}+\varepsilon^{\mathrm{fs}}, \quad \varepsilon^{\mathrm{ms}}=-\left\langle \tau_{ij}\right\rangle \left\langle\overline{S}_{ij} \right\rangle, \quad \varepsilon^{\mathrm{fs}}=-\left\langle \tau_{ij}'' \overline{S}_{ij}'' \right\rangle.
\end{equation}
In (\ref{eq:esgs_decomposition}), $\varepsilon^{\mathrm{ms}}$ quantifies the dissipation of mean resolved kinetic energy $\frac{1}{2}\langle \overline{u}_{i} \rangle \langle \overline{u}_{i} \rangle$ through interactions with the subgrid scales, while $\varepsilon^{\mathrm{fs}}$ is its counterpart for the fluctuating, or turbulent resolved kinetic energy $\frac{1}{2}\langle \overline{u}_{i}''  \overline{u}_{i}'' \rangle$. A correct evaluation of $\varepsilon^{\mathrm{ms}}$ and $\varepsilon^{\mathrm{fs}}$ may help to ensure the fidelity of the predicted mean flow and second-order statistics, respectively.

\begin{figure}
\centering
\begin{subfigure}{.28\linewidth}
\centering
\includegraphics[width=1.\textwidth]{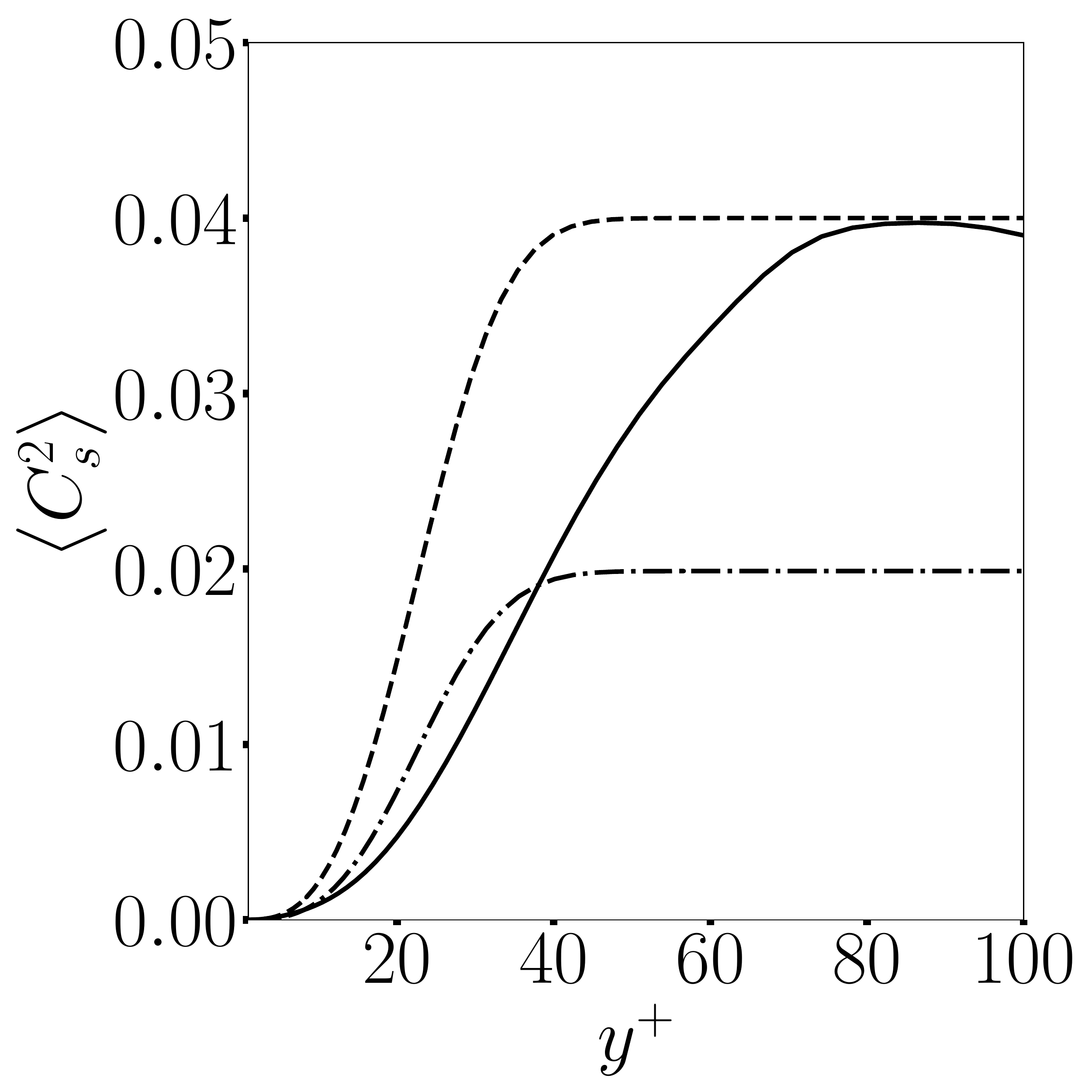}
\caption{Smagorinsky coefficient profile $C_s(y)$}
\end{subfigure}
\begin{subfigure}{.28\linewidth}
\centering
\includegraphics[width=1.\textwidth]{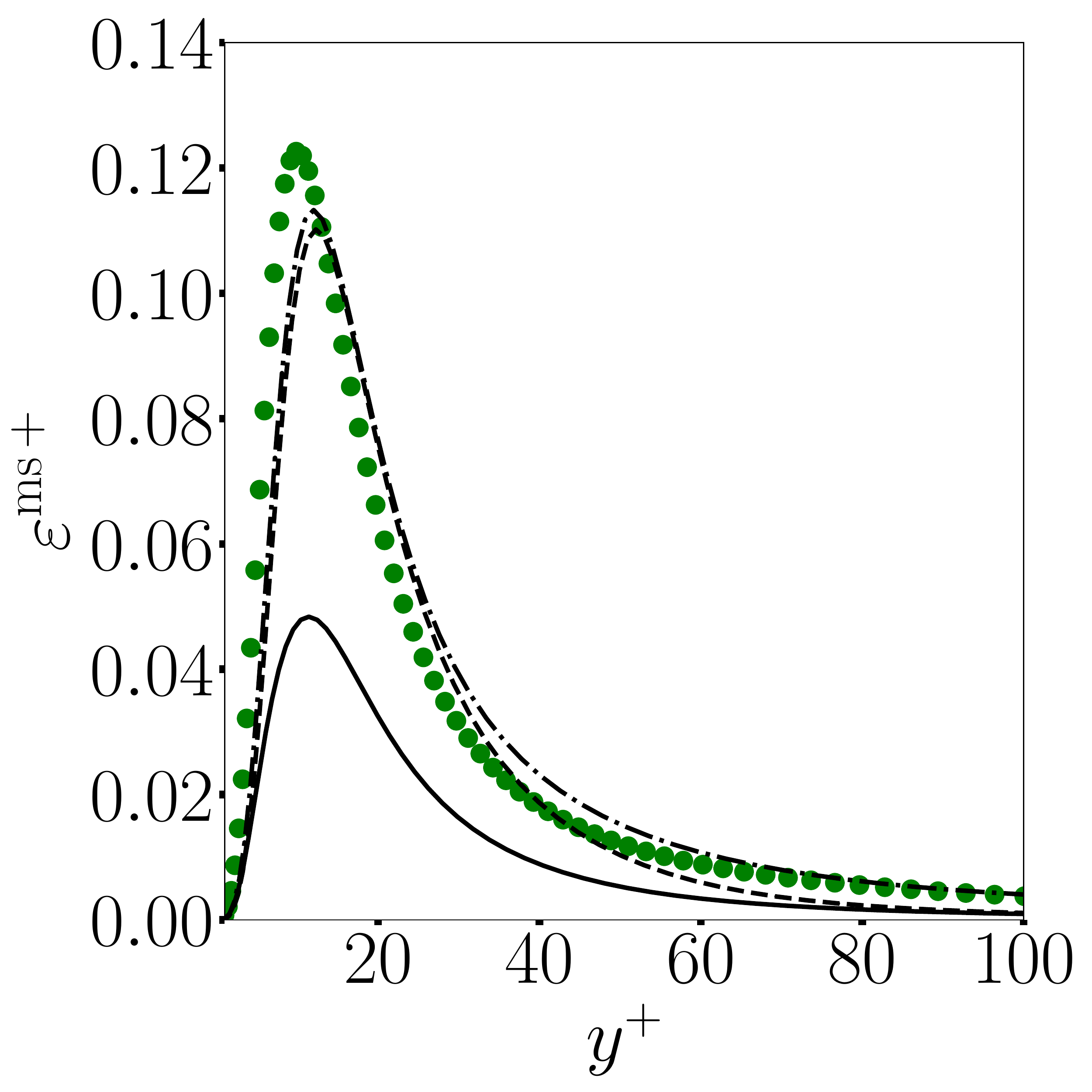}
\caption{$\varepsilon^{\mathrm{ms}}(y)$ profile}
\end{subfigure}
\begin{subfigure}{.28\linewidth}
\centering
\includegraphics[width=1.\textwidth]{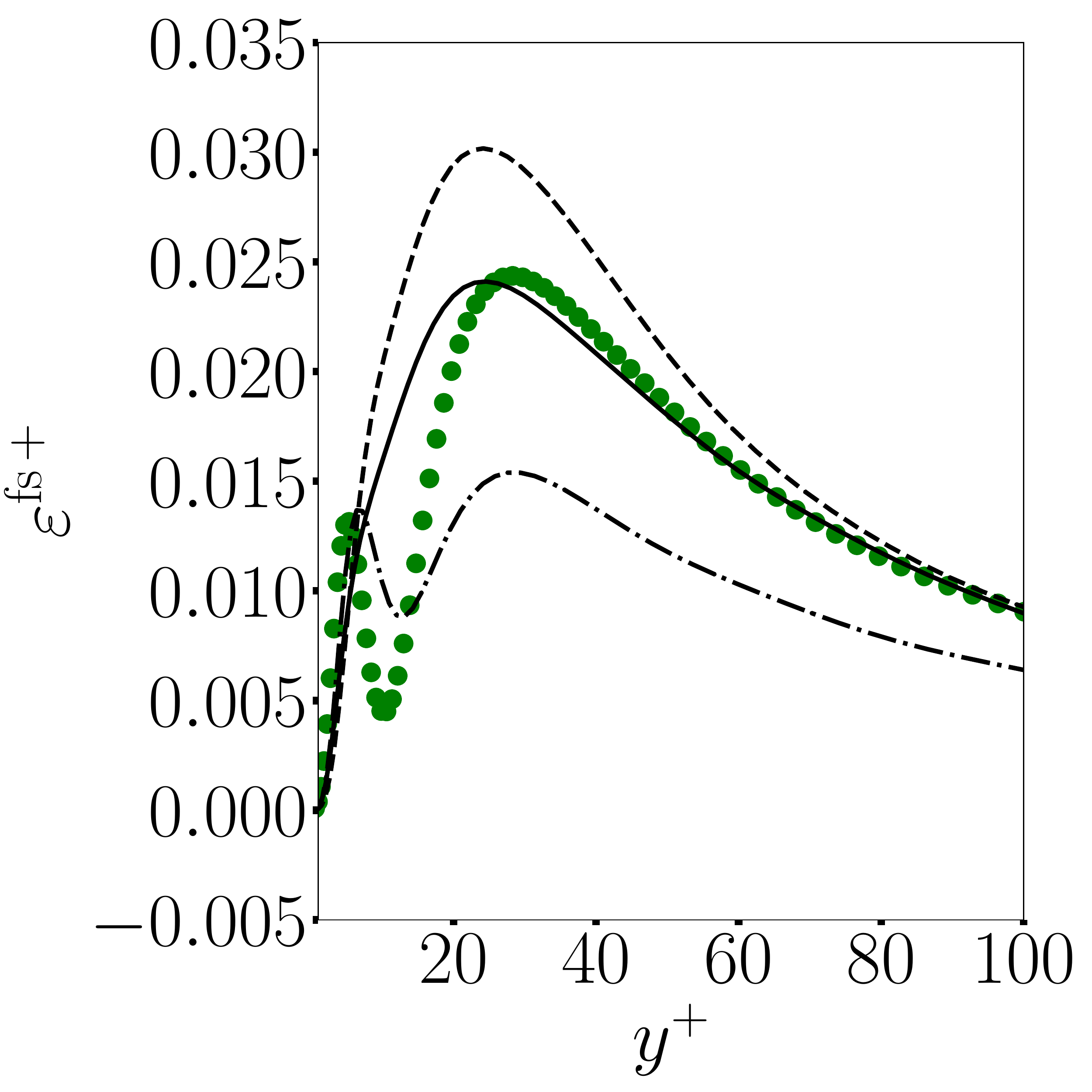}
\caption{$\varepsilon^{\mathrm{fs}}(y)$ profile}
\label{adjchnl:assimilated}
\end{subfigure}
\caption{\label{fig:esgs}Smagorinsky coefficient $C_\mathrm{s}$ and subgrid dissipations $\varepsilon^{\mathrm{ms}}$ and $\varepsilon^{\mathrm{fs}}$ in (\ref{eq:esgs_decomposition}) evaluated using the Smagorinsky model (\dashed), the dynamic model (\fullblack), the mixed model (\chainblack) and filtered DNS (\mycircle{black!50!green}) at $Re_{\tau}=590$. Grid LES590c is used for the LES calculations.}
\end{figure}

Figure \ref{fig:esgs} reports these two dissipation rates from the various subgrid models in \S\ref{sec:subgrid_models} at $Re_{\tau}=590$ using the grid LES590c in table \ref{tab:grid_resolutions}. These LES predictions are compared with the results of a DNS that is performed on the grid DNS590 and filtered in the homogeneous directions with a top-hat filter. Comparisons between the Smagorinsky and dynamic models may help in assessing the influence of the coefficient $C_{\mathrm{s}}$ on $\varepsilon^{\mathrm{ms}}$ and $\varepsilon^{\mathrm{fs}}$. In the near-wall region ($y^{+} < 50$) and slightly beyond, the dynamic model predicts a value for $C_{\mathrm{s}}$ that is significantly lower than the profile (\ref{eq:smagorinsky_constant}) chosen for the Smagorinsky model. As a result, the subgrid dissipation rates that are predicted by the dynamic model are overall reduced compared to the Smagorinsky model. The latter provides a satisfactory estimation of the subgrid dissipation of mean kinetic energy $\varepsilon^{\mathrm{ms}}$, while this same dissipation rate is significantly underestimated by the dynamic model. The converse is true for the prediction of the subgrid dissipation of turbulent kinetic energy $\varepsilon^{\mathrm{fs}}$, which is more accurately predicted by the dynamic model and is overestimated with the Smagorinsky model. Accordingly, the results in figure \ref{fig:esgs} suggest that it may be difficult to identify an optimal profile for $C_{\mathrm{s}}$ that would correctly estimate both $\varepsilon^{\mathrm{ms}}$ and $\varepsilon^{\mathrm{fs}}$, and thus both the mean flow and second-order statistics. These results are in accordance with the discussion in \cite{Hartel1998_jfm}.  
Note that, similarly, the mixed model only matches one, but not both $\varepsilon^{\mathrm{ms}}$ and $\varepsilon^{\mathrm{fs}}$ that are predicted from the filtered DNS. Its estimation of $\varepsilon^{\mathrm{ms}}$ is satisfactory; While its prediction of $\varepsilon^{\mathrm{fs}}$ is poor, it has the desirable feature of reproducing a non-monotic behavior close to the wall, which originates from the ability of the mixed model to reproduce backward energy transfers that occur around $y^{+}=10$.

The above discussion motivates the consideration of additional features of Smagorinsky-like models to adjust. Specifically, such features must enable independent control of the mean and the fluctuations of the subgrid tensor $\tau_{ij}$, which are involved in $\varepsilon^{\mathrm{ms}}$ and $\varepsilon^{\mathrm{fs}}$, respectively. We here choose to continue to directly adjust the fluctuations of $\tau_{ij}$ through the coefficient $C_{\mathrm{s}}$, and add a stationary contribution to $\tau_{ij}$ to control its mean. Due to the present flow symmetries, only $\langle\tau_{12}\rangle$ is involved in $\varepsilon^{\mathrm{ms}}$ and can directly affect the mean flow. As such, we aim to adjust $\langle\tau_{12}\rangle$ only, which may be performed by modifying the LES equations (\ref{eq:filtered_NS_1}) according to
\begin{equation}\label{eq:sigma}
\frac{\partial \overline{u}_{i}}{\partial t}+\frac{\partial }{\partial x_{j}}(\overline{u}_{i}\overline{u}_{j})=-\frac{1}{\rho}\frac{\partial \overline{p}}{\partial x_{i}}+2\nu\frac{\partial \overline{S}_{ij}}{\partial x_{j}}-\frac{\partial \tau_{ij}}{\partial x_{j}}-\frac{d \sigma}{d y}\delta_{i1},
\end{equation}
where the supplementary stationary contribution $\frac{d \sigma}{ dy}(y)$ acts only on the (mean) streamwise velocity. Introducing $\sigma$ alters the mean of the subgrid tensor according to $\langle \tau_{12}\rangle \rightarrow \langle \tau_{12}\rangle+\sigma$. We will thus rely on the coefficient $C_{\mathrm{s}}$ and/or the forcing $\sigma$ to optimize LES predictions through the data assimilation procedure that is described in \S\ref{sec:DA_methodology}.

\section{Data assimilation method for optimizing subgrid models}\label{sec:DA_methodology}

\subsection{Control vectors, observations and objective}

The main goal of this study is to faithfully reproduce, using LES, reference statistics. These statistical data are here evaluated from DNS but may, in principle, be available from experiments. The present approach, whose outcome will be referred to as data assimilated LES (DA-LES), relies on adjustment of the Smagorinsky model (\ref{eq:smagorinsky_model})-(\ref{eq:smagorinsky_constant}) through the coefficient $C_{\mathrm{s}}$ and/or the forcing $\sigma$ in (\ref{eq:sigma}). These quantities, which here depend on the wall-normal coordinate $y$ only, are discretized each with $N_{y}$ grid points in the wall-normal direction. Together they form the so-called control vector $\boldsymbol{\gamma}$, which is thus comprised of either $N_{y}$ or $2N_{y}$ elements, depending on whether only $C_{\mathrm{s}}$ or $\sigma$ is considered alone or if both quantities are included in $\boldsymbol{\gamma}$. Appropriate vectors $\boldsymbol{\gamma}$ are here inferred from reference statistical data, such as the mean flow or the mean subgrid tensor, and whose nature will vary in the following data assimilation experiments. These reference data, or observations, are gathered in the vector $\boldsymbol{m}$. We will observe either one and two statistical quantities over the whole channel height, at all wall-normal grid points. Accordingly, the size of $\boldsymbol{m}$ will be either $N_{y}$ or $2N_{y}$. It should be noted that our data assimilation procedure does not require observations to be available at all wall-normal grid points and can be performed with sparse data whose positions do not coincide with grid points, although they were here chosen to coincide for simplicity.

In the Bayesian formulation of data assimilation \citep{vanLeeuwen1996_mwr,Wikle2007_phyD,Lewis2006_cambridge}, both the control vector $\boldsymbol{\gamma}$ and the observations $\boldsymbol{m}$ are considered as random vectors, in the sense of aleatory uncertainties. In this framework, data assimilation identifies the control vector $\boldsymbol{\gamma}$ with maximum likelihood conditioned on the available observations $\boldsymbol{m}$. Relying on the standard Gaussian assumption, this amounts to minimizing the cost function  
\begin{equation}\label{eq:cost_function}
J=\frac{1}{2} \left\|\boldsymbol{\gamma} - \boldsymbol{\gamma}^{\mathrm{f}} \right\|_{\boldsymbol{\mathrm{B}}^{-1}}^{2}    +\frac{1}{2}\left\|\boldsymbol{m} - \boldsymbol{h}(\boldsymbol{\gamma}) \right\|_{\boldsymbol{\mathrm{R}}^{-1}}^{2}.
\end{equation}
The first term reflects the knowledge on the control vector prior to the consideration of the observations in the form of an estimate $\boldsymbol{\gamma}^{\mathrm{f}}$. At the beginning of the data assimilation procedure, the estimate $\boldsymbol{\gamma}^{\mathrm{f}}$ corresponds to the standard Smagorinsky model in \S\ref{sec:Smagorinsky model}, i.e.\, using $C_{\mathrm{s}}$ as given in equation (\ref{eq:smagorinsky_constant}) and $\sigma=0$. This estimate will be then updated in the iterative procedure that is detailed below and summarized in Algorithm \ref{tab:EnVar_algo}. This contribution to the cost also includes the covariance matrix $\boldsymbol{\mathrm{B}}$ which is associated to $\boldsymbol{\gamma}^{\mathrm{f}}$, with the notation $\left\| \bullet \right\|^{2}_{\boldsymbol{\mathrm{B}}^{-1}}=\bullet ^{\mathrm{T}}\boldsymbol{\mathrm{B}}^{-1}\bullet$. The second term in (\ref{eq:cost_function}) quantifies the discrepancies between the LES predictions and the considered observations $\boldsymbol{m}$ with covariance matrix $\boldsymbol{\mathrm{R}}$. The operator $\boldsymbol{h}$ maps the control vector space to the observation space. In the present case, it includes the LES computation and the extraction of the statistical quantities to compare with $\boldsymbol{m}$. If observation locations did not coincide with grid points, $\boldsymbol{h}$ would also include an evaluation/interpolation scheme from observation to grid point locations. The methodology to minimize the cost function $J$ is described in \ref{sec:EnVar}, while the choice of the covariance matrices $\boldsymbol{\mathrm{B}}$ and $\boldsymbol{\mathrm{R}}$ is discussed in \ref{sec:DA_parameters} along with other data assimilation parameters.

\subsection{Ensemble-based variational approach}\label{sec:EnVar}

The minimization of the cost function $J$ in (\ref{eq:cost_function}) is here performed following an ensemble-based variational (EnVar) approach \citep{Liu2008_mwr}, where the control vector $\boldsymbol{\gamma}$ is searched in a subspace spanned by an ensemble of realizations which are representative of the prior statistics. In this framework, $\boldsymbol{\gamma}$ is expressed as
\begin{equation}\label{eq:ensemble_representation}
\boldsymbol{\gamma}=\boldsymbol{\gamma}^{\mathrm{f}}+\boldsymbol{\mathrm{E}}\boldsymbol{w}, \quad \boldsymbol{\mathrm{B}}\simeq \frac{1}{N_{\mathrm{POD}}-1}\boldsymbol{\mathrm{E}}\boldsymbol{\mathrm{E}}^{\mathrm{T}},
\vspace{-0.5em}
\end{equation}
where $\boldsymbol{w}$ becomes the new control vector in the minimization process, and $\boldsymbol{\mathrm{E}}$ contains a set of $N_{\mathrm{POD}}$ suitable basis vectors which may be related to the prior covariance matrix $\boldsymbol{\mathrm{B}}$. The matrix $\boldsymbol{\mathrm{E}}$ is built as follows: We first form the matrix $\boldsymbol{\mathrm{Q}}$ as
\begin{equation}\label{eq:Q_matrix}
\boldsymbol{\mathrm{Q}}=\left( \boldsymbol{\gamma}^{(1)}-\boldsymbol{\gamma}^{\mathrm{f}},\boldsymbol{\gamma}^{(2)}-\boldsymbol{\gamma}^{\mathrm{f}}, \cdots , \boldsymbol{\gamma}^{(N_{\mathrm{ens}})}-\boldsymbol{\gamma}^{\mathrm{f}} \right)
\end{equation}
where $\boldsymbol{\gamma}^{(1)}$, $\boldsymbol{\gamma}^{(2)}$, $\cdots$, $\boldsymbol{\gamma}^{(N_{\mathrm{ens}})}$ are $N_{\mathrm{ens}}$ realizations of $\boldsymbol{\gamma}$, or ensemble members following a normal distribution with mean $\boldsymbol{\gamma}^{\mathrm{f}}$ and covariance $\boldsymbol{\mathrm{B}}$. Using the LES solver, the observations corresponding to each of these realizations are evaluated and assembled in the matrix $\boldsymbol{\mathrm{H}}$ defined as
\begin{equation}\label{eq:H_matrix}
\boldsymbol{\mathrm{H}}=\left( \boldsymbol{h}(\boldsymbol{\gamma}^{(1)})-\boldsymbol{h}(\boldsymbol{\gamma}^{\mathrm{f}}),\boldsymbol{h}(\boldsymbol{\gamma}^{(2)})-\boldsymbol{h}(\boldsymbol{\gamma}^{\mathrm{f}}), \cdots , \boldsymbol{h}(\boldsymbol{\gamma}^{(N_{\mathrm{ens}})})-\boldsymbol{h}(\boldsymbol{\gamma}^{\mathrm{f}}) \right).
\vspace{-0.5em}
\end{equation}
Evaluating $\boldsymbol{\mathrm{H}}$ is thus by far the most demanding part of the present methodology, as it requires the realization of $N_{\mathrm{ens}}+1$ LES calculations to compute the flow for each ensemble member and the estimate $\boldsymbol{\gamma}^{\mathrm{f}}$. 
Following \cite{Tian2011_tellus}, a proper orthogonal decomposition (POD) representation of the ensemble in observation space is performed by solving the following eigenvalue problem
\begin{equation}\label{eq:POD_1}
\frac{1}{N_{\mathrm{ens}}}\boldsymbol{\mathrm{H}}^{\mathrm{T}}\boldsymbol{\mathrm{H}}\boldsymbol{v}^{(i)}=\lambda^{(i)}\boldsymbol{v}^{(i)}, \quad i \in \left\{ 1,2,\cdots,N_{\mathrm{ens}}\right\},
\end{equation}
which corresponds to the method of snapshots \citep{Sirovich1987_qam}. The $N_{\mathrm{ens}}$ eigenvalues $\lambda^{(i)}$ are sorted by decreasing value, and a set of eigenvectors $\boldsymbol{v}^{(i)}$ is formed according to
\begin{equation}\label{eq:POD_2}
\boldsymbol{\mathrm{V}}=\left( \boldsymbol{v}^{(1)},\boldsymbol{v}^{(2)},\cdots, \boldsymbol{v}^{(N_{\mathrm{POD}})}\right),\quad
N_{\mathrm{POD}}= \min \left\{ k  \middle|  \frac{\sum_{i=1}^{k}  \lambda^{(i)}  } { \sum_{i=1}^{N_{\mathrm{ens}}}  \lambda^{(i)} }\geqslant \epsilon_{\mathrm{POD}} \right\},
\end{equation}
where the parameter $0<\epsilon_{\mathrm{POD}}\leq 1$ is used to control the size and smoothness of the POD basis. The matrix $\boldsymbol{\mathrm{E}}$ in (\ref{eq:ensemble_representation}) is finally obtained through
\begin{equation}\label{eq:E_POD}
\boldsymbol{\mathrm{E}}=\boldsymbol{\mathrm{Q}}\boldsymbol{\mathrm{V}}.
\end{equation} 
Using (\ref{eq:E_POD}) and after linearization around $\boldsymbol{\gamma}^{\mathrm{f}}$, the cost function $J$ becomes
\begin{equation}\label{eq:cost_function_envar}
\tilde{J}=\frac{1}{2}(N_{\mathrm{POD}}-1)\boldsymbol{w}^{\mathrm{T}}\boldsymbol{w}+  \frac{1}{2}\left\| \boldsymbol{\mathrm{H}}_{\mathrm{POD}}\boldsymbol{w}+\boldsymbol{h}(\boldsymbol{\gamma}^{\mathrm{f}}) - \boldsymbol{m} \right\|_{\boldsymbol{\mathrm{R}}^{-1}}^{2}, \quad \boldsymbol{\mathrm{H}}_{\mathrm{POD}}=\boldsymbol{\mathrm{H}}\boldsymbol{\mathrm{V}}.
\end{equation}
Compared to the straightforward choice $\boldsymbol{\mathrm{E}}=\boldsymbol{\mathrm{Q}}$, the POD step yields orthogonal vectors to match the observations and improves the well-posedness of the data assimilation problem. Since $\tilde{J}$ in (\ref{eq:cost_function_envar}) is quadratic, the minimizer, or assimilated vector, $\boldsymbol{w}^{\mathrm{a}}$ is easily obtained from
\begin{equation}\label{eq:minimizing_w}
  \boldsymbol{w}^{\mathrm{a}}=\left( (N_{\mathrm{POD}}-1)\boldsymbol{\mathrm{I}}+
  \boldsymbol{\mathrm{H}}_{\mathrm{POD}}^{\mathrm{T}}\boldsymbol{\mathrm{R}}^{-1}\boldsymbol{\mathrm{H}}_{\mathrm{POD}}
  \right)^{-1}  \boldsymbol{\mathrm{H}}_{\mathrm{POD}}^{\mathrm{T}}\left(  \boldsymbol{m}-\boldsymbol{h}(\boldsymbol{\gamma}^{\mathrm{f}})  \right),
\end{equation}
and the original assimilated control vector is therefore  $\boldsymbol{\gamma}^{\mathrm{a}}=\boldsymbol{\gamma}^{\mathrm{f}}+\boldsymbol{\mathrm{E}}\boldsymbol{w}^{\mathrm{a}}$. DA-LES thus refers to the control vector $\boldsymbol{\gamma}^{\mathrm{a}}$ and the associated predictions. As the original cost function in (\ref{eq:cost_function}) is a nonlinear least-square problem, it may be beneficial to consider $\boldsymbol{\gamma}^{\mathrm{a}}$ as a new estimate $\boldsymbol{\gamma}^{\mathrm{f}}$ and to perform a new assimilation cycle. If so, we may exploit the fact that the above procedure not only provides an optimal control vector $\boldsymbol{\gamma}^{\mathrm{a}}$, but also the covariance matrix $\boldsymbol{\mathrm{P}}^{\mathrm{a}}$ of $\boldsymbol{\gamma}^{\mathrm{a}}$, which may be evaluated as
\begin{equation}\label{eq:posterior_statistics}
\boldsymbol{\mathrm{P}}^{\mathrm{a}}=\boldsymbol{\mathrm{Q}} \boldsymbol{\mathrm{L}}^{-1}\boldsymbol{\mathrm{Q}}^{\mathrm{T}}, \quad \boldsymbol{\mathrm{L}}=(N_{\mathrm{ens}}-1)\boldsymbol{\mathrm{I}}+
  \boldsymbol{\mathrm{H}}^{\mathrm{T}}\boldsymbol{\mathrm{R}}^{-1}\boldsymbol{\mathrm{H}},
\end{equation}
where $\boldsymbol{\mathrm{L}}$ corresponds to the Hessian matrix of the quadratic cost function for the choice $\boldsymbol{\mathrm{E}}=\boldsymbol{\mathrm{Q}}$. Using (\ref{eq:posterior_statistics}), a new ensemble of $N_{\mathrm{ens}}$ realizations of the control vector can be obtained and the matrix $\boldsymbol{\mathrm{Q}}$ updated as
\begin{equation}\label{eq:ens_iter}
\boldsymbol{\mathrm{Q}}\leftarrow\sqrt{N_{\mathrm{ens}}-1}\boldsymbol{\mathrm{Q}}\boldsymbol{\mathrm{L}}^{-\frac{1}{2}}\boldsymbol{\mathrm{U}},
\end{equation}
where $\boldsymbol{\mathrm{U}}$ is a random mean-preserving orthonormal matrix. If this process is repeated $N_{\mathrm{it}}$ iterations, the total computational cost that is associated to the data assimilation procedure can be evaluated as $N_{\mathrm{CFD}}=N_{\mathrm{it}}\times (N_{\mathrm{ens}}+1)$ LES calculations. The whole procedure is summarized in Algorithm \ref{tab:EnVar_algo}.

\begin{algorithm}[h]
	\SetAlgoLined
	\textbf{Step 1}: Initialization\;
	\Indp 
	Initialize the control vector $\boldsymbol{\gamma}^{\mathrm{f}}$ based on the standard Smagorinsky model (\ref{eq:smagorinsky_model})-(\ref{eq:smagorinsky_constant}) and set the covariance matrices $\boldsymbol{\mathrm{B}}$ (\ref{eq:cov_init}) and $\boldsymbol{\mathrm{R}}$  (\ref{eq:observation_covariance})\;
	\Indm
	\textbf{Step 2}: Generation of initial ensemble\;
	\Indp
    Generate $N_{\mathrm{ens}}$ realizations of $\boldsymbol{\gamma}$ following a normal distribution with mean $\boldsymbol{\gamma}^{\mathrm{f}}$ and covariance matrix $\boldsymbol{\mathrm{B}}$ to form the matrix $\boldsymbol{\mathrm{Q}}$ in (\ref{eq:Q_matrix})\;
    \Indm
	\textbf{Step 3}: Large eddy simulations\;
	\Indp
    Perform $N_{\mathrm{ens}}+1$ LES calculations to obtain the matrix $\boldsymbol{\mathrm{H}}$ in (\ref{eq:H_matrix})\;
    \Indm
	\textbf{Step 4}: Proper Orthogonal decomposition\;
	\Indp
    Perform POD step (\ref{eq:POD_1})-(\ref{eq:E_POD})\;
    \Indm
	\textbf{Step 5}: Assimilation\;
	\Indp
    Linearize the cost function (\ref{eq:cost_function}) around $\boldsymbol{\gamma}^{\mathrm{f}}$ and evaluate the assimilated control vector $\boldsymbol{\gamma}^{\mathrm{a}}$ through (\ref{eq:minimizing_w})\;
    \Indm
	\textbf{Step 6}: Iteration cycle\;
	\Indp
    If performing a new assimilation cycle, set $\boldsymbol{\gamma}^{\mathrm{a}} \rightarrow \boldsymbol{\gamma}^{\mathrm{f}}$ and generate a new ensemble using (\ref{eq:ens_iter}). Return to step 3. The total number of iterations is denoted as $N_{\mathrm{it}}$.
    \caption{Summary of the EnVar procedure for data assimilated LES (DA-LES)}.
	\label{tab:EnVar_algo}
\end{algorithm}

\subsection{Ensemble generation and data assimilation parameters}\label{sec:DA_parameters}

The prior covariance matrix $\boldsymbol{\mathrm{B}}$ that is used to generate the first ensemble (step 2 in Algorithm \ref{tab:EnVar_algo}) is  a usual squared exponential function. In the case where the control vector $\boldsymbol{\gamma}$ is formed by either the coefficient $C_{\mathrm{s}}$ or the forcing $\sigma$ only, the components of $\boldsymbol{\mathrm{B}}$ may be expressed as
\begin{equation}\label{eq:cov_init}
  B_{ij} = s_{b}(y_i)s_{b}(y_j)\mathrm{exp}\left(-\left(\frac{f(y_i)-f(y_j)}{l_\mathrm{c}}\right)^2\right).
\end{equation}
where $s_{b}(y_i)$ is the standard deviation at the wall-normal location $y_i$ of a quantity $b$ which refers to either $C^2_{\mathrm{s}}$ or $\sigma$. The consideration of $C^2_{\mathrm{s}}$ instead of $C_{\mathrm{s}}$ was preferred because the former is the effective coefficient that determines the intensity of the Smagorinsky model (\ref{eq:smagorinsky_model}). The standard deviation is chosen as $s_b(y_i)=\frac{1}{10}|b^{\mathrm{f}}(y_i)|$, where $b^{\mathrm{f}}$ refers to the first-guess value for $b$. Since the first guess of the forcing $\sigma$ is $\sigma^{\mathrm{f}}=0$, the value of $\langle \tau_{12}\rangle$ is added to $b^{\mathrm{f}}(y_i)$ for that term. It may be noticed that this choice for $s_b(y_i)$ corresponds to a relatively large level of uncertainty. Significantly increasing this standard deviation would result in realizations of $\boldsymbol{\gamma}$ that correspond to variations of the same order of magnitude or higher than the first-guess $\boldsymbol{\gamma}^{\mathrm{f}}$ itself, which is not desirable in terms of numerical stability and convergence of the data assimilation procedure.
In order to take into account the wall-normal change in the lengthscale associated with variations in the flow statistics, in particular the mean flow, the squared exponential in (\ref{eq:cov_init}) does not directly involve $y$ but a transformed variable $f(y)=\log_{10}(y^{*})$. The term $l_\mathrm{c}$ in (\ref{eq:cov_init}) may thus be interpreted as a correlation length in logarithmic scale, and is chosen as unity. Preliminary tests confirmed that this value was appropriate to obtain smooth realizations of $\boldsymbol{\gamma}$, while variations around the latter did not entail significant changes in the outcome of the data assimilation procedure.
In the case where both $C_{\mathrm{s}}$ and $\sigma$ are included in $\boldsymbol{\gamma}$, they are considered as independent random variables, i.e.\,cross-correlations between these quantities are set to zero.

The first ensemble of realizations of $\boldsymbol{\gamma}$ (step 2 in Algorithm \ref{tab:EnVar_algo}), which is stored in the matrix $\boldsymbol{\mathrm{Q}}$ in (\ref{eq:Q_matrix}), is constructed such that it spans the same subspace as the $N_{\mathrm{ens}}$ dominant eigenvectors of $\boldsymbol{\mathrm{B}}$ in (\ref{eq:cov_init}) according to
\begin{equation}\label{eq:B_eigendecomposition}
   \boldsymbol{\mathrm{Q}} = \sqrt{N_{\mathrm{ens}}-1}\boldsymbol{\mathrm{W}}_{\mathrm{ens}}\boldsymbol{\Lambda}_{\mathrm{ens}}^{\frac{1}{2}}\boldsymbol{\mathrm{U}}, \qquad \boldsymbol{\mathrm{B}} = \boldsymbol{\mathrm{W}}\boldsymbol{\Lambda}\boldsymbol{\mathrm{W}}^{\mathrm{T}}
\end{equation}
where the matrices $\boldsymbol{\mathrm{W}}$ and $\boldsymbol{\Lambda}$ are formed by the eigenvectors and eigenvalues of $\boldsymbol{\mathrm{B}}$, respectively, while $\boldsymbol{\mathrm{W}}_{\mathrm{ens}}$ and $\boldsymbol{\Lambda}_{\mathrm{ens}}$ denote their restriction to the $N_{\mathrm{ens}}$ dominant modes. As in (\ref{eq:ens_iter}), $\boldsymbol{\mathrm{U}}$ refers to a random mean-preserving orthonormal matrix. Subsequent ensembles are then obtained through (\ref{eq:ens_iter}) (step 6 in Algorithm \ref{tab:EnVar_algo}).

In order to make the realizations of $\boldsymbol{\gamma}$ as appropriate as possible to match the reference statistics in $\boldsymbol{m}$, the ensemble members that are obtained by the above procedure are supplemented with a contribution which is inspired from the nudging technique \citep{Hoke1976_mwr}. When the mean flow is observed and if $\boldsymbol{\gamma}$ includes the forcing $\sigma$, realizations of the latter are generated as follows
\begin{equation}
    \sigma^{(i)}=\sigma^{(i),\boldsymbol{\mathrm{B}}}+\sigma^{(i),\mathrm{nud}}.
\end{equation}
The contribution $\sigma^{(i),\boldsymbol{\mathrm{B}}}$ is obtained as described above, based either on (\ref{eq:ens_iter}) or (\ref{eq:cov_init})-(\ref{eq:B_eigendecomposition}), while $\sigma^{(i),\mathrm{nud}}$ corresponds to a nudging-like term of the form
\begin{equation}\label{eq:sigma_nudging}
    \sigma^{(i),\mathrm{nud}}(y)=-C_{\mathrm{nud}}^{(i)}\theta(y), \qquad \theta(y)=\int_{0}^{y}\left(    U_{\mathrm{DNS}}(y{'})-U_{\mathrm{LES}}(y{'})  \right)dy{'}, 
\end{equation}
where $U_{\mathrm{DNS}}$ refers to the reference mean flow that is obtained by DNS and included in the observations $\boldsymbol{m}$, while $U_{\mathrm{LES}}$ is the mean flow that is predicted by LES with the previous estimate of the control vector $\boldsymbol{\gamma}^{\mathrm{f}}$.
The nudging yields the contribution $-d\sigma^{(i),\mathrm{nud}}/d y=C_{\mathrm{nud}}^{(i)}(U_{\mathrm{DNS}}-U_{\mathrm{LES}})$ in the augmented LES momentum equations (\ref{eq:sigma}), and thus facilitates reproducing the reference DNS mean flow by data assimilation, which was the primary aim when introducing the forcing $\sigma$ in \S\ref{sec:minimal_correction_LES}. The nudging coefficient $C_{\mathrm{nud}}^{(i)}$ is sampled from a uniform distribution in the range $[0,D]$, where the upper bound $D$ is chosen so that, on average, the intensity of the nudging contribution $\sigma^{(i),\mathrm{nud}}$ is consistent with the uncertainties in $\sigma$ prescribed through (\ref{eq:ens_iter}) or (\ref{eq:cov_init}). Specifically, $D$ ensures that 
\begin{equation}
    0.5D|\theta_{\mathrm{max}}|=s_{\mathrm{max}},
\end{equation}
where $\theta_{\mathrm{max}}$ corresponds to the maximum value of $\theta(y)$ in (\ref{eq:sigma_nudging}), while $s_{\mathrm{max}}$ refers to the maximum value of the standard deviation of $\sigma$.

The covariance matrix $\boldsymbol{\mathrm{R}}$, which weights the observation term in the cost function $J$ in (\ref{eq:cost_function}), is chosen to be a diagonal matrix according to
\begin{equation}\label{eq:observation_covariance}
    R_{ij}=0.01\|\boldsymbol{m}-\boldsymbol{h}(\boldsymbol{\gamma}^{\mathrm{f}})\|_{\boldsymbol{\mathrm{M}}}^{2}M^{-1}_{ij}, \qquad M_{ij}=\Delta y_{i}\delta_{ij},
\end{equation}
where Einstein summation convention does not apply in the second equality. The diagonal mass matrix $\boldsymbol{\mathrm{M}}$ accounts for the discretization in the wall-normal direction. 
Equation (\ref{eq:observation_covariance}) amounts to a statement that the observation term in the cost function can be reduced by two orders of magnitude, at which point it becomes commensurate with the initial variance which models measurement noise.
Preliminary tests confirmed that this was appropriate to match the observations $\boldsymbol{m}$, while still ensuring that the prior term is non-negligible and effectively regularizing.

An important parameter is the size $N_{\mathrm{ens}}$ of the ensemble of simulations that are performed in the EnVar approach. While it is desirable to set $N_{\mathrm{ens}}$ as small as possible due to computational cost, $N_{\mathrm{ens}}$ should be sufficiently large to capture the dominant directions of uncertainties in the control-vector space. It is suggested from previous studies \citep{Mons2016_jcp} that EnVar techniques may be efficient even when the dimension of the control vector is large compared the size of the ensemble by a ratio of the order of $10$ to $1{,}000$.  In the present case, the dimension of the control vector is at most $2N_{y}=640$ for the grid LES1000. Accordingly, following \citep{Mons2016_jcp,Mons2017_jweia,Jahanbakhshi2019_jfm,Mons2019_jcp,Jahanbakhshi2021_jfm,buchta2021_jfm}, ensembles of sizes $N_{\mathrm{ens}}=20$ are sufficiently large. As will be discussed in \S\ref{sec:further_assessment}, $N_{\mathrm{ens}}=10$ may be adopted. However, $N_{\mathrm{ens}}=20$ was selected to ensure convergence of the data assimilation procedure and the robustness of the reported results. One last parameter is $\epsilon_{\mathrm{POD}}$, which determines the size of the POD basis in (\ref{eq:POD_2}) and is set to $0.999$. This choice retains a vast majority of the most energetic modes, while still removing highly oscillatory and unphysical ones.

\begin{table}
\begin{center}
\begin{tabular}{ cccccccc } 
 \hline
 Case & section & $\mathrm{Re}_{\tau}$ & $\boldsymbol{\gamma}$& grid & $\boldsymbol{m}$ & $N_{\mathrm{it}}$ & $N_{\mathrm{CFD}}$\\ 
 \hline
1 & \ref{sec:cases_1_2} &~590 &$C_s(y)$& LES590f & $\left<\tau_{12}\right>$ & 4 & 84\\ 
2 & \ref{sec:cases_1_2} &~590 &$C_s(y)$ & LES590f & $\mathrm{d}U/\mathrm{d}y$, $\left<\tau_{12}\right>$ & 1 & 21\\ 
3 & \ref{sec:cases_3_3c} &~590 &$C_s(y)$, $\sigma(y)$& LES590f & $\left<\overline{u}''\overline{v}''\right>$, $\left<\tau_{12}\right>$& 3 & 63\\ 
3c & \ref{sec:cases_3_3c} &~590 & $C_s(y)$, $\sigma(y)$& LES590c & $\left<\overline{u}''\overline{v}''\right>$, $\left<\tau_{12}\right>$& 3 & 63\\ 
\hline
4 & \ref{sec:results_DA_statistics_Re_590_case4} &~590 & $\sigma(y)$ & LES590f & $\mathrm{d}U/\mathrm{d}y$& 2 & 42\\
5 & \ref{sec:results_DA_statistics_Re_590_cases5_5c} &~590 & $C_s(y)$, $\sigma(y)$ & LES590f & $\mathrm{d}U/\mathrm{d}y$, $\left<u''u''\right>$ & 2 & 42\\
5c & \ref{sec:results_DA_statistics_Re_590_cases5_5c} &~590 & $C_s(y)$, $\sigma(y)$ & LES590c & $\mathrm{d}U/\mathrm{d}y$, $\left<u''u''\right>$ & 2 & 42\\
6 & \ref{sec:results_DA_statistics_Re_1000} & 1{,}000 & $C_s(y)$, $\sigma(y)$ & LES1000 & $\mathrm{d}U/\mathrm{d}y$, $\left<u''u''\right>$& 2 & 42\\
\hline
\end{tabular}
\caption{\label{tab:cases}Summary of the data assimilation experiments that are performed in this study. For each case, the table reports the friction Reynolds number $\mathrm{Re}_{\tau}$, the control vector $\boldsymbol{\gamma}$, the grid designation (see table \ref{tab:grid_resolutions} for resolution), the observations $\boldsymbol{m}$, the number of iterations $N_{\mathrm{it}}$, and the total number of LES $N_{\mathrm{CFD}}=N_{\mathrm{it}}\times (N_{\mathrm{ens}}+1)$ when $N_{\mathrm{ens}}=20$.  Cases 5, 5c and 6 are also revisited in \S\ref{sec:further_assessment} with alternative data assimilation parameters ($N_{\mathrm{ens}}=10$ and $N_{\mathrm{it}}=1$, so $N_{\mathrm{CFD}}=11$), and in \S\ref{sec:results_DA_statistics_uncertainties}.}
\end{center}
\end{table}

\section{Data assimilation based on filtered observations}\label{sec:results_DA_filtered}

\subsection{Optimizing the coefficient $C_{\mathrm{s}}$ (cases 1 and 2)}\label{sec:cases_1_2}

In this section, the data assimilation procedure from \S\ref{sec:DA_methodology} is employed to examine the benefits and limitations of observing the mean subgrid stress tensor to optimize the Smagorinsky model. These data assimilation experiments are therefore in the spirit of a priori testing \citep{Sagaut2006_springer}. In contrast, in \S\ref{sec:results_DA_statistics} data assimilation will be performed with observations of statistical quantities such as the mean flow or the Reynolds stresses, and thus will provide a posteriori assessment of the enhancement to LES predictions.

\begin{figure}
\centering
\begin{subfigure}{.28\linewidth}
\centering
\includegraphics[width=1.\textwidth]{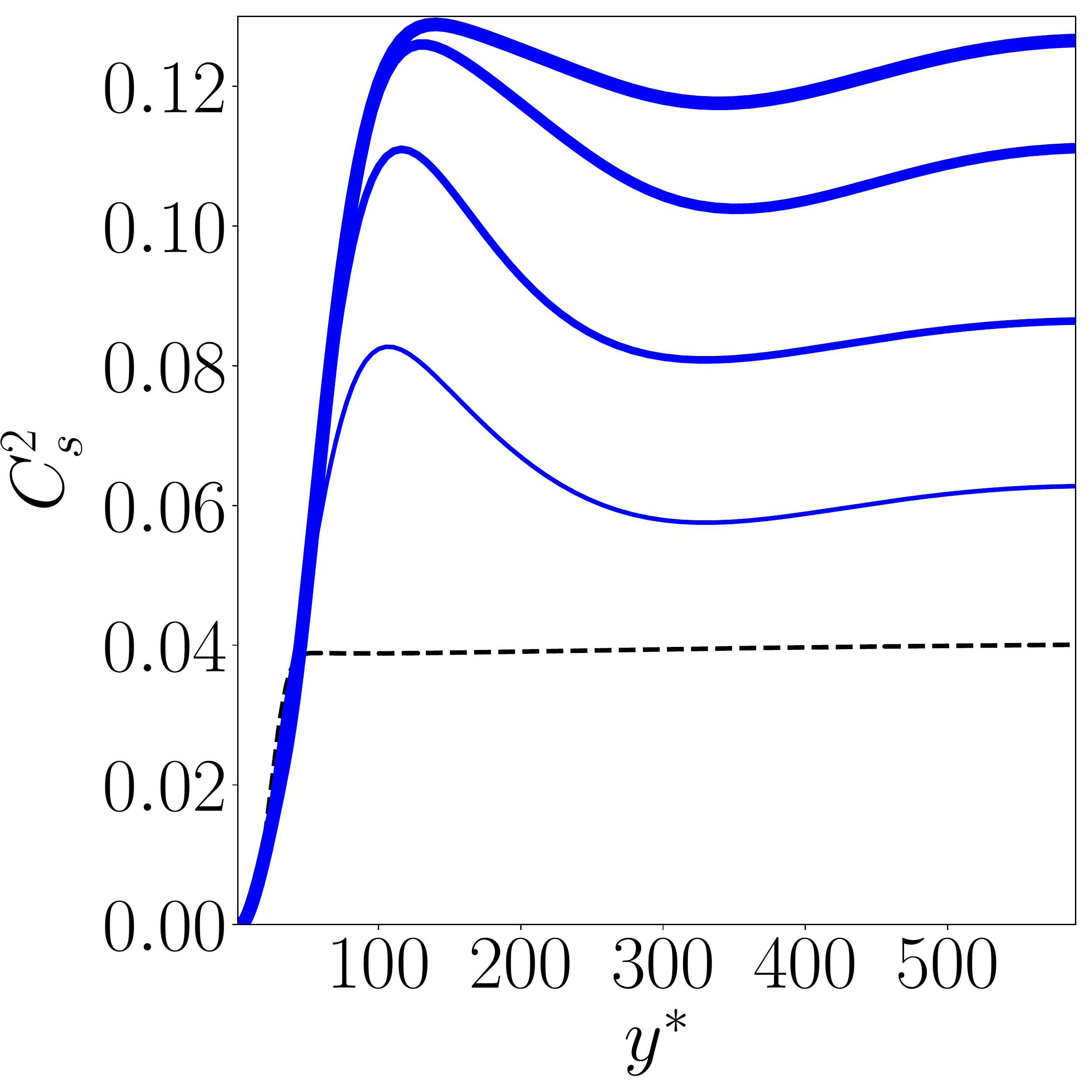}
\caption{\label{fig:Yifan_case1_R_C}Control vector $C_s(y)$}
\end{subfigure}%
\begin{subfigure}{.28\linewidth}
\centering
\includegraphics[width=1.\textwidth]{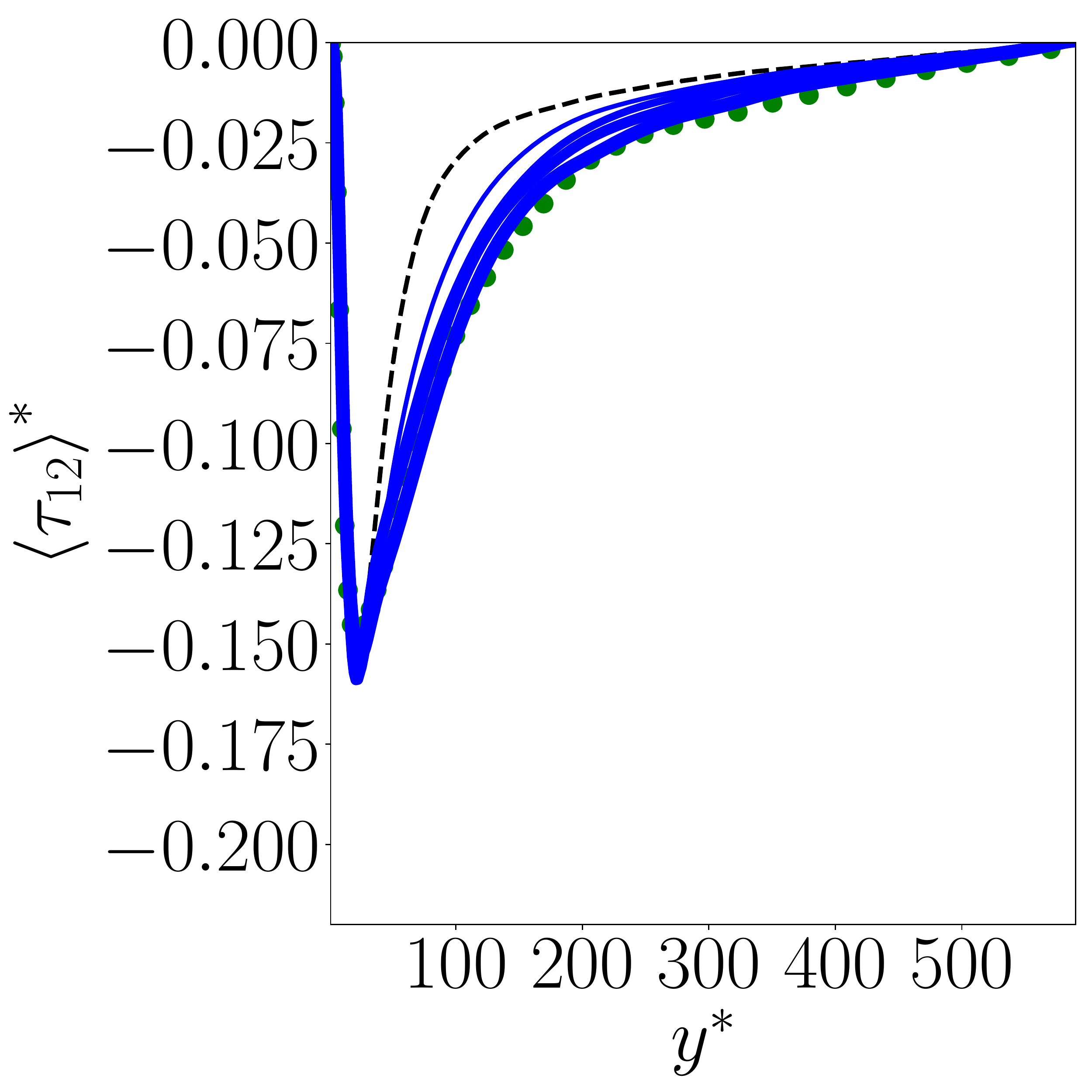}
\caption{\label{fig:Yifan_case1_R_tau12}Subgrid stress $\left<\tau_{12}\right>$}
\end{subfigure}\\
\begin{subfigure}{.28\linewidth}
\centering
\includegraphics[width=1.\textwidth]{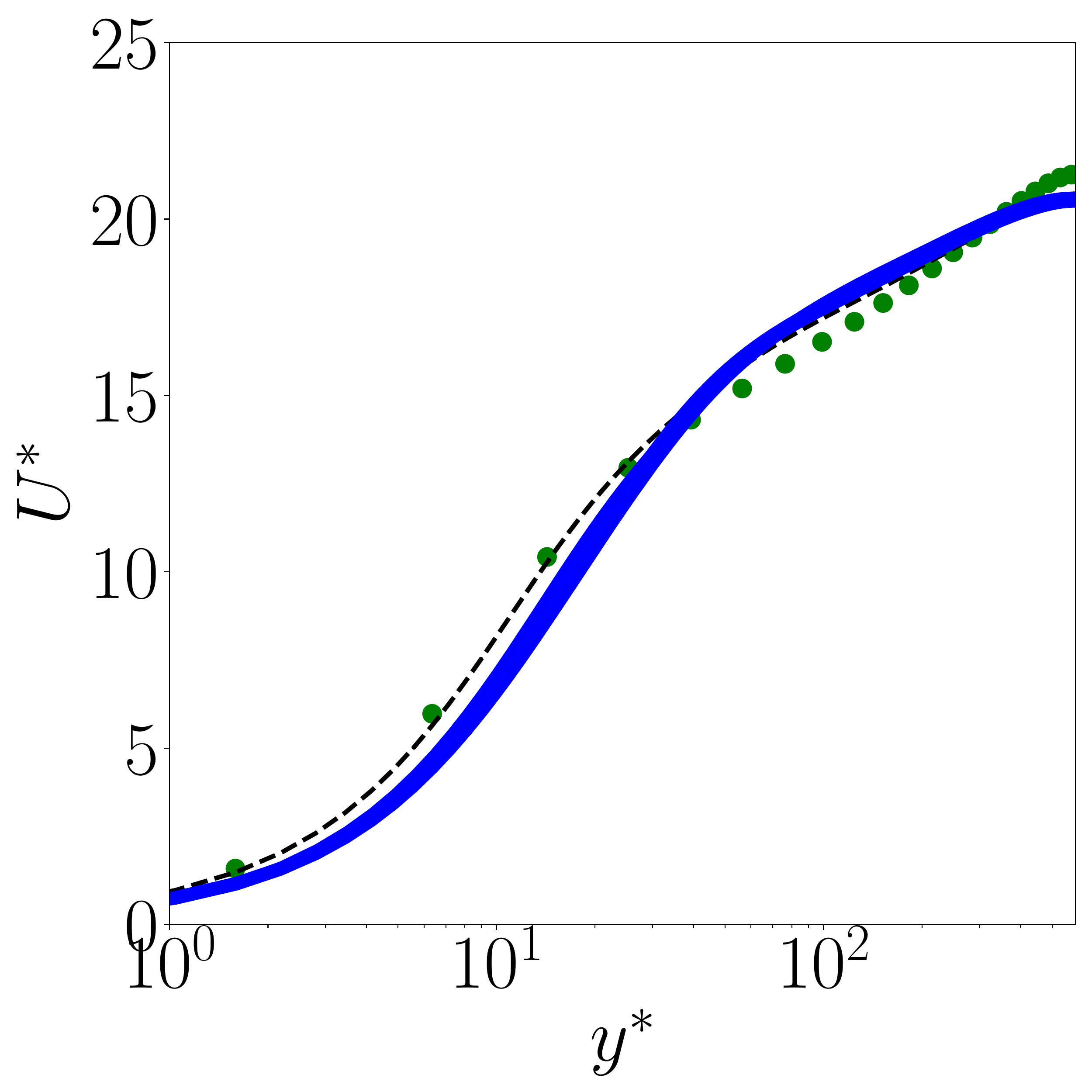}
\caption{\label{fig:Yifan_case1_R_U}Mean profile $U(y)$}
\end{subfigure}
\begin{subfigure}{.28\linewidth}
\centering
\includegraphics[width=1.\textwidth]{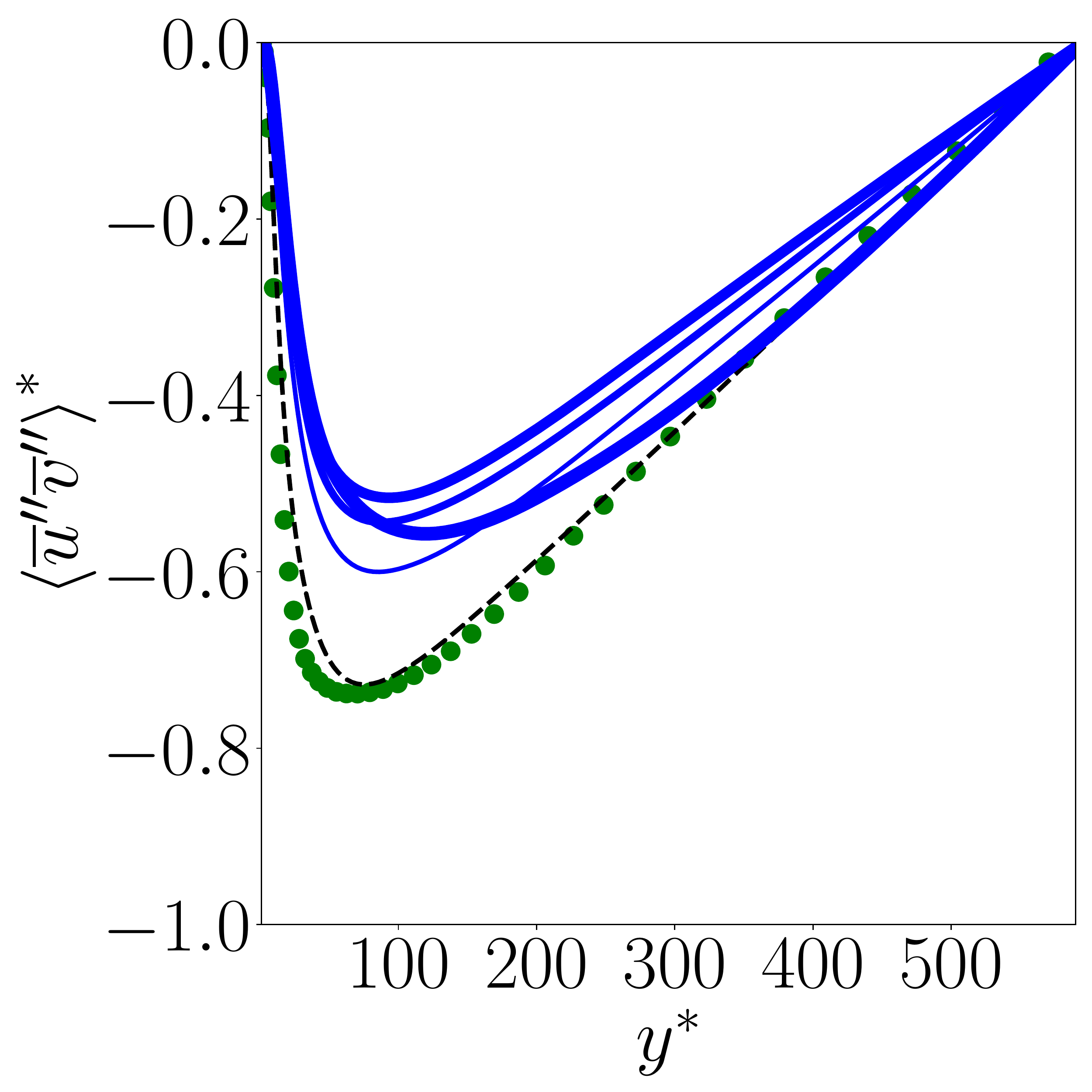}
\caption{\label{fig:Yifan_case1_R_uvr}Resolved shear stress  $\left<\overline{u}''\overline{v}''\right>$}
\end{subfigure}%
\caption{\label{fig:Yifan_case1}Control vector and flow statistics of case 1 for DNS (\mycircle{black!50!green} \mycircle{black!40!green} \mycircle{black!40!green}), the Smagorinsky model (\dashed) and for DA-LES (\full, from thin to thick for increasing iteration of the data assimilation procedure).}
\end{figure}

\begin{figure}
\centering
\begin{subfigure}{.28\linewidth}
\centering
\includegraphics[width=1.\textwidth]{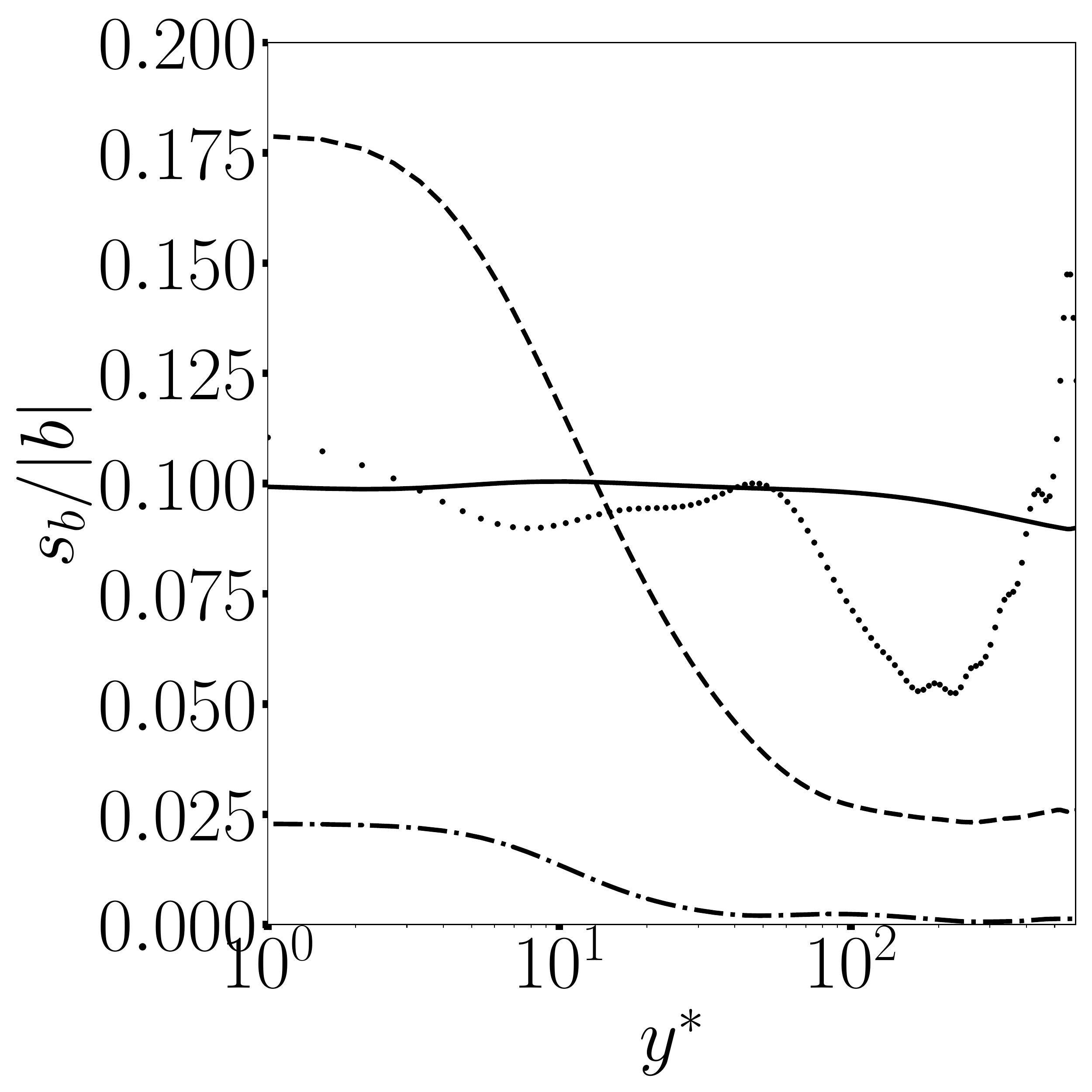}
\caption{\label{fig:case1_uncertainties_std_mean}}
\end{subfigure}\\
\begin{subfigure}{.36\linewidth}
\centering
\includegraphics[width=1\textwidth]{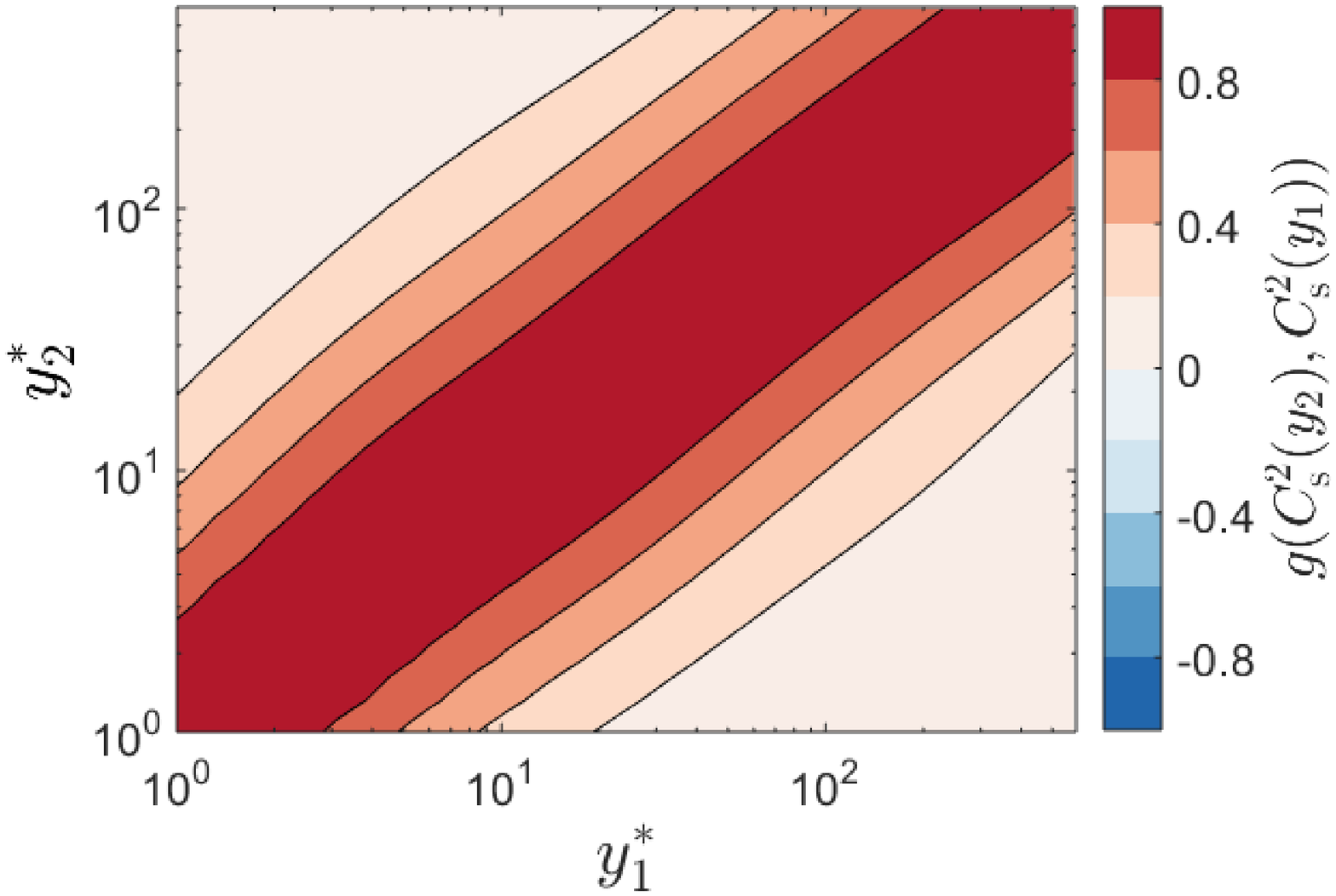}
\caption{\label{fig:case1_uncertainties_g_C_C}}
\end{subfigure}
\begin{subfigure}{.36\linewidth}
\centering
\includegraphics[width=1\textwidth]{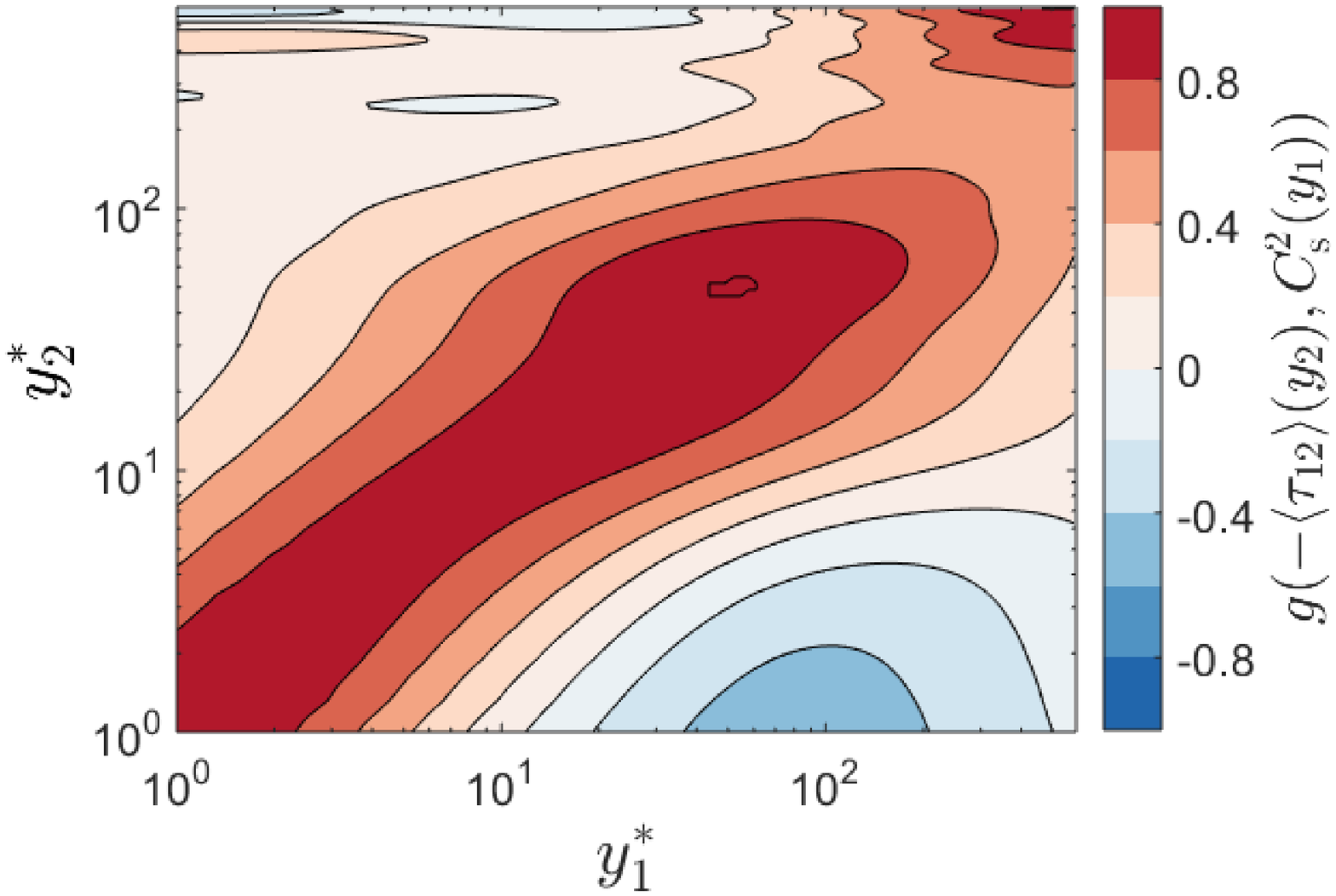}
\caption{\label{fig:case1_uncertainties_g_tau12_C}}
\end{subfigure}\\
\begin{subfigure}{.36\linewidth}
\centering
\includegraphics[width=1\textwidth]{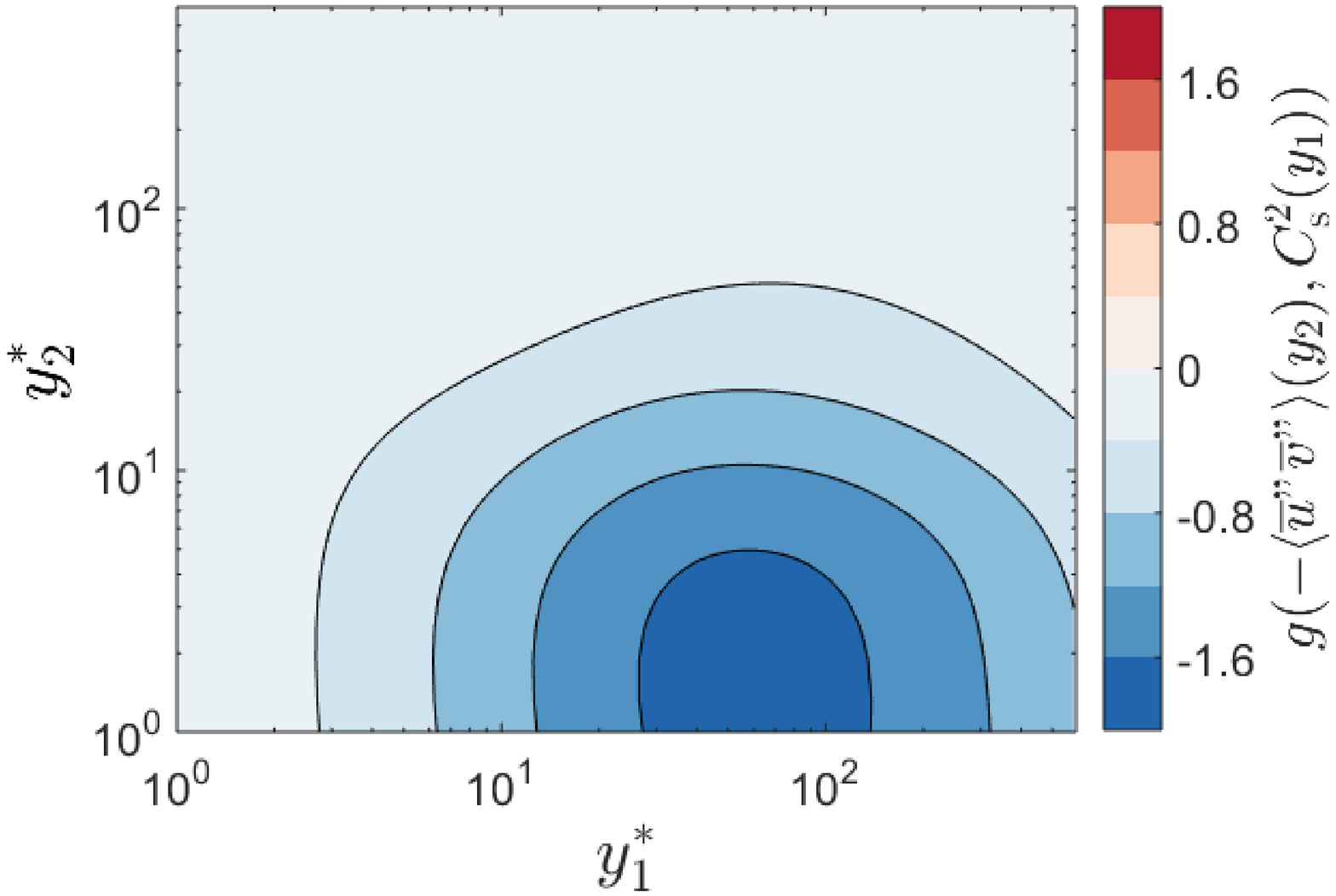}
\caption{\label{fig:case1_uncertainties_g_uvr_C}}
\end{subfigure}
\begin{subfigure}{.36\linewidth}
\centering
\includegraphics[width=1.\textwidth]{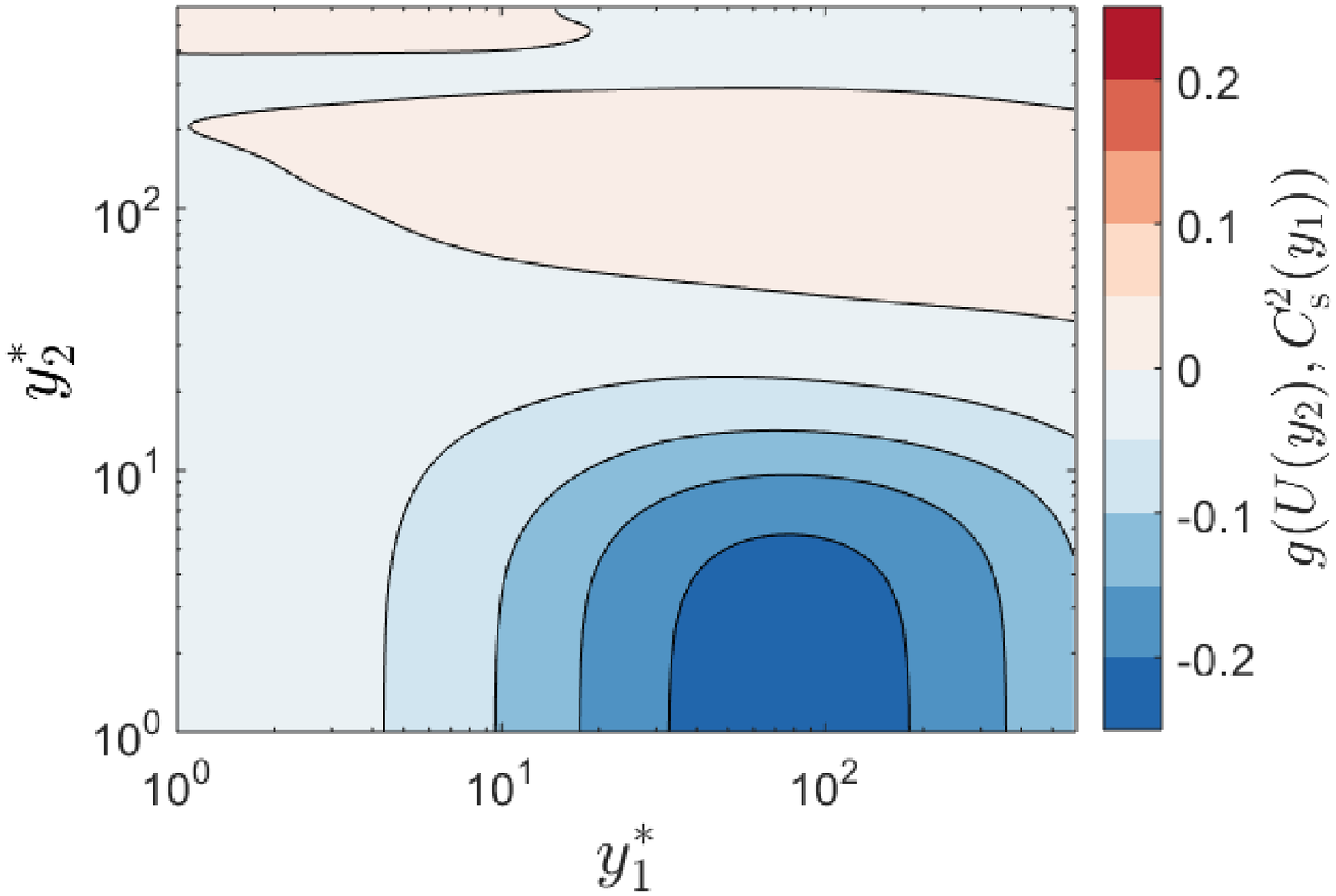}
\caption{\label{fig:case1_uncertainties_g_U_C}}
\end{subfigure}
\caption{\label{fig:uncertainties_case1}Statistics of the first ensemble of case 1: (a) standard deviation $s_b$ of a quantity $b$ normalized by its magnitude at the same $y$-location where $b$ refers to $C^2_{\mathrm{s}}$ (\fullblack), $\left<\tau_{12}\right>$ (\dotted), $\left< \overline{u}''\overline{v}''\right>$ (\dashed) or $U$ (\chainblack); (b)-(e) nondimensional gradient $g(b(y_2),C^2_{\mathrm{s}}(y_1))$ in (\ref{eq:pseudo_gradient_def}) where $b$ refers to $C^2_{\mathrm{s}}$ (b), $(-\left<\tau_{12}\right>)$ (c), $(-\left< \overline{u}''\overline{v}''\right>)$ (d) or $U$ (e).}
\end{figure}

We first consider the data assimilation experiment denoted case $1$ in table \ref{tab:cases}. As in the majority of the calculations, the targeted friction Reynolds number is $Re_{\tau}=590$. We here rely on the finest LES grid in table \ref{tab:grid_resolutions} (LES590f) for this value of $Re_{\tau}$. The aim of this first data assimilation experiment is to improve the prediction of the mean subgrid stress tensor. As discussed in \S\ref{sec:minimal_correction_LES}, only the mean of the subgrid shear stress $\tau_{12}$ can directly affect the mean flow $U$, which may also be inferred from the governing equation for $U$ obtained by averaging (\ref{eq:sigma}), 
\begin{equation}\label{eq:meanflow_equation}
\nu\frac{\mathrm{d}^2U}{\mathrm{d}y^2} - \frac{\mathrm{d}P}{\mathrm{d}x} - \frac{\mathrm{d}}{\mathrm{d}y}\left( \left< \overline{u}''\overline{v}''\right> + \left<\tau_{12}\right>+\sigma \right)  = 0, 
\end{equation}
where $\mathrm{d}P/\mathrm{d}x$ is the fixed mean pressure gradient, and $\left< \overline{u}''\overline{v}''\right>$ is the mean resolved shear stress. 
Furthermore, it is known that the Smagorinsky model predicts essentially zero diagonal subgrid stresses \citep{Domaradzki1997_pof}. Accordingly, the observation vector $\boldsymbol{m}$ is here formed by $\left<\tau_{12}\right>$ alone evaluated from filtering data from DNS (see \S\ref{sec:minimal_correction_LES}). In order to ensure the validity of the discussion regarding the functional limitations of the Smagorinsky model in \S\ref{sec:minimal_correction_LES}, the control vector $\boldsymbol{\gamma}$ is the $C_{\mathrm{s}}(y)$ profile only, and the forcing $\sigma$ is set as zero in all calculations. The results for this first data assimilation experiment are illustrated in figures \ref{fig:Yifan_case1} and \ref{fig:uncertainties_case1}. In these figures, as in the rest of this section, non-dimensionalization is based on the reference friction velocity obtained from DNS. This choice safeguards against misinterpretation of the results, in particular when the mean flow varies primarily near the wall \citep{Majander2002_ijnmf}, as will be the case here.

The first guess of the data assimilation procedure, namely the Smagorinsky model (\ref{eq:smagorinsky_model})-(\ref{eq:smagorinsky_constant}), is illustrated with dashed lines in figure \ref{fig:Yifan_case1}. Below $y^{*}< 20$, the agreement in the mean subgrid stress $\left<\tau_{12}\right>$ between the Smagorinsky model and the reference DNS results is rather satisfactory, as reported in figure \ref{fig:Yifan_case1_R_tau12}. On the other hand, the intensity of $\left<\tau_{12}\right>$ appears underestimated beyond $y^{*}= 20$. We will focus on the ability of the data assimilation procedure to correct this defect through the adjustment of the coefficient $C_{\mathrm{s}}$.

In the first main iteration of the data assimilation procedure, based on (\ref{eq:cov_init})-(\ref{eq:B_eigendecomposition}), $N_{\mathrm{ens}}$ realizations of the profile $C_{\mathrm{s}}(y)$ are generated around the standard Smagorinsky proposal (\ref{eq:smagorinsky_constant}). LES calculations are performed for each of these profiles (steps 1 through 3 in Algorithm \ref{tab:EnVar_algo}). Before proceeding further to analyze the assimilated $C_{\mathrm{s}}(y)$ profile that is the outcome of the remaining steps of the data assimilation procedure, the outputs of these LES calculations are examined in order to determine the sensitivity of the LES statistics to variations in $C_{\mathrm{s}}(y)$. The sensitivity of a quantity $b$ at wall-normal location $y_2$ with respect to variations in the coefficient $C_{\mathrm{s}}$ at $y_1$ is evaluated using 
\begin{equation}\label{eq:pseudo_gradient_def}
    g(b(y_2),C^2_{\mathrm{s}}(y_1))= \frac{\mathrm{cov}(b(y_2),C^2_{\mathrm{s}}(y_1))}{s^2_{C^2_{\mathrm{s}}}(y_1)}\left|\frac{\mathbb{E}(C^2_{\mathrm{s}}(y_1))}{\mathbb{E}(b(y_2))}\right|. 
\end{equation}
In this expression, $\mathrm{cov}$, $\mathbb{E}$ and $s^2$ are the covariance, expectation and variance operators, respectively, which are approximated through ensemble averaging. Equation (\ref{eq:pseudo_gradient_def}) for $g(b(y_2),C^2_{\mathrm{s}}(y_1))$ can be interpreted as a non-dimensional gradient based on the following first-order approximation that relates the variation $\Delta b(y_2)$ to the variation $\Delta C^2_{\mathrm{s}}(y_1)$ that induces the former
\begin{equation}\label{eq:pseudo_gradient_application}
    \frac{\Delta b(y_2)}{|b(y_2)|}\simeq g(b(y_2),C^2_{\mathrm{s}}(y_1))\frac{\Delta C^2_{\mathrm{s}}(y_1)}{|C^2_{\mathrm{s}}(y_1)|}.
\end{equation}

The standard deviation and the non-dimensional gradient in (\ref{eq:pseudo_gradient_def}) evaluated from the first ensemble of case 1 are reported in figure \ref{fig:uncertainties_case1} for the different statistical quantities involved in the mean-flow equation (\ref{eq:meanflow_equation}). Figures \ref{fig:case1_uncertainties_std_mean} and \ref{fig:case1_uncertainties_g_C_C} confirm the efficacy of the ensemble generation procedure that is detailed in \S\ref{sec:DA_parameters}. Despite the moderate size of the ensemble ($N_{\mathrm{ens}}$=20), the realizations of $C^2_{\mathrm{s}}$ recover the statistics prescribed through the covariance matrix $\boldsymbol{\mathrm{B}}$ in (\ref{eq:cov_init}): the value for the standard deviation $s_{C^2_{\mathrm{s}}} = \frac{1}{10}C^2_{\mathrm{s}}$ is satisfactorily recovered over the whole channel (solid line in figure \ref{fig:case1_uncertainties_std_mean}). In addition, the shape of the non-dimensional gradient $g(C^2_{\mathrm{s}}(y_2),C^2_{\mathrm{s}}(y_1))$ in figure \ref{fig:case1_uncertainties_g_C_C} is in agreement with the analytical result $g(C^2_{\mathrm{s}}(y_2),C^2_{\mathrm{s}}(y_1))=\mathrm{exp}(-((f(y_1)-f(y_2))/l_\mathrm{c})^2)$ which can be deduced from the definitions (\ref{eq:cov_init}) and (\ref{eq:pseudo_gradient_def}). Also note that no spurious long-distance correlation have been introduced. These qualities of the ensemble can be attributed to the sampling scheme (\ref{eq:B_eigendecomposition}) based on the dominant eigenvectors of $\boldsymbol{\mathrm{B}}$, and provide confidence in the reported statistics.

Now considering the other results in figure \ref{fig:uncertainties_case1}, it appears that the outputs of the LES simulations are not all similarly affected by changes in $C_{\mathrm{s}}(y)$. The mean subgrid shear stress $(-\left<\tau_{12}\right>)$ is well correlated with $C^2_{\mathrm{s}}(y)$ as illustrated in figure \ref{fig:case1_uncertainties_g_tau12_C}, i.e.\,the reported non-dimensional gradient is positive, which results from the proportional relationship between the two (c.f.\,equation (\ref{eq:smagorinsky_model})). The influence of $C_{\mathrm{s}}$ on $\left<\tau_{12}\right>$ is also mostly local;  one exception is a region very close to the wall ($y^*_2 < 10$) where $(-\left<\tau_{12}\right>)$ appears slightly anti-correlated with variations in $C^2_{\mathrm{s}}$ around $y^*_1= 100$, which is further discussed below. 
In contrast, figure \ref{fig:case1_uncertainties_g_uvr_C} shows that the sensitivity of the resolved shear stress $(-\left< \overline{u}''\overline{v}''\right>)$ to $C_{\mathrm{s}}$ is mostly non-local. The associated non-dimensional gradient is overall negative, as increasing $C_{\mathrm{s}}$ enhances the dissipation of resolved turbulent kinetic energy $\varepsilon^{\mathrm{fs}}$ in (\ref{eq:esgs_decomposition}). Only variations in $C^2_{\mathrm{s}}$ in the range $30 \leq y^*_1 \leq 150$ seem to have a significant impact on $(-\left< \overline{u}''\overline{v}''\right>)$ and only for $y^*_2 < 30$. The same applies for the mean flow $U$, although the amplitude of $g(U(y_2),C^2_{\mathrm{s}}(y_1))$ is significantly smaller compared to the previously discussed non-dimensional gradients. This relatively low sensitivity is also reflected in the small values for the standard deviation of $U$ as reported in figure \ref{fig:case1_uncertainties_std_mean}, which becomes almost negligible beyond $y^*=30$. Figure \ref{fig:case1_uncertainties_g_U_C} also indicates that increasing $C_{\mathrm{s}}$ induces a decrease of $U$, and thus of $\mathrm{d} U/\mathrm{d} y$, close to the wall, which arises from the relationship between $(-\left< \overline{u}''\overline{v}''\right>)$ and $U$ through the mean-flow equation (\ref{eq:meanflow_equation}). This might explain the slight anti-correlation between $(-\left<\tau_{12}\right>)$ and $C^2_{\mathrm{s}}$ that was noted above, as a decrease in $\mathrm{d} U/\mathrm{d} y$ should result in lower values for the contribution $\left\langle \overline{S}_{ij}(2\overline{S}_{kl}\overline{S}_{kl})^{\frac{1}{2}}\right\rangle$ when averaging the Smagorinsky form (\ref{eq:smagorinsky_model}). In summary, the present analysis suggests that, apart from $\left<\tau_{12}\right>$, only the flow statistics in the near-wall region below $y^*=30$ may be significantly impacted by variations in $C_{\mathrm{s}}$, and only if the latter occur in the range $30 \leq y^* \leq 150$. This result and its consequences will be revisited during discussions of the data assimilation experiments.

After having investigated the sensitivity of the flow to $C_{\mathrm{s}}(y)$ using the first ensemble from case 1, we turn to the assimilated states that were computed using the subsequent steps of the data assimilation procedure. The outcome from the first ensemble (steps 4-5 in Algorithm \ref{tab:EnVar_algo}) is reported with the thinnest blue lines in figure \ref{fig:Yifan_case1}. While enhancing the prediction of the subgrid shear stress $\left<\tau_{12}\right>$ compared to the standard Smagorinsky model, the stress remains underestimated compared to the filtered DNS. To further improve this result, three more iterations of the data assimilation procedure were performed (step 6 in Algorithm \ref{tab:EnVar_algo}). As reported in figure \ref{fig:Yifan_case1_R_tau12}, the data assimilation procedure successfully recovers the reference profile for the mean subgrid shear stress $\left<\tau_{12}\right>$, the final DA-LES results being illustrated with the thickest blue line. It appears from figure \ref{fig:Yifan_case1_R_C} that this objective has been reached through a large increase in $C_{\mathrm{s}}$ over the entire channel height. In particular, the value of $C_{\mathrm{s}}$ above $y^{+}=30$ has become significantly larger than the theoretical value for isotropic turbulence $C_{\mathrm{s}}^2 \sim 0.04$ for the final assimilated state. This large $C_{\mathrm{s}}$ is indicative of the inadequacy of considering the reference $\left<\tau_{12}\right>$ as the sole objective of the data assimilation procedure while only adjusting $C_{\mathrm{s}}$. This view is confirmed by the examination of figures \ref{fig:Yifan_case1_R_U}-\ref{fig:Yifan_case1_R_uvr} where we report the mean flow and the mean resolved shear stress profiles. The improvement in terms of $\left<\tau_{12}\right>$ for DA-LES has been obtained at the expense of a deterioration in the predicted mean flow, which is significantly underestimated in the viscous wall region, while it was correctly evaluated by the standard Smagorinsky model. The large increase in $C_{\mathrm{s}}$ has induced a similar enhancement in the dissipation of resolved turbulent kinetic energy $\varepsilon^{\mathrm{fs}}$ in (\ref{eq:esgs_decomposition}), inducing a decrease in $(-\left< \overline{u}''\overline{v}''\right>)$ and thus in $U$ close to the wall. It can also be noted based on figure \ref{fig:Yifan_case1_R_U} that the mean flow has barely been altered above $y^*=30$ despite the large variations in $C_{\mathrm{s}}$ relative to its first-guess profile. All these results are in accordance with the discussion of figure \ref{fig:uncertainties_case1}.

\begin{figure}
\centering
\begin{subfigure}{.28\linewidth}
\centering
\includegraphics[width=1.\textwidth]{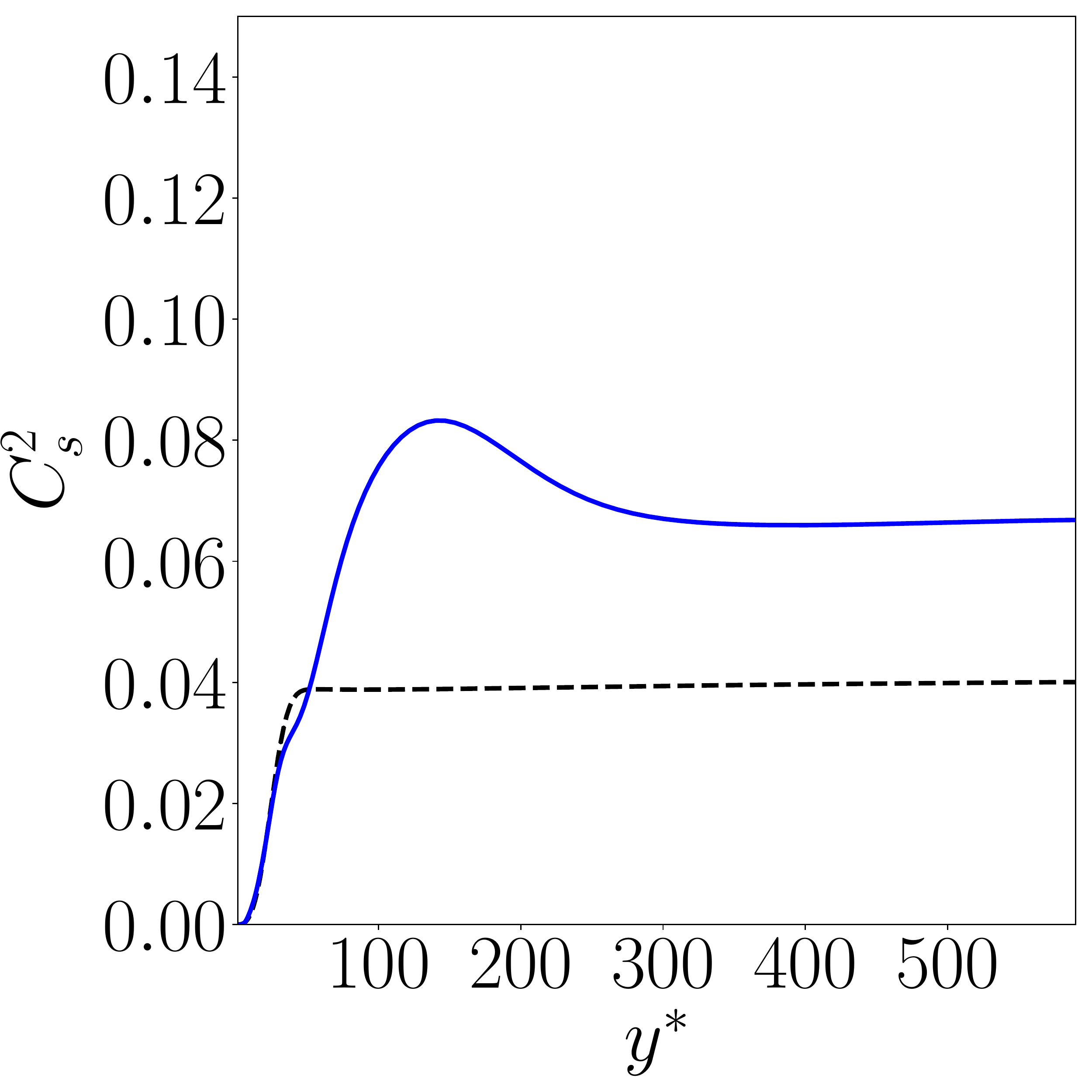}
\caption{Control vector $C_s(y)$}
\end{subfigure}%
\begin{subfigure}{.28\linewidth}
\centering
\includegraphics[width=1.\textwidth]{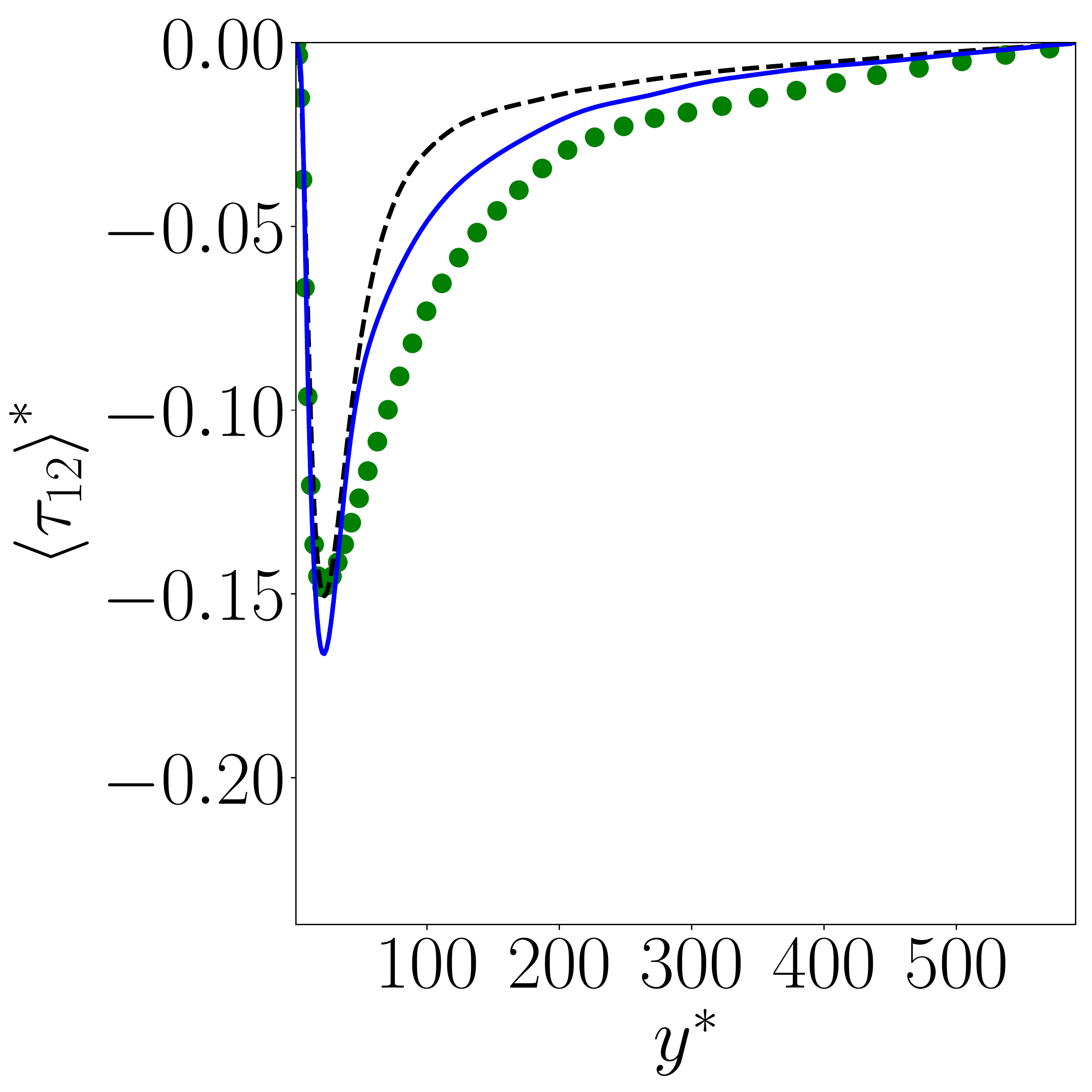}
\caption{Subgrid stress $\left<\tau_{12}\right>$}
\end{subfigure}\\
\begin{subfigure}{.28\linewidth}
\centering
\includegraphics[width=1.\textwidth]{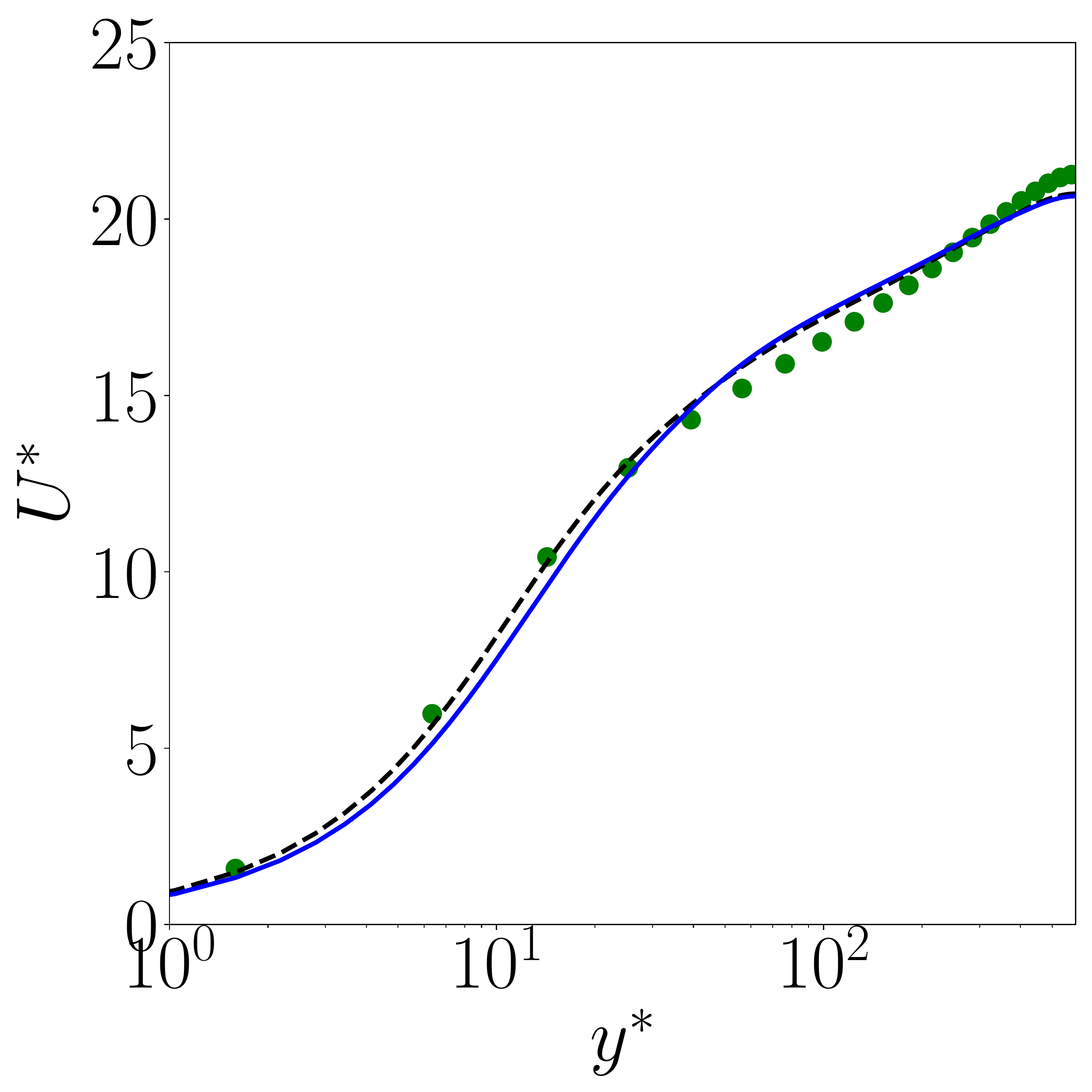}
\caption{Mean profile $U(y)$}
\end{subfigure}
\begin{subfigure}{.28\linewidth}
\centering
\includegraphics[width=1.\textwidth]{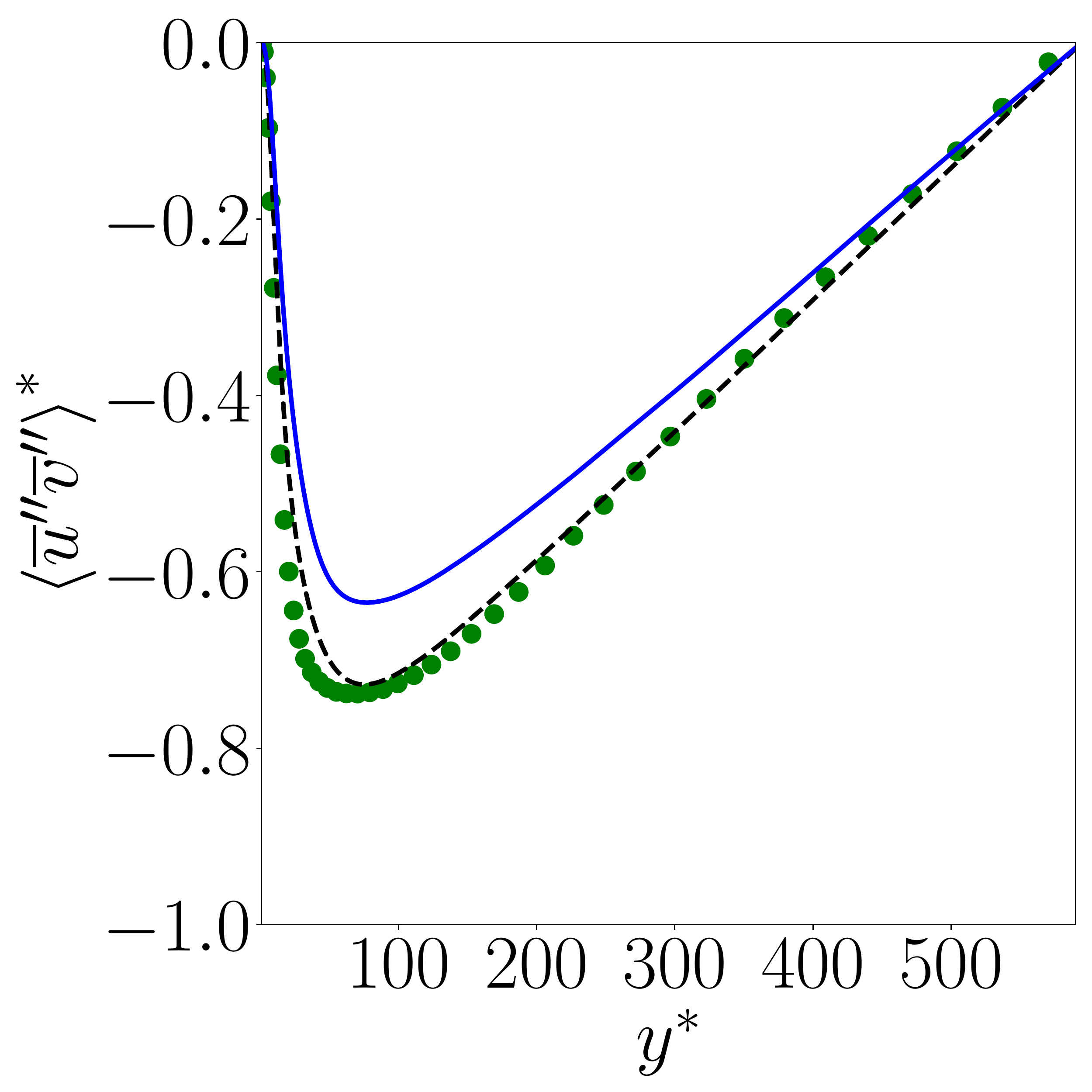}
\caption{Resolved shear stress $\left<\overline{u}''\overline{v}''\right>$}
\end{subfigure}%
\caption{\label{fig:Yifan_case2}Control vector and flow statistics of case 2 for DNS (\mycircle{black!50!green} \mycircle{black!40!green} \mycircle{black!40!green}), the Smagorinsky model (\dashed) and for final DA-LES (\full).}
\end{figure}

Since use of $\left<\tau_{12}\right>$ alone as observation in the data assimilation procedure is detrimental to the accuracy of estimation of other quantities, we consider adding other types of observations. In the data assimilation experiment 2, the observations $\boldsymbol{m}$ include both the subgrid shear stress $\left<\tau_{12}\right>$ and information regarding the mean flow $U$. 
Specifically, the gradient $\mathrm{d} U/\mathrm{d} y$ is used in the cost function $J$ in (\ref{eq:cost_function}) so that, when discretization in the wall-normal direction is taken into account through the covariance matrix $\boldsymbol{\mathrm{R}}$ in (\ref{eq:observation_covariance}), observations close to the wall are of similar magnitude as those towards the channel center. 
In addition, the components of $\boldsymbol{m}$ are normalized so that the contributions of $\left<\tau_{12}\right>$ and $\mathrm{d} U/\mathrm{d} y$ have the same weight in the cost function. The results for this data assimilation experiment are reported in figure \ref{fig:Yifan_case2}. Only one iteration of the data assimilation procedure was performed and it was sufficient to highlight the difficulties in matching both $\left<\tau_{12}\right>$ and $\mathrm{d} U/\mathrm{d} y$ when only adjusting $C_{\mathrm{s}}$. The mean flow for DA-LES slightly deteriorates compared to the standard Smagorinsky model, but to a lesser extent compared to the previously discussed case 1. On the other hand, some improvement is achieved in matching $\left<\tau_{12}\right>$, even if the discrepancies from the reference profile remain significant. The difficulties in satisfactorily correcting the LES predictions in this data assimilation experiment may be quantified thanks to the linearized cost function $\tilde{J}$ in (\ref{eq:cost_function_envar}). Since $\tilde{J}$ is quadratic, its gradient with respect to the control vector $\boldsymbol{w}$, which arises from the ensemble representation of the original control vector $\boldsymbol{\gamma}$ in (\ref{eq:ensemble_representation}), is straightforward to compute. 
Specifically, we separate the observation term in $\tilde{J}$ into $\tilde{J}_{\left<\tau_{12}\right>}$ for observing $\left<\tau_{12}\right>$ and $\tilde{J}_{dU/dy}$ for observing $\mathrm{d} U/\mathrm{d} y$; we then compute their gradients with respect to $\boldsymbol{w}$ at $\boldsymbol{w}=\boldsymbol{0}$ (i.e.\,at the first guess). The normalized scalar product of the gradients is
\begin{equation}\label{eq:gradients_J_tau12_dUdy}
    \frac{\nabla \tilde{J}_{dU/dy} \cdot \nabla \tilde{J}_{\left<\tau_{12}\right>}}{\left\Vert\nabla \tilde{J}_{dU/dy}\right\Vert\left\Vert \nabla \tilde{J}_{\left<\tau_{12}\right>}\right\Vert} = -0.19,
\end{equation}
which confirms that matching the mean flow and the subgrid shear stress are competitive objectives. The same outcomes applies if, instead of adding the observation of $dU/dy$, the resolved shear stress $\left< \overline{u}''\overline{v}''\right>$ had been considered in conjunction with $\left<\tau_{12}\right>$, and a similar analysis would have led to
\begin{equation}\label{eq:gradients_J_tau12_ruv}
    \frac{\nabla \tilde{J}_{\left< \overline{u}''\overline{v}''\right>} \cdot \nabla \tilde{J}_{\left<\tau_{12}\right>}}{\left\Vert\nabla \tilde{J}_{\left< \overline{u}''\overline{v}''\right>}\right\Vert\left\Vert \nabla \tilde{J}_{\left<\tau_{12}\right>}\right\Vert} = -0.15.
\end{equation}
Equations (\ref{eq:gradients_J_tau12_dUdy}) and (\ref{eq:gradients_J_tau12_ruv}) indicate that by targeting $\left<\tau_{12}\right>$ in the data assimilation problem, it is not possible to simultaneously improve the prediction of the other terms in the governing equation for the mean flow (\ref{eq:meanflow_equation}). This point is in accordance with previous discussions on the sensitivity of the LES statistics with respect to changes in $C_{\mathrm{s}}$ and the fact that $\left< \overline{u}''\overline{v}''\right>$ and $U$ are relatively well predicted by the standard Smagorinsky model, in particular close to the wall. The present data assimilation has thus only found the best compromise between an accurate prediction of $\left<\tau_{12}\right>$ and a correct estimation of the other terms in the mean flow equation (\ref{eq:meanflow_equation}). 

The present difficulty in correctly reproducing multiple statistical quantities when only adjusting $C_{\mathrm{s}}$ seems somewhat analogous to findings in \citep{franceschini2020_prf} in the context of data assimilation with RANS models when preserving the Boussinesq hypothesis. In that case, the authors noted that the considered model correction may be too conservative and prevents reaching many possible flow states, 
making a correct finely-detailed estimation of the observed reference flow difficult. In a similar way, in the present context, correctly estimating the different contributions in the mean-flow equation (\ref{eq:meanflow_equation}) requires more flexibility in the adjustment of the subgrid model than offered by variations in the coefficient $C_{\mathrm{s}}$ alone.

\subsection{Optimizing the coefficient $C_{\mathrm{s}}$ and the forcing $\sigma$ (cases 3 and 3c)}\label{sec:cases_3_3c}

\begin{figure}
\centering
\begin{subfigure}{.28\linewidth}
\centering
\includegraphics[width=1.\textwidth]{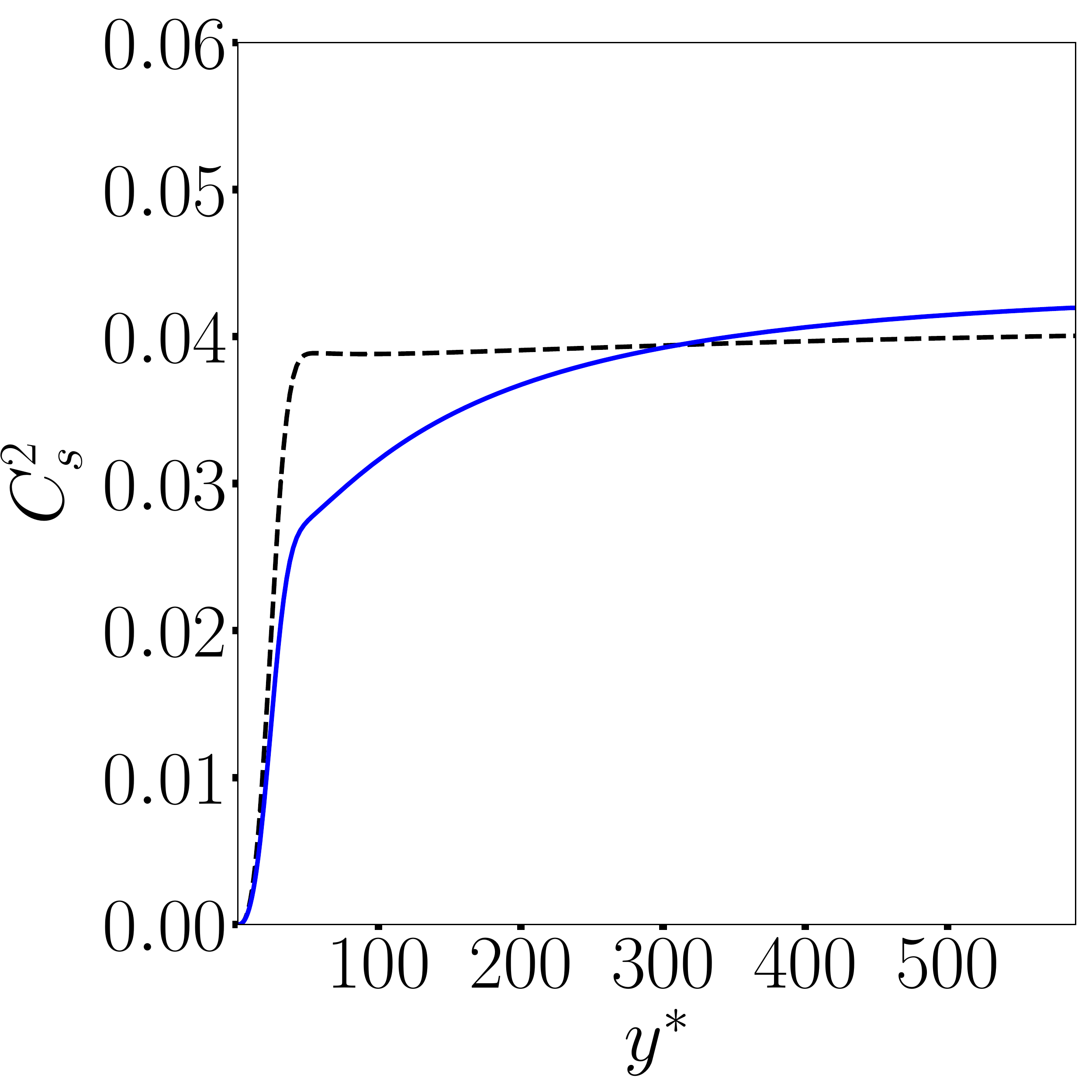}
\caption{\label{fig:Yifan_case3_R_C}Control vector $C_s(y)$}
\end{subfigure}%
\begin{subfigure}{.28\linewidth}
\centering
\includegraphics[width=1.\textwidth]{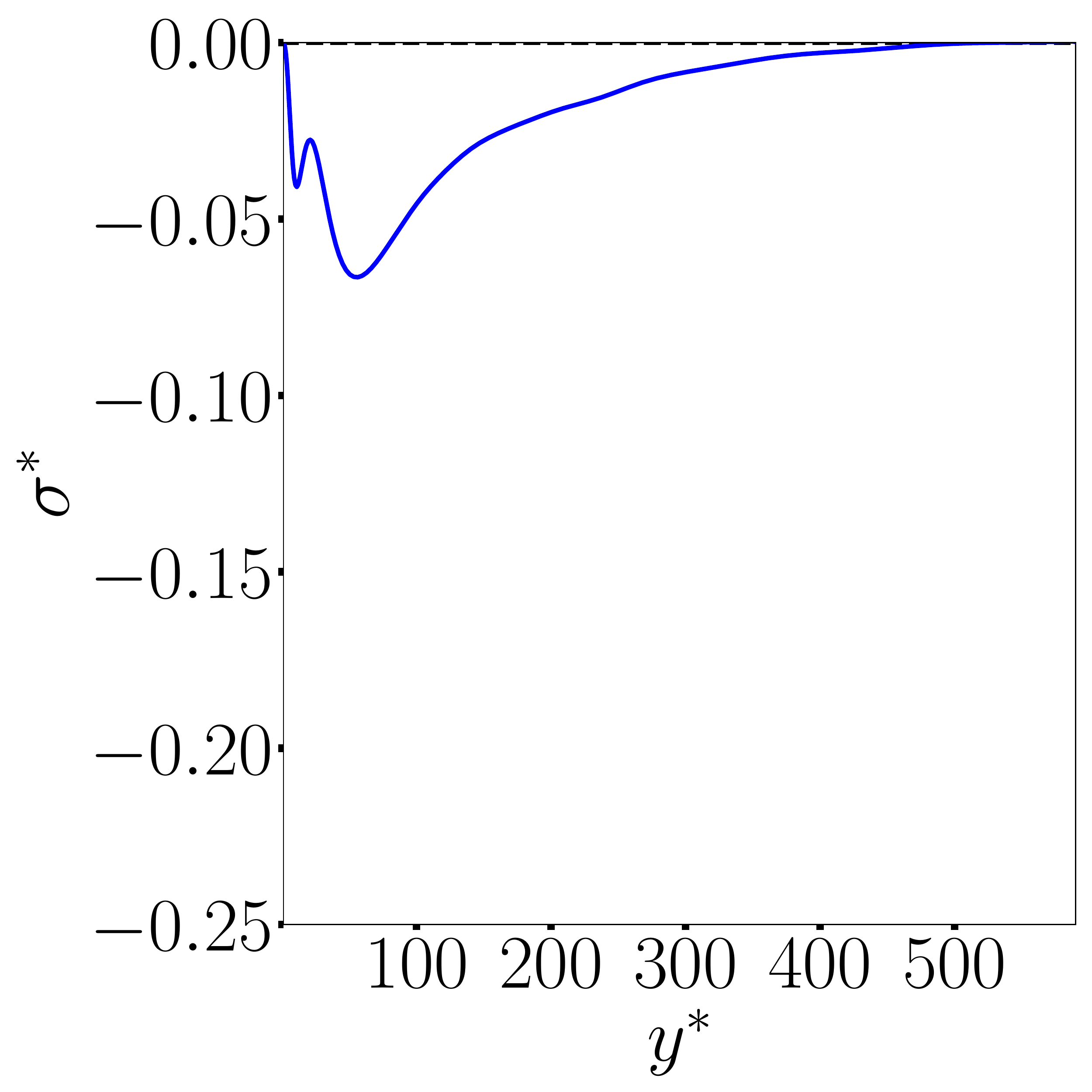}
\caption{\label{fig:Yifan_case3_R_sigma}Control vector $\sigma(y)$}
\end{subfigure}
\begin{subfigure}{.28\linewidth}
\centering
\includegraphics[width=1.\textwidth]{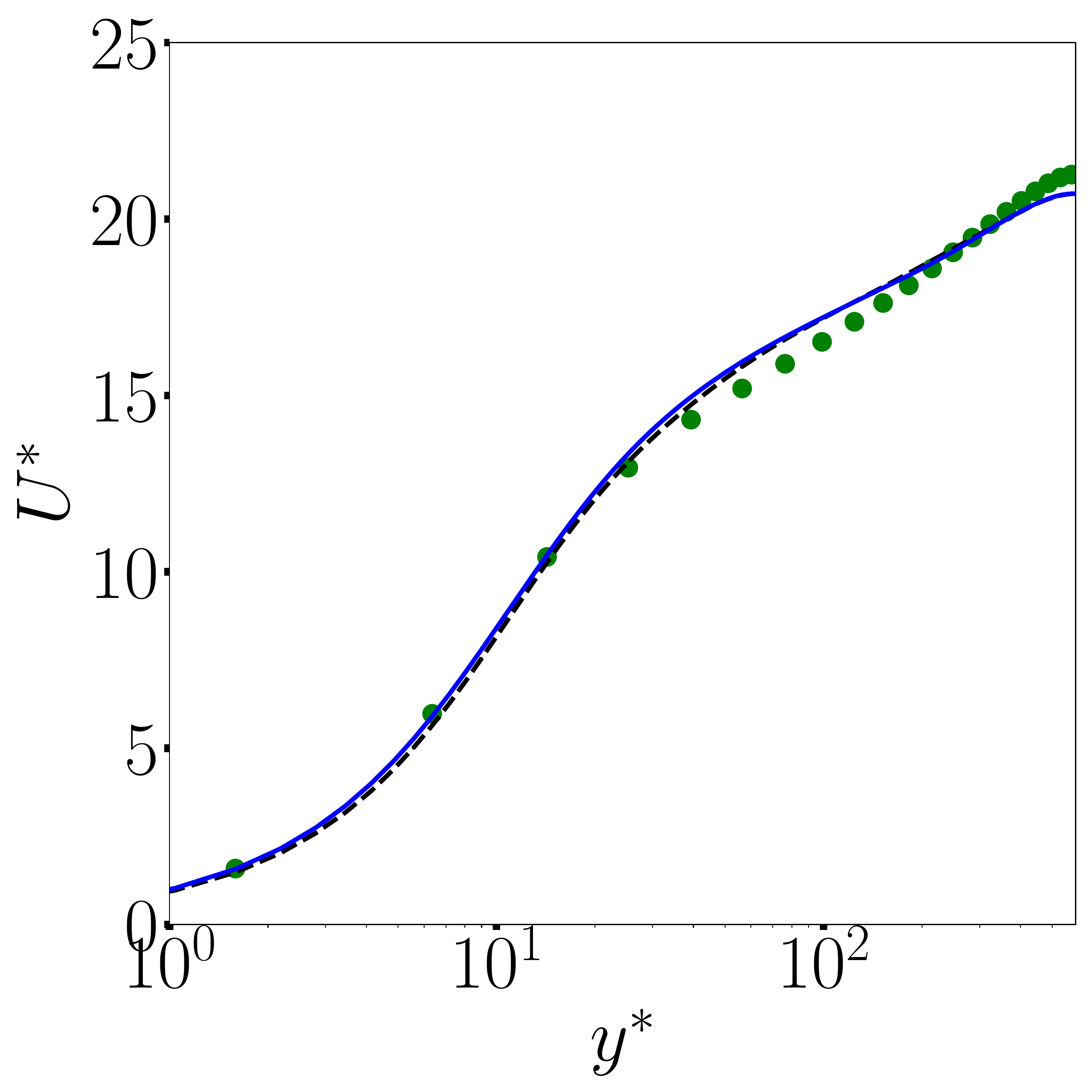}
\caption{\label{fig:Yifan_case3_R_U}Mean profile $U(y)$}
\end{subfigure}
\begin{subfigure}{.28\linewidth}
\centering
\includegraphics[width=1.\textwidth]{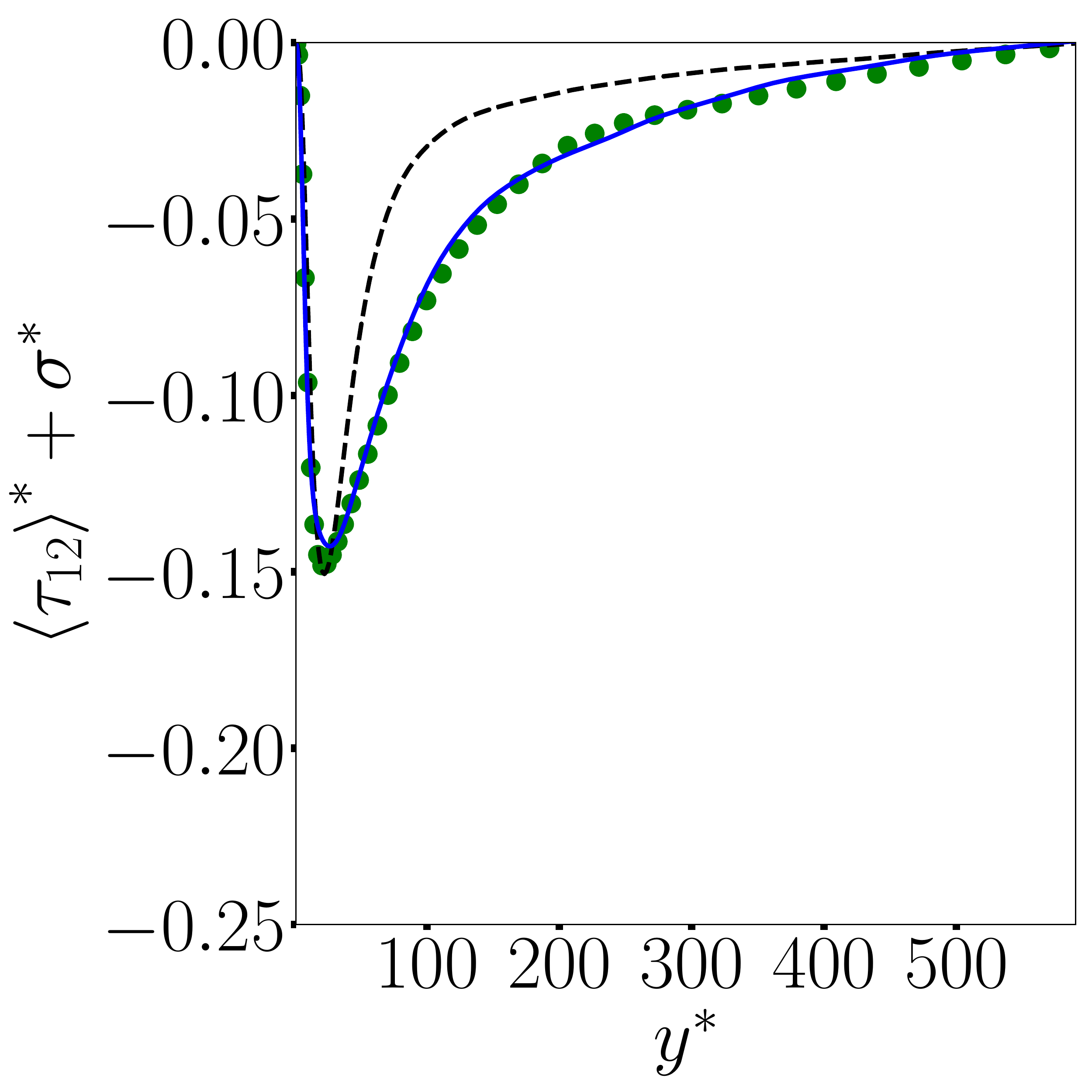}
\caption{\label{fig:Yifan_case3_R_tau12}Subgrid stress $\left<\tau_{12}\right>$}
\end{subfigure}
\begin{subfigure}{.28\linewidth}
\centering
\includegraphics[width=1.\textwidth]{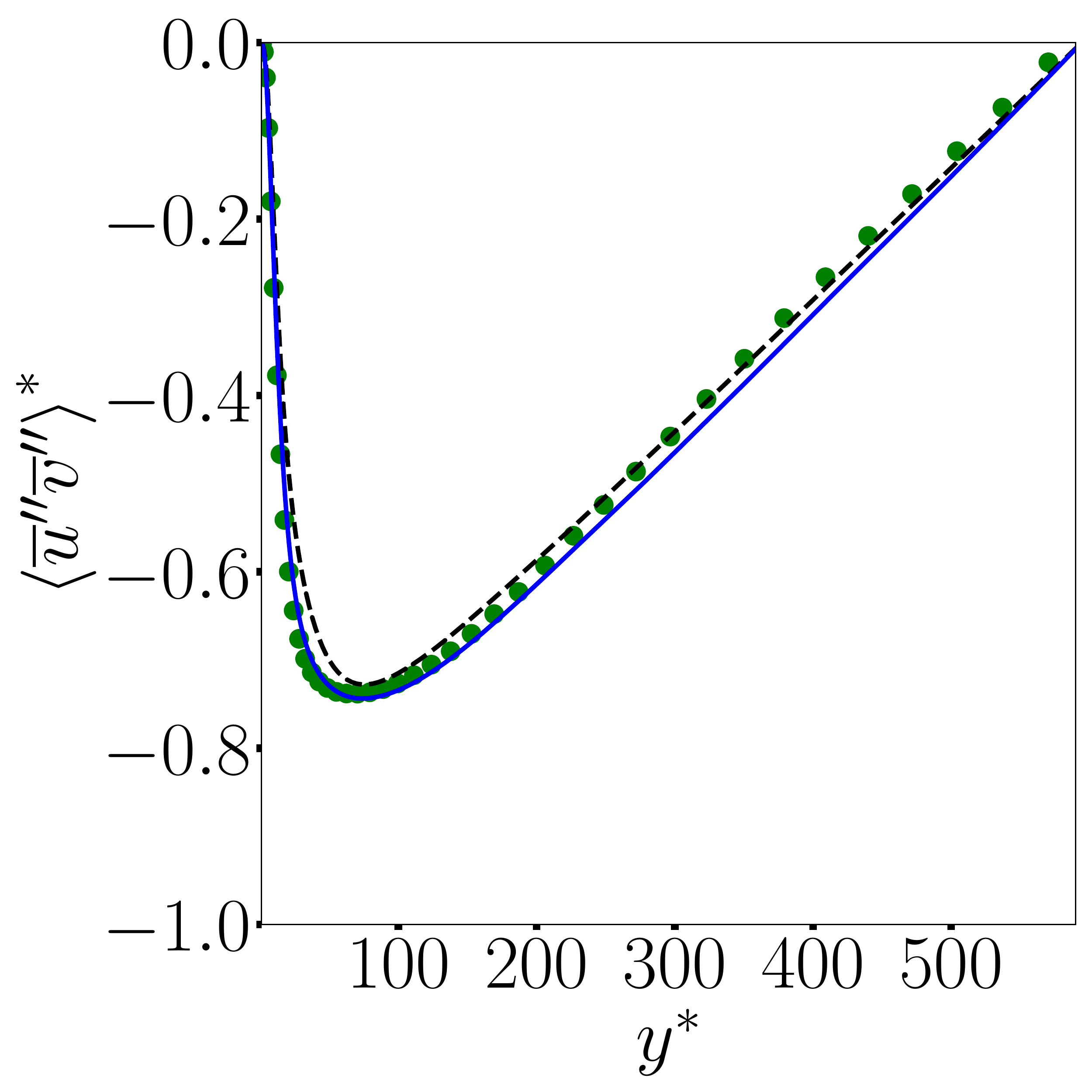}
\caption{\label{fig:Yifan_case3_R_uvr}Resolved shear stress $\left<\overline{u}''\overline{v}''\right>$}
\end{subfigure}%
\begin{subfigure}{.28\linewidth}
\centering
\includegraphics[width=1.\textwidth]{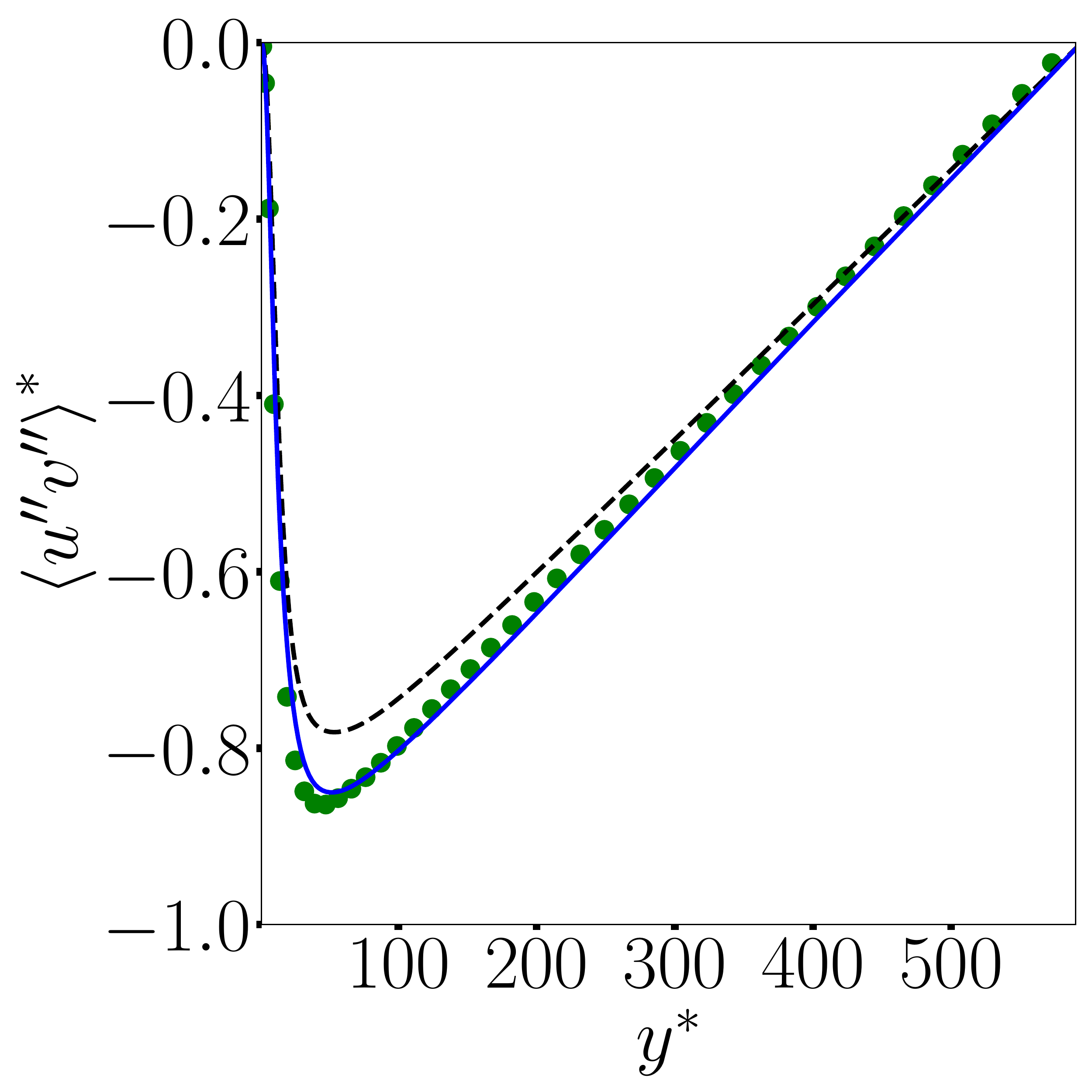}
\caption{\label{fig:Yifan_case3_R_uv}Total shear stress $\left<u''v''\right>$}
\end{subfigure}%
\caption{\label{fig:Yifan_case3}Control vector and flow statistics of case 3 for DNS (\mycircle{black!50!green} \mycircle{black!40!green} \mycircle{black!40!green}), the Smagorinsky model (\dashed) and for final DA-LES (\full).}
\end{figure}

\begin{figure}
\centering
\begin{subfigure}{.28\linewidth}
\centering
\includegraphics[width=1.\textwidth]{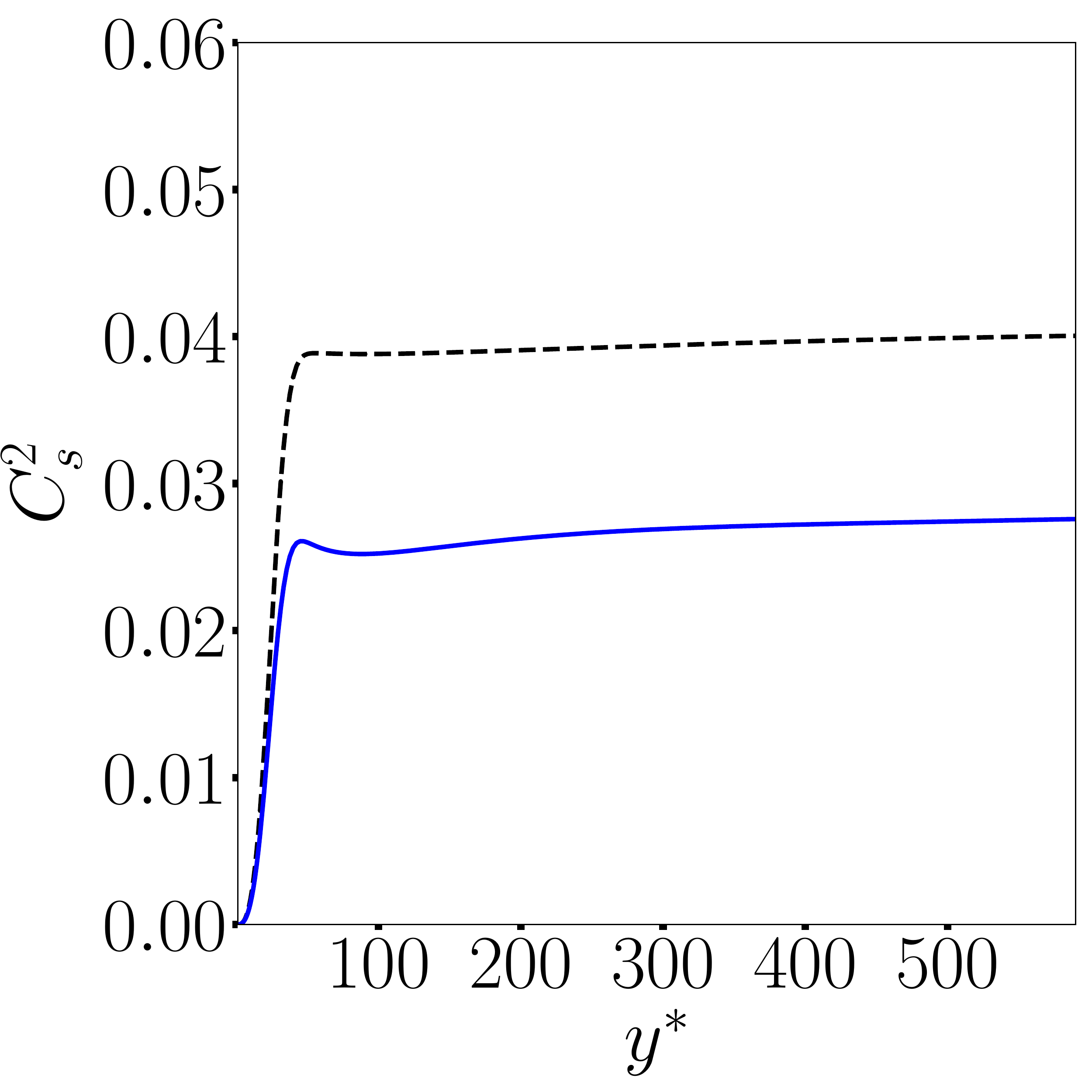}
\caption{Control vector $C_s(y)$}
\end{subfigure}%
\begin{subfigure}{.28\linewidth}
\centering
\includegraphics[width=1.\textwidth]{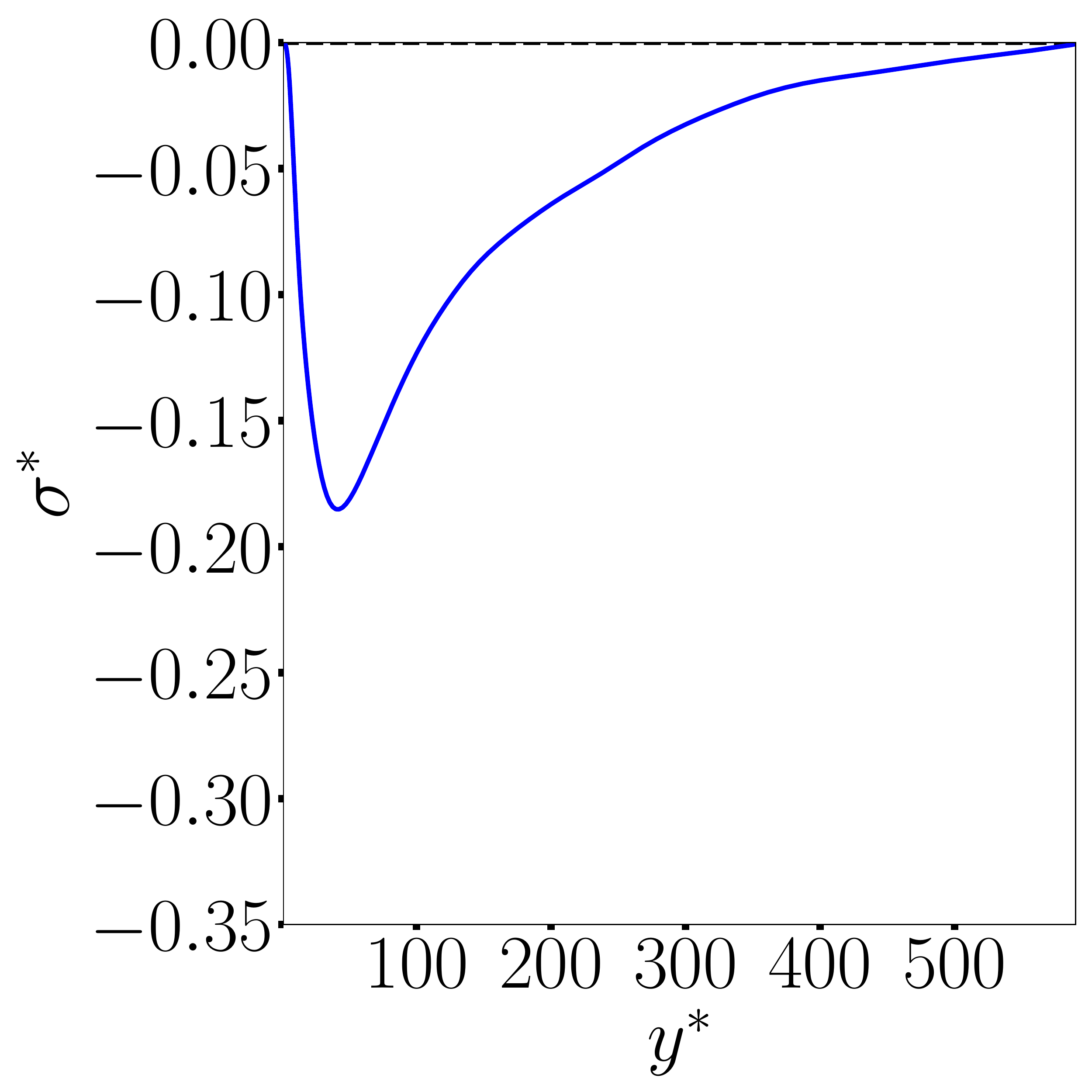}
\caption{Control vector $\sigma(y)$}
\end{subfigure}
\begin{subfigure}{.28\linewidth}
\centering
\includegraphics[width=1.\textwidth]{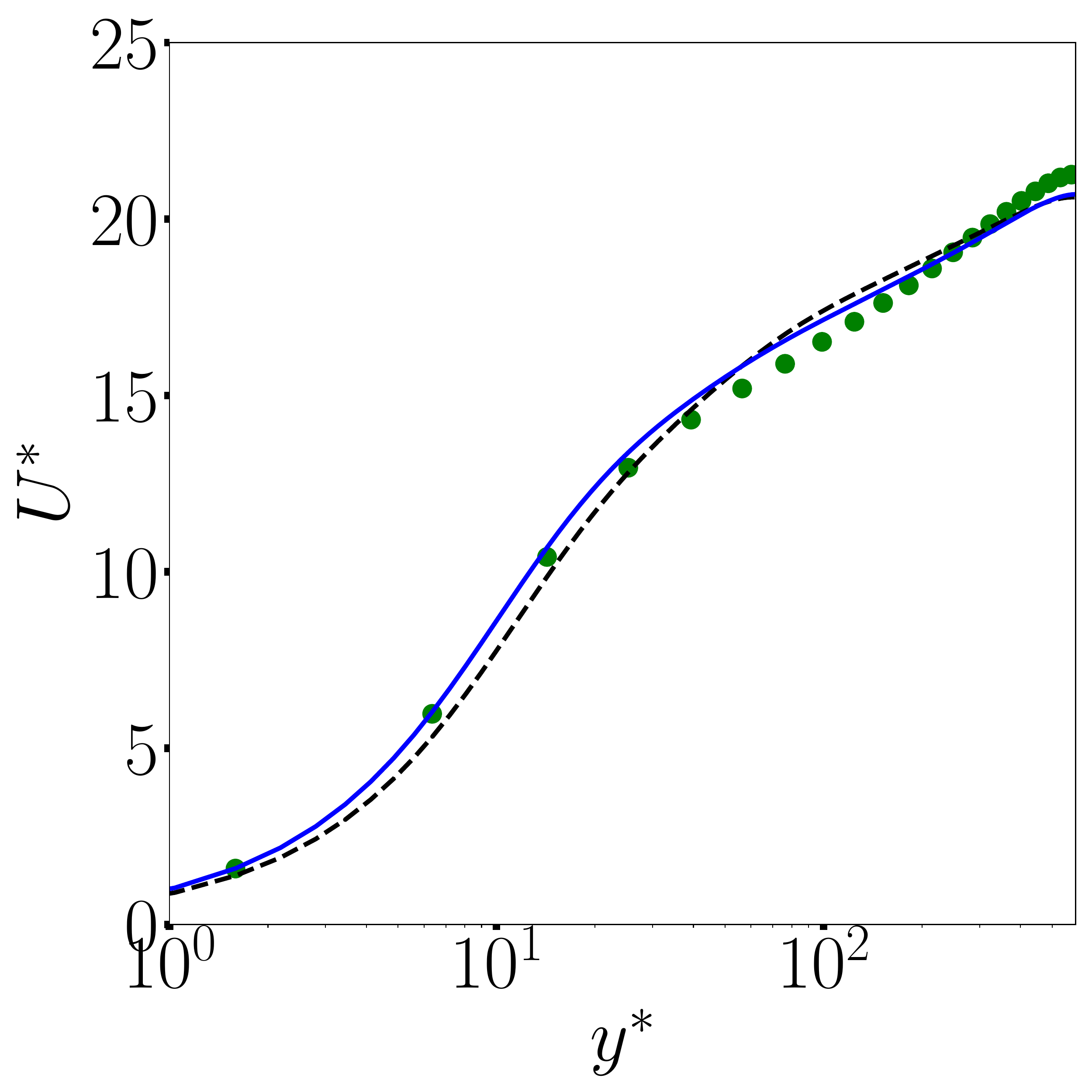}
\caption{Mean profile $U(y)$}
\end{subfigure}
\begin{subfigure}{.28\linewidth}
\centering
\includegraphics[width=1.\textwidth]{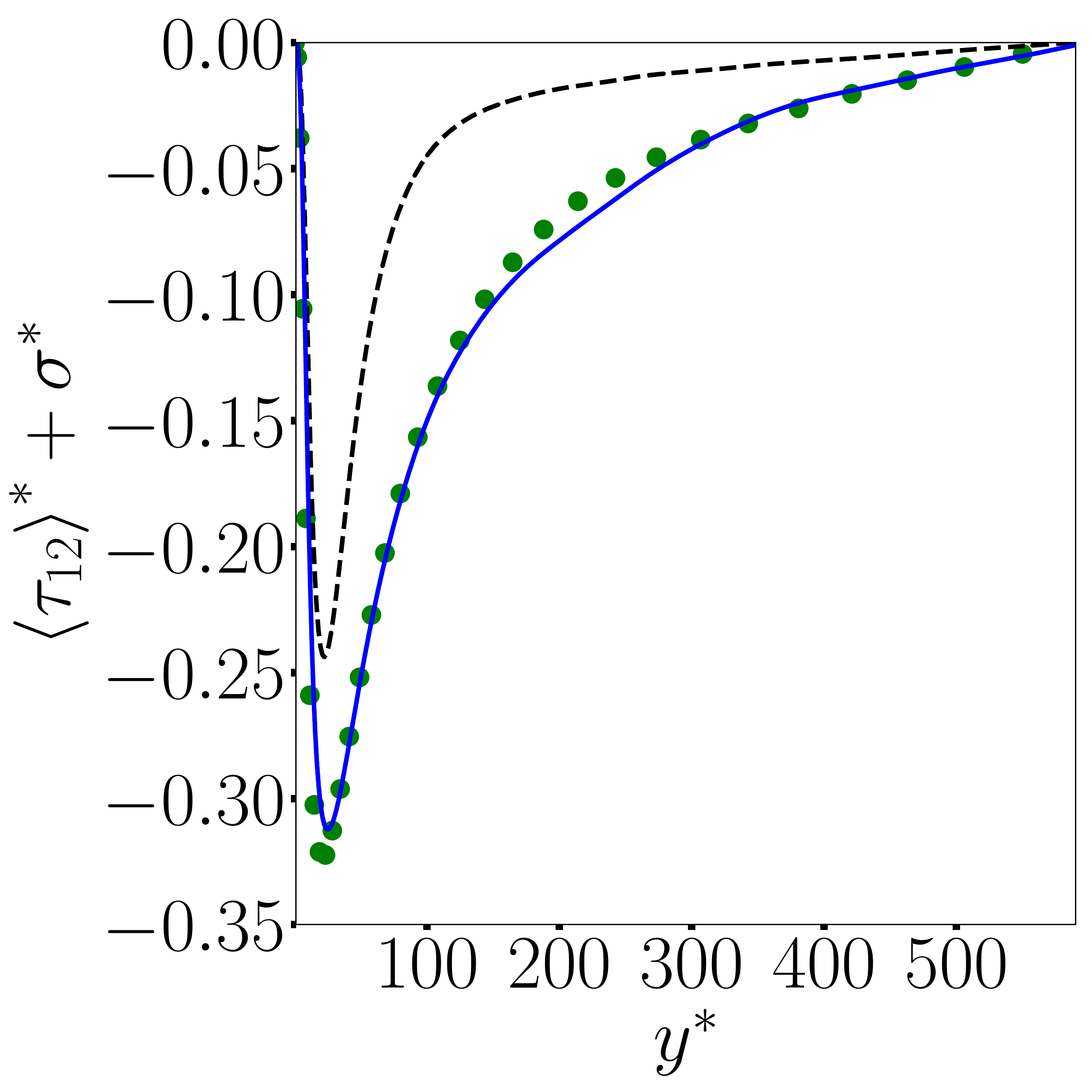}
\caption{Subgrid stress $\left<\tau_{12}\right>$}
\end{subfigure}%
\begin{subfigure}{.28\linewidth}
\centering
\includegraphics[width=1.\textwidth]{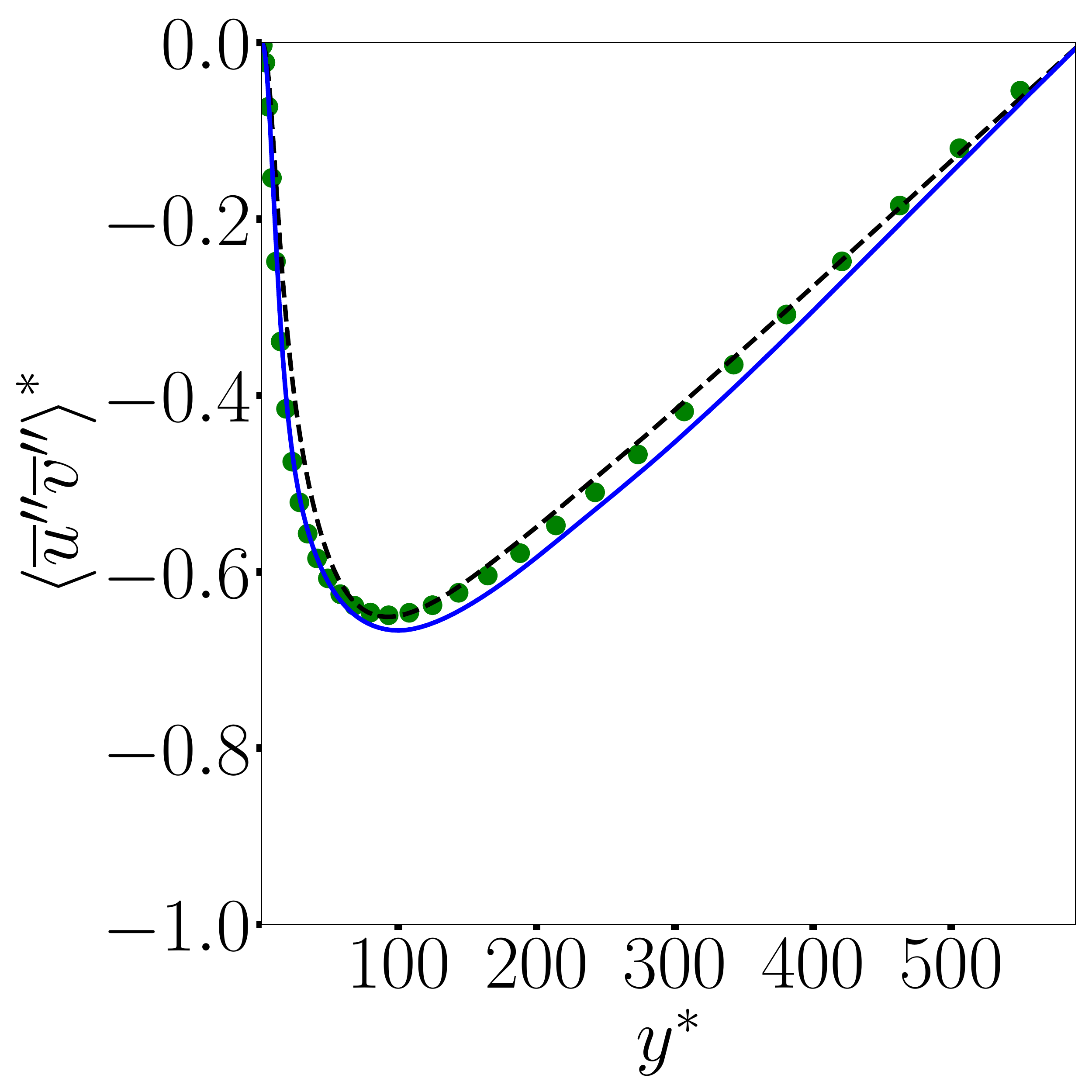}
\caption{Resolved shear stress $\left<\overline{u}''\overline{v}''\right>$}
\end{subfigure}
\begin{subfigure}{.28\linewidth}
\centering
\includegraphics[width=1.\textwidth]{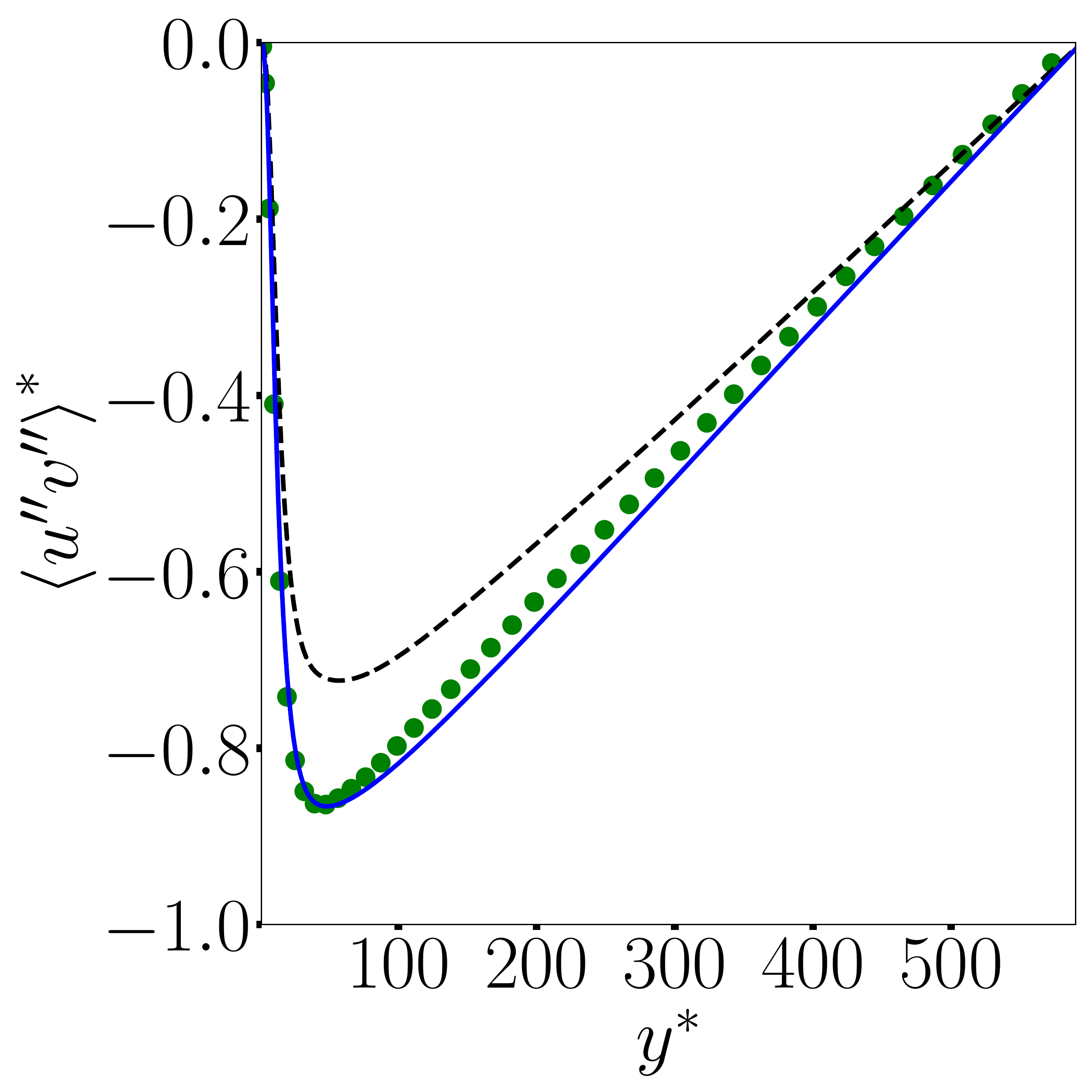}
\caption{\label{fig:Yifan_case3c_R_uv}Total shear stress $\left<u''v''\right>$}
\end{subfigure}%
\caption{\label{fig:Yifan_case3c}Control vector and flow statistics of case 3c for DNS (\mycircle{black!50!green} \mycircle{black!40!green} \mycircle{black!40!green}), the Smagorinsky model (\dashed) and for final DA-LES (\full).}
\end{figure}

In accord with the discussion in \S\ref{sec:minimal_correction_LES}, the above results suggest the need of a more exhaustive adjustment of the subgrid tensor, such as a separate control of its mean and fluctuations. We now consider data assimilation experiment 3 where the control vector $\boldsymbol{\gamma}$ is formed by the coefficient $C_{\mathrm{s}}$ and the forcing $\sigma$ in the momentum equations (\ref{eq:sigma}). The observations are the mean subgrid and resolved shear stresses $\left<\tau_{12}\right>$ and $\left< \overline{u}''\overline{v}''\right>$, which appeared as competitive objectives when only adjusting $C_{\mathrm{s}}$ according to (\ref{eq:gradients_J_tau12_ruv}). Figure \ref{fig:Yifan_case3} illustrates the results for this data assimilation experiment, and directly reports profiles for the final assimilated state, which was obtained after 3 iterations of the data assimilation procedure.
As illustrated in figures \ref{fig:Yifan_case3_R_tau12}-\ref{fig:Yifan_case3_R_uvr}, DA-LES has successfully recovered both the observed subgrid and resolved shear stresses, thanks to incorporating the forcing $\sigma$ in the control vector. As the intensity of the mean resolved shear stress $\left<\overline{u}''\overline{v}''\right>$ is slightly underestimated close to the wall with the standard Smagorinsky model, the value of the coefficient $C_{\mathrm{s}}$ is decreased for DA-LES in this region, inducing a decrease in subgrid dissipation. In order to compensate for this reduction in $C_{\mathrm{s}}$ and to match the reference mean subgrid shear stress $\left<\tau_{12}\right>$, whose intensity is also underestimated by the standard Smagorinsky model, the data assimilation procedure has resulted in significant negative values for the forcing $\sigma$.

Despite the fact that DA-LES is able to correctly recover both the subgrid and resolved part of the shear stress in the present case, and thus the total shear stress $\left<u''v''\right>= \left< \overline{u}''\overline{v}''\right> + \left<\tau_{12}\right>+\sigma$ as confirmed by figure \ref{fig:Yifan_case3_R_uv}, it appears from figure \ref{fig:Yifan_case3_R_U} that the discrepancies in the assimilated mean-velocity profile are still large  beyond $y^+= 30$ and have not been significantly decreased compared to the standard Smagorinsky model. While this may seem in contradiction with the governing equation for the mean flow (\ref{eq:meanflow_equation}), it is well known that discretization errors in LES may overwhelm the subgrid model, in particular for second-order methods and when the filter size is the same as the grid spacing \citep{Vreman1994_cnme,Ghosal1996_jcp}, as in the present case. The importance of discretization errors in LES was investigated in \cite{Majander2002_ijnmf} in the case of turbulent channel flow with similar numerical techniques as the present solver. Accordingly, the LES governing equations (\ref{eq:sigma}) are not exactly realized and could include supplementary contributions that interfere with the accuracy of prediction. The present data assimilation experiment still provides the best achievable mean-flow prediction when both the mean and fluctuations of the subgrid tensor of the Smagorinsky model (\ref{eq:smagorinsky_model}) yield accurate total shear stress and its subgrid and resolved components.

The findings from case 3 are confirmed by data assimilation experiment 3c which adopts a coarser grid LES590c. The corresponding results are reported in figure \ref{fig:Yifan_case3c}. DA-LES almost perfectly reproduces both the subgrid and resolved shear stresses through an overall decrease in the value of $C_{\mathrm{s}}$ and the introduction of a strongly negative forcing $\sigma$. Nonetheless, despite the correct estimation of the total shear stress, the mean flow from DA-LES is only mildly improved compared to the use of the Smagorinsky model, and the remaining discrepancies are larger than in the previous case, which can be attributed to the increase in discretization errors that are induced by the use of a coarser grid. However, one has to remain cautious when interpreting this trend due to the non-monotonic decrease of discretization errors in simulations of turbulent channel flows \citep{Meyers2007_pof}. 
The results of data assimilation experiments 3 and 3c still suggest that rather than improving the estimation of the subgrid tensor, it is more beneficial to directly target statistical quantities of interest. In doing so, discretization errors would be fully taken into account during the assimilation procedure, and the LES is directly optimized for accurate prediction of the target statistics. This approach is the subject of \S\ref{sec:results_DA_statistics}.

\section{Data assimilation using target flow statistics}\label{sec:results_DA_statistics}

\subsection{Recovering reference statistics from mean-flow observations (case 4)}\label{sec:results_DA_statistics_Re_590_case4}

\begin{figure}
\centering
\begin{subfigure}{.28\linewidth}
\centering
\includegraphics[width=1.\textwidth]{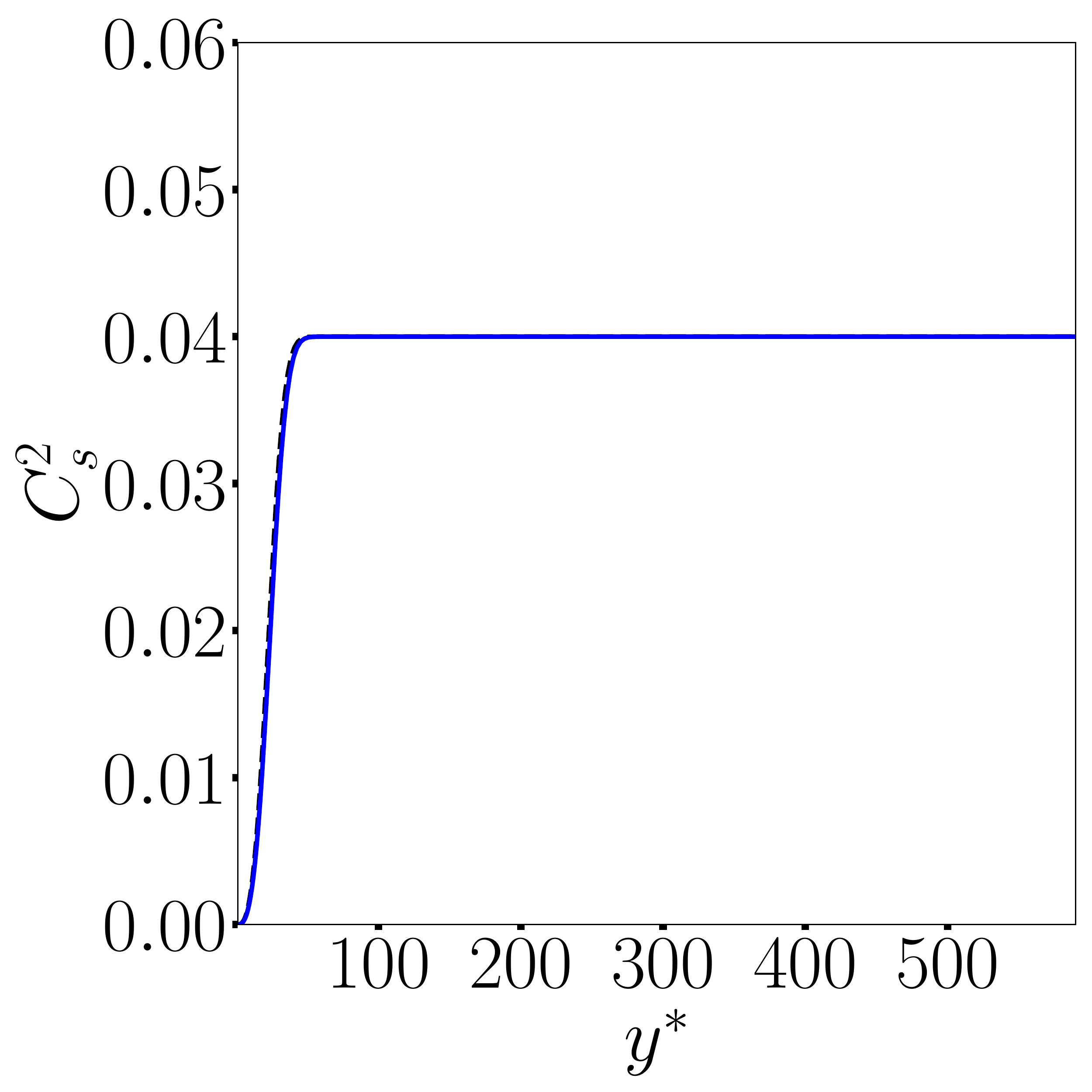}
\caption{\label{fig:Vincent_case4_R_C}Control vector $C_\mathrm{s}(y)$}
\end{subfigure}%
\begin{subfigure}{.28\linewidth}
\centering
\includegraphics[width=1.\textwidth]{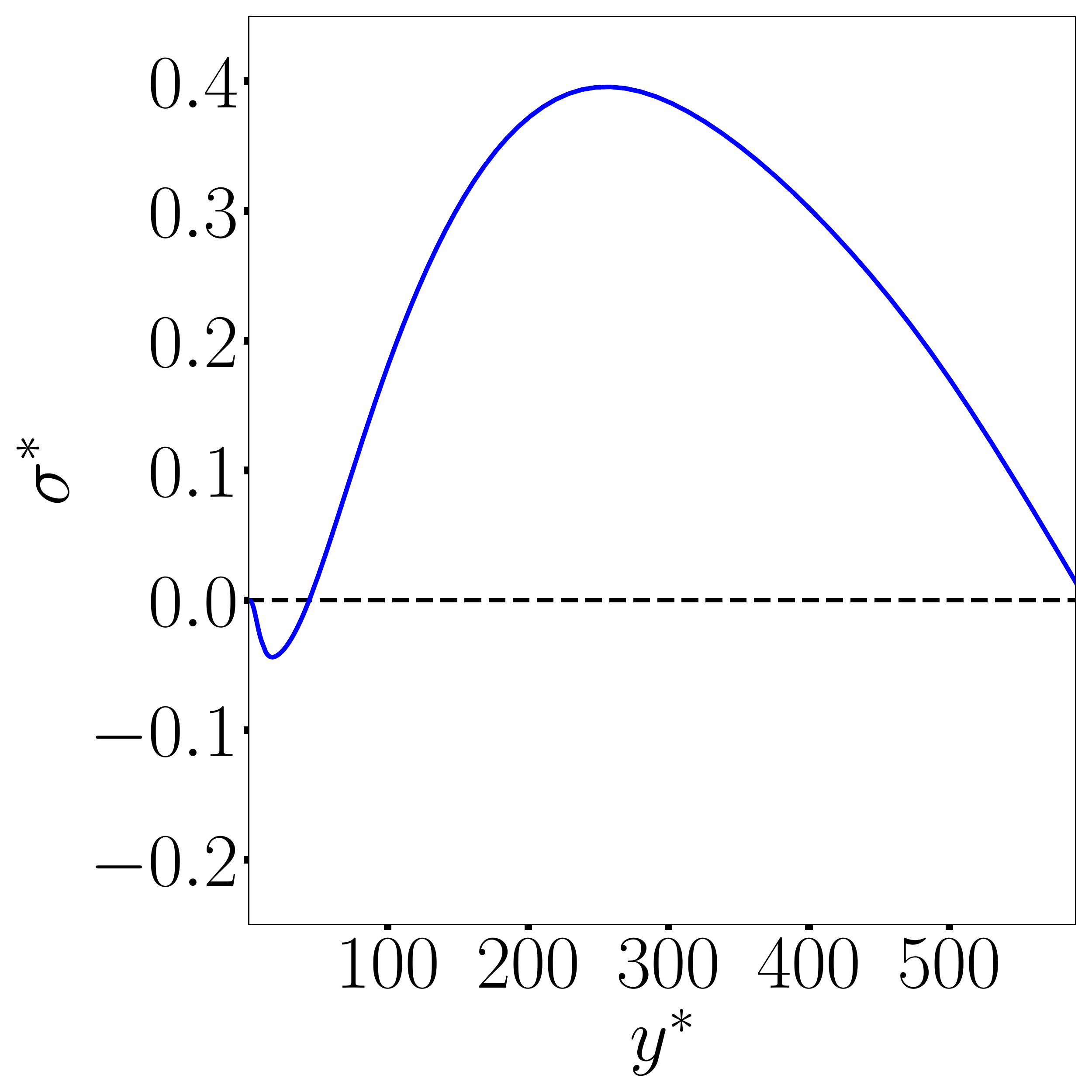}
\caption{\label{fig:Vincent_case4_R_sigma}Control vector $\sigma(y)$}
\end{subfigure}
\begin{subfigure}{.28\linewidth}
\centering
\includegraphics[width=1.\textwidth]{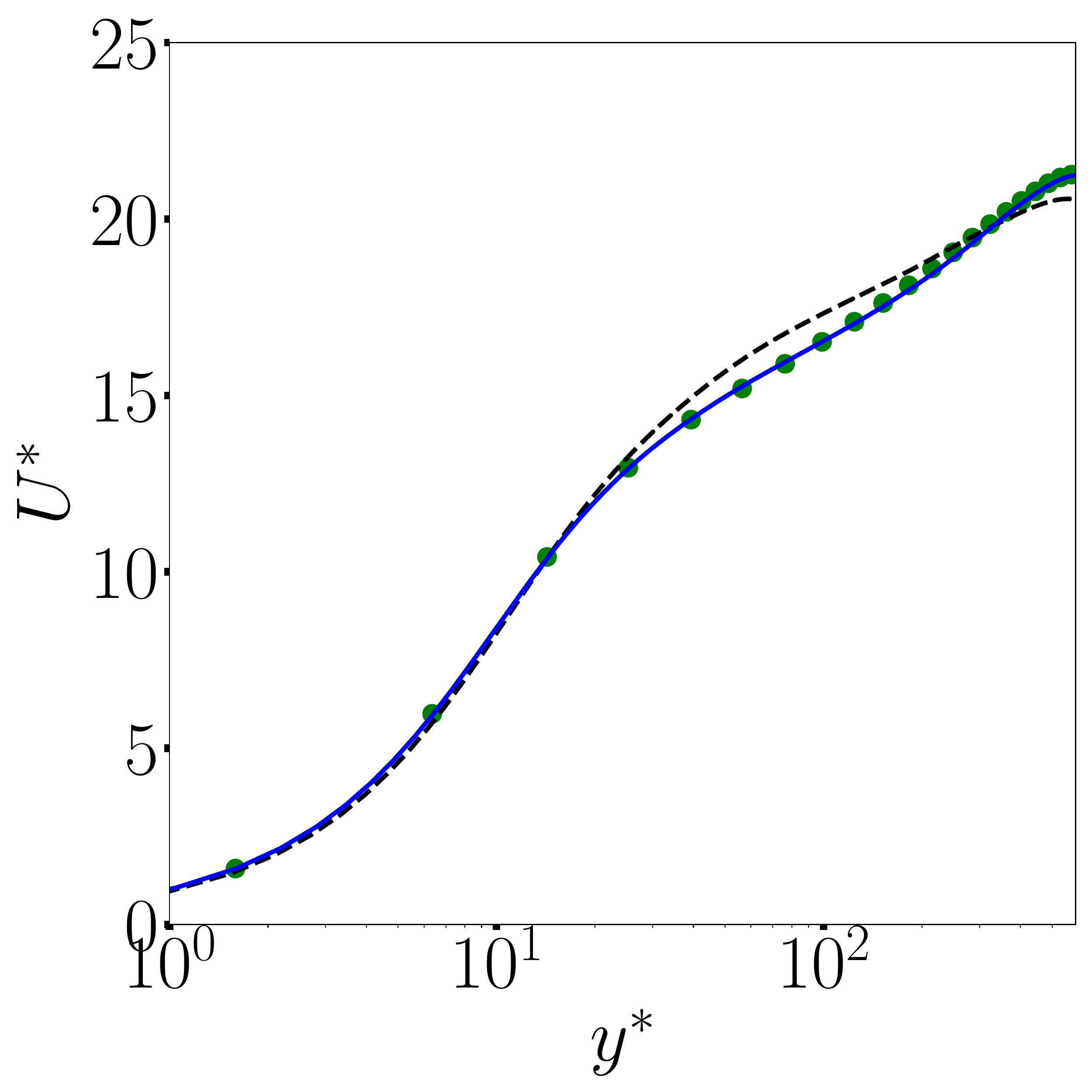}
\caption{\label{fig:Vincent_case4_R_U}Mean profile $U(y)$}
\end{subfigure}
\begin{subfigure}{.28\linewidth}
\centering
\includegraphics[width=1.\textwidth]{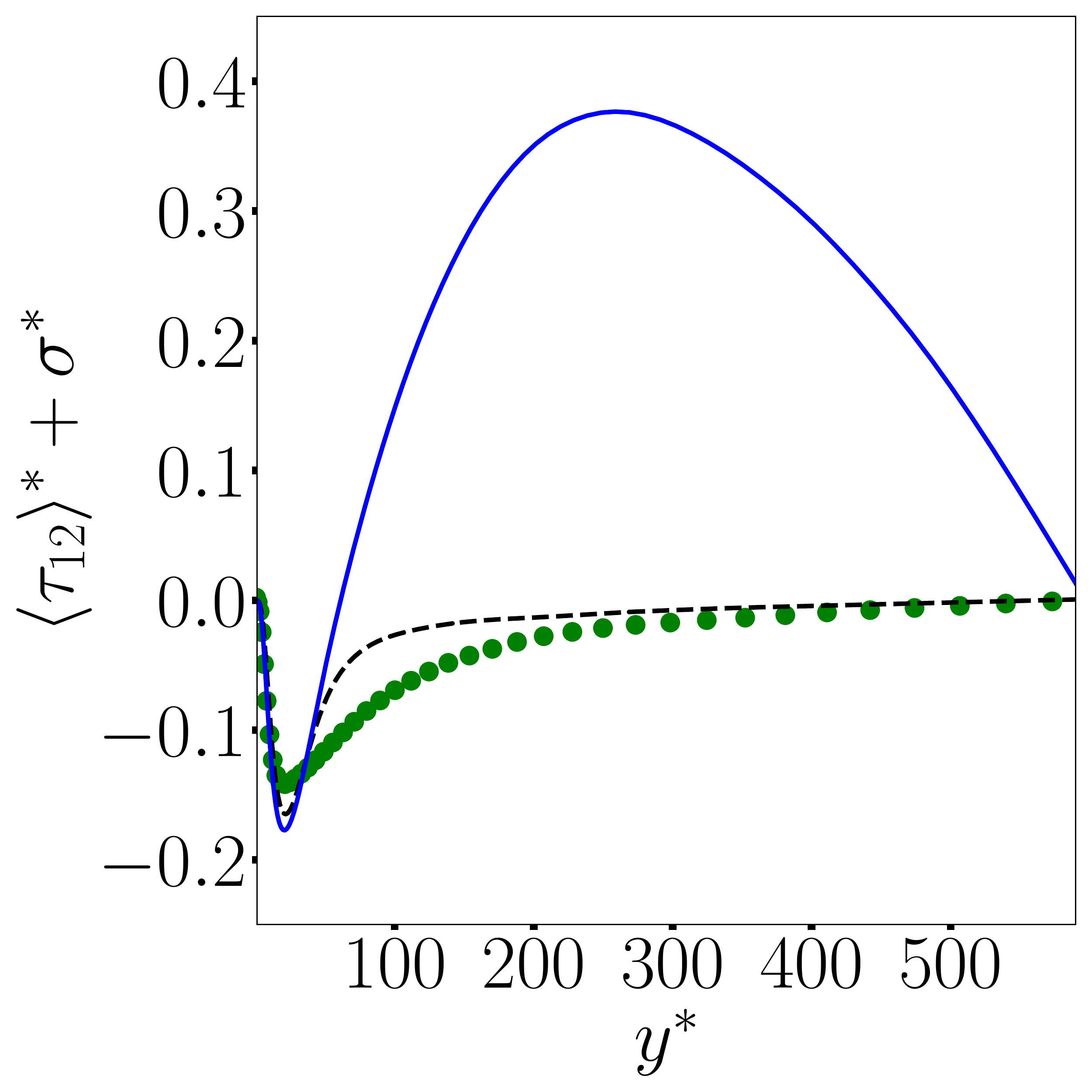}
\caption{\label{fig:Vincent_case4_R_tau12}Subgrid stress $\left<\tau_{12}\right>$}
\end{subfigure}%
\begin{subfigure}{.28\linewidth}
\centering
\includegraphics[width=1.\textwidth]{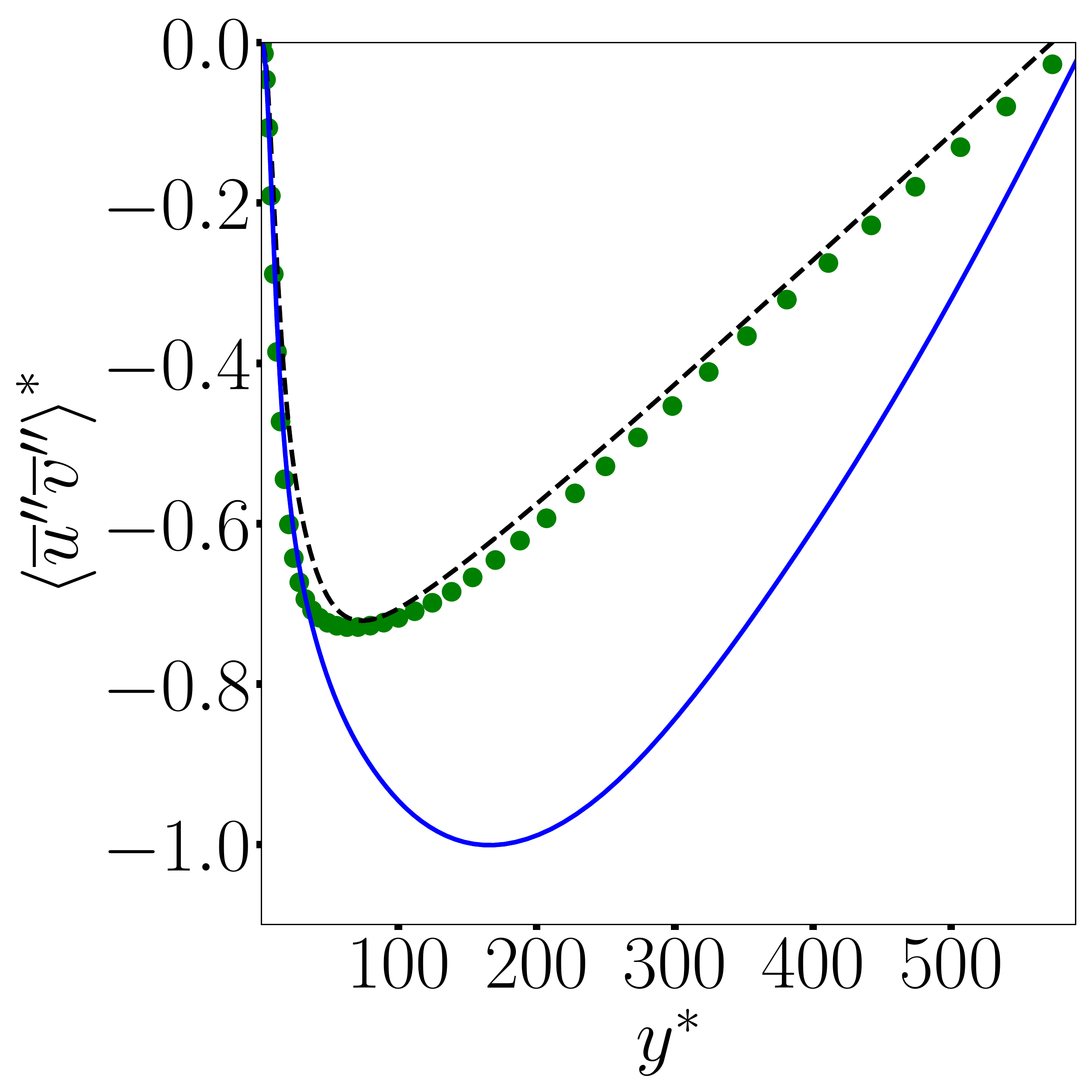}
\caption{\label{fig:Vincent_case4_R_uvr}Resolved shear stress $\left<\overline{u}''\overline{v}''\right>$}
\end{subfigure}
\begin{subfigure}{.28\linewidth}
\centering
\includegraphics[width=1.\textwidth]{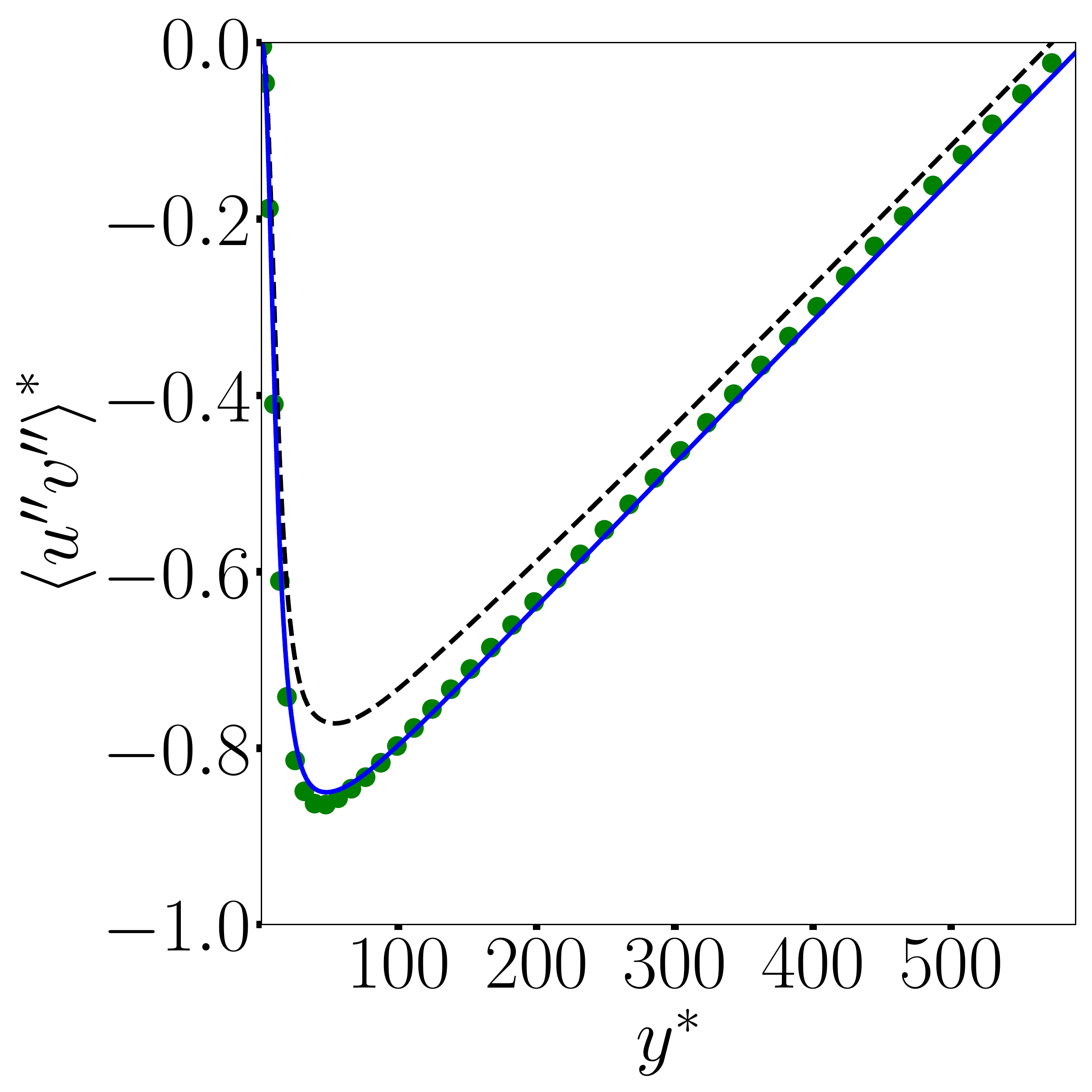}
\caption{\label{fig:Vincent_case4_R_uv}Total shear stress $\left<u''v''\right>$}
\end{subfigure}%
\caption{\label{fig:Vincent_case4}Control vector and flow statistics of case 4 for DNS (\mycircle{black!50!green} \mycircle{black!40!green} \mycircle{black!40!green}), the Smagorinsky model (\dashed) and for final DA-LES (\full).}
\end{figure}

\begin{figure}
\centering
\begin{subfigure}{.28\linewidth}
\centering
\includegraphics[width=1.\textwidth]{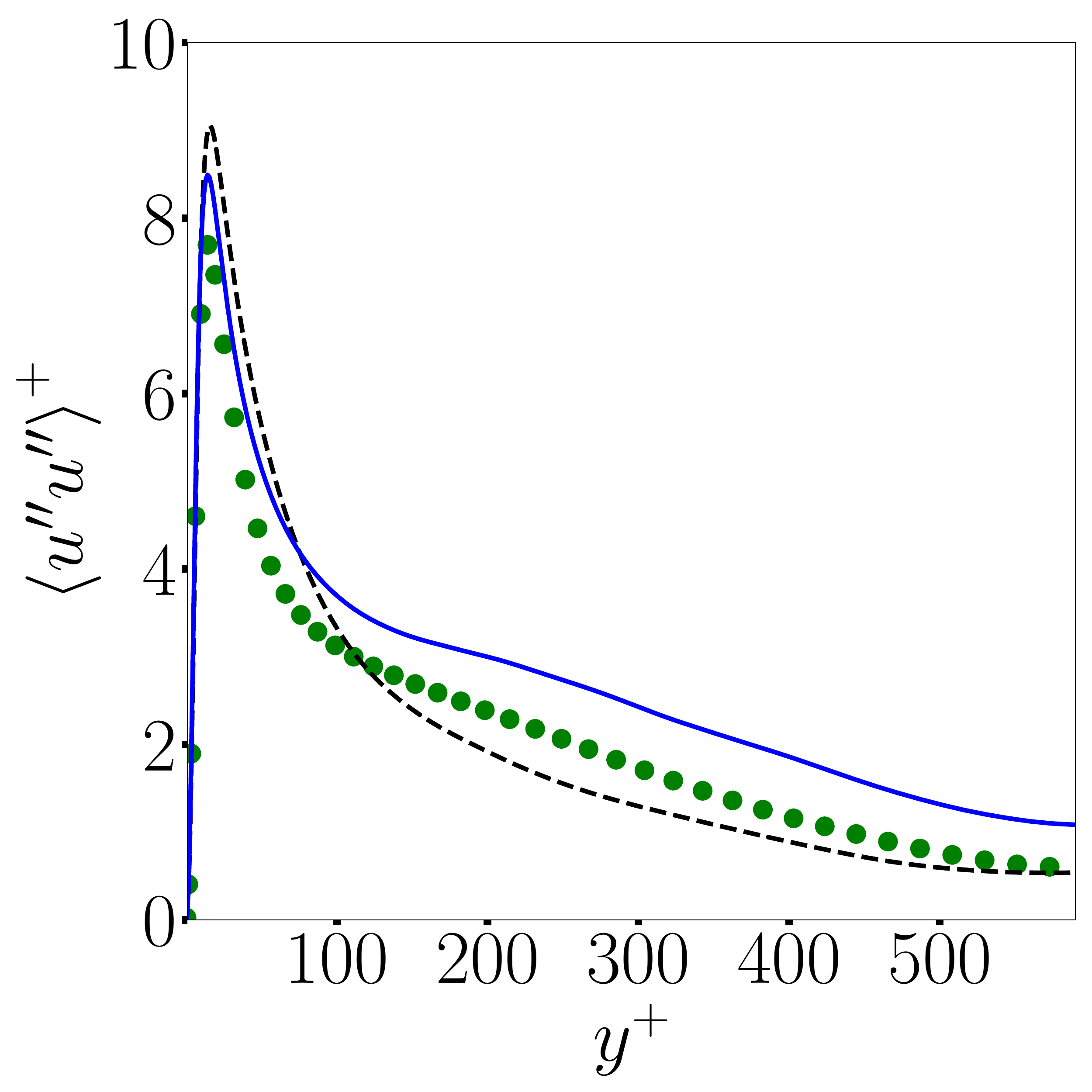}
\caption{\label{fig:Vincent_case4_R_uu}Reynolds stress $\left<u''u''\right>$}
\end{subfigure}%
\begin{subfigure}{.28\linewidth}
\centering
\includegraphics[width=1.\textwidth]{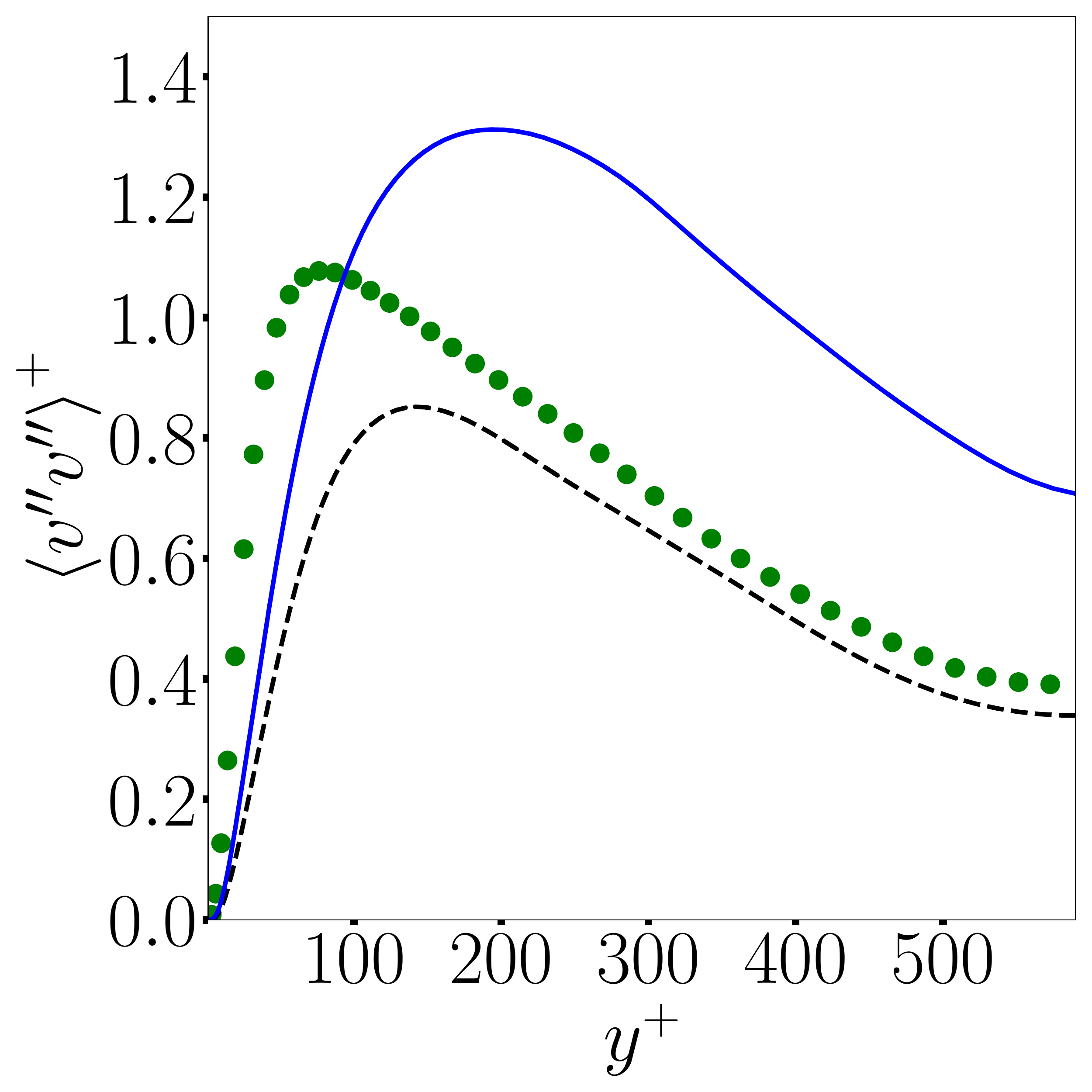}
\caption{\label{fig:Vincent_case4_R_vv}Reynolds stress $\left<v''v''\right>$}
\end{subfigure}
\begin{subfigure}{.28\linewidth}
\centering
\includegraphics[width=1.\textwidth]{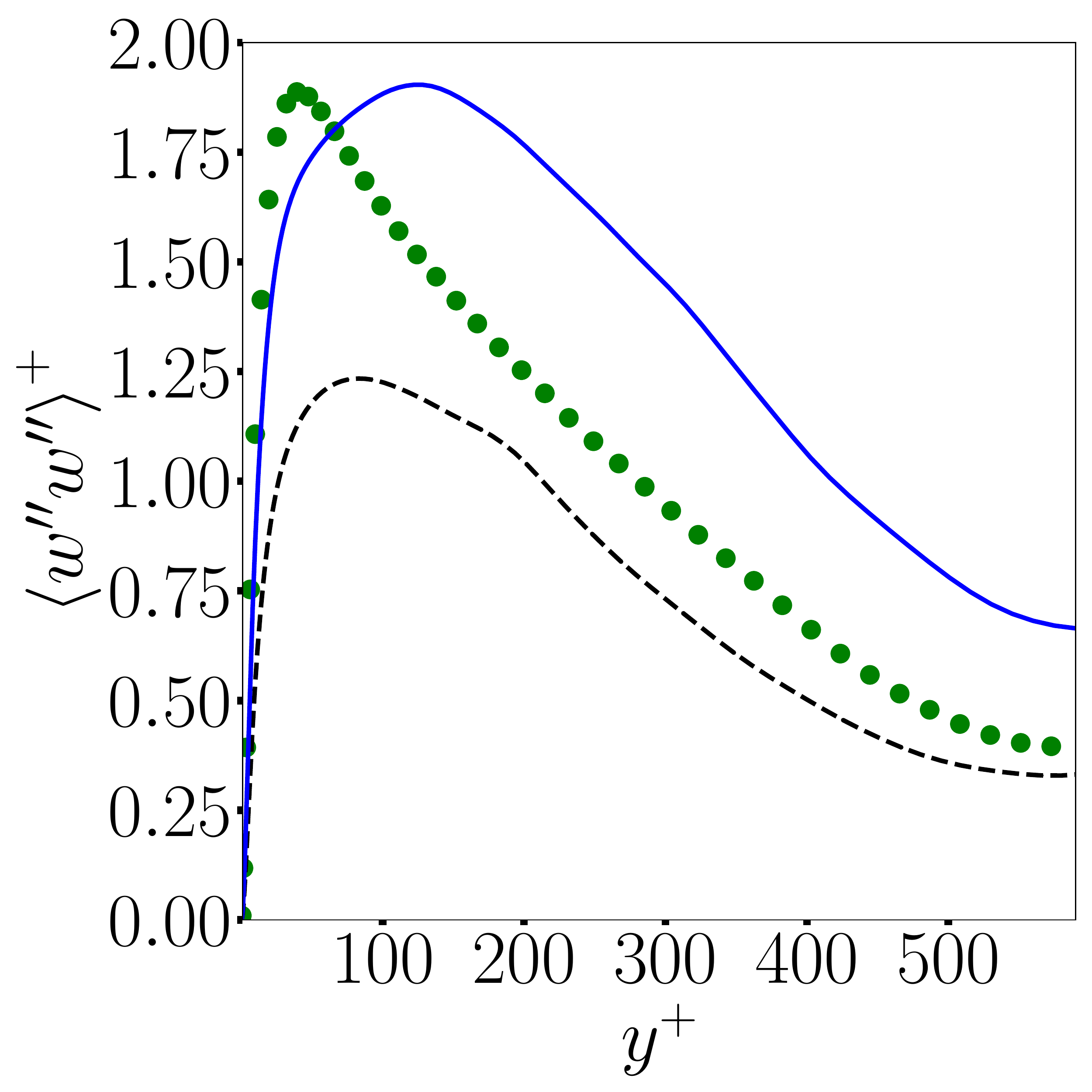}
\caption{\label{fig:Vincent_case4_R_ww}Reynolds stress $\left<w''w''\right>$}
\end{subfigure}
\begin{subfigure}{.28\linewidth}
\centering
\includegraphics[width=1.\textwidth]{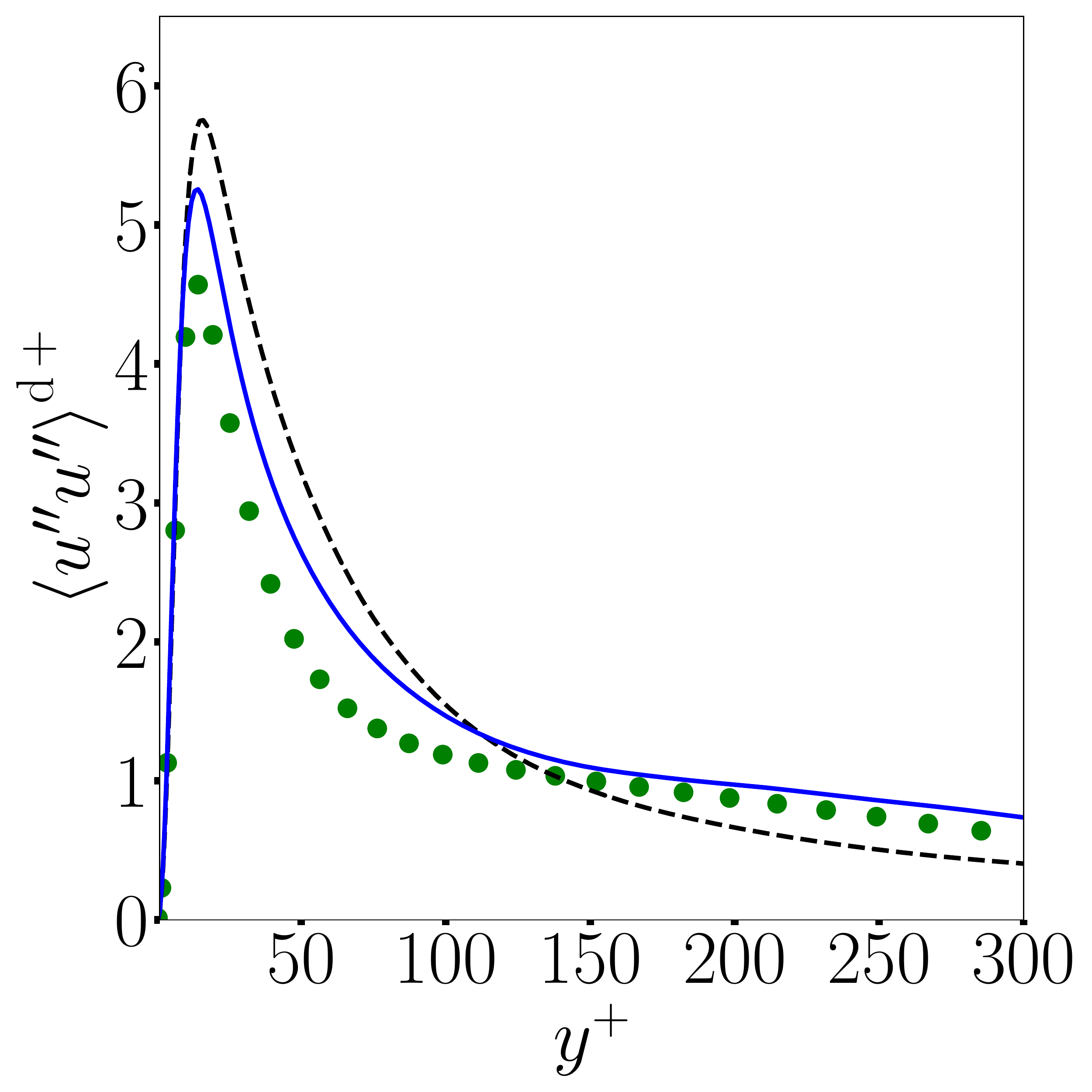}
\caption{\label{fig:Vincent_case4_R_duu}Deviatoric Reynolds stress $\left<u''u''\right>^{\mathrm{d}}$}
\end{subfigure}%
\begin{subfigure}{.28\linewidth}
\centering
\includegraphics[width=1.\textwidth]{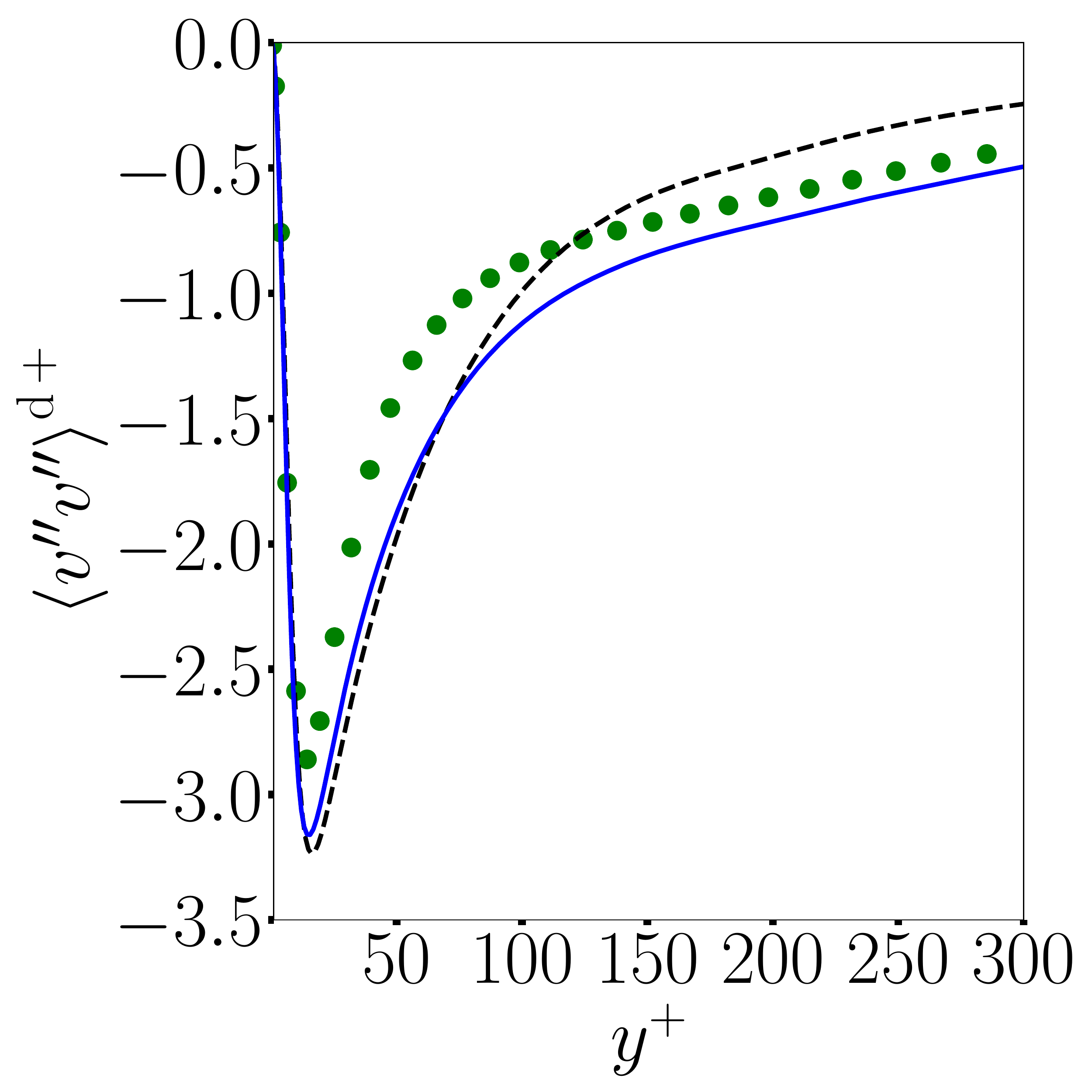}
\caption{\label{fig:Vincent_case4_R_dvv}Deviatoric Reynolds stress $\left<v''v''\right>^{\mathrm{d}}$}
\end{subfigure}
\begin{subfigure}{.28\linewidth}
\centering
\includegraphics[width=1.\textwidth]{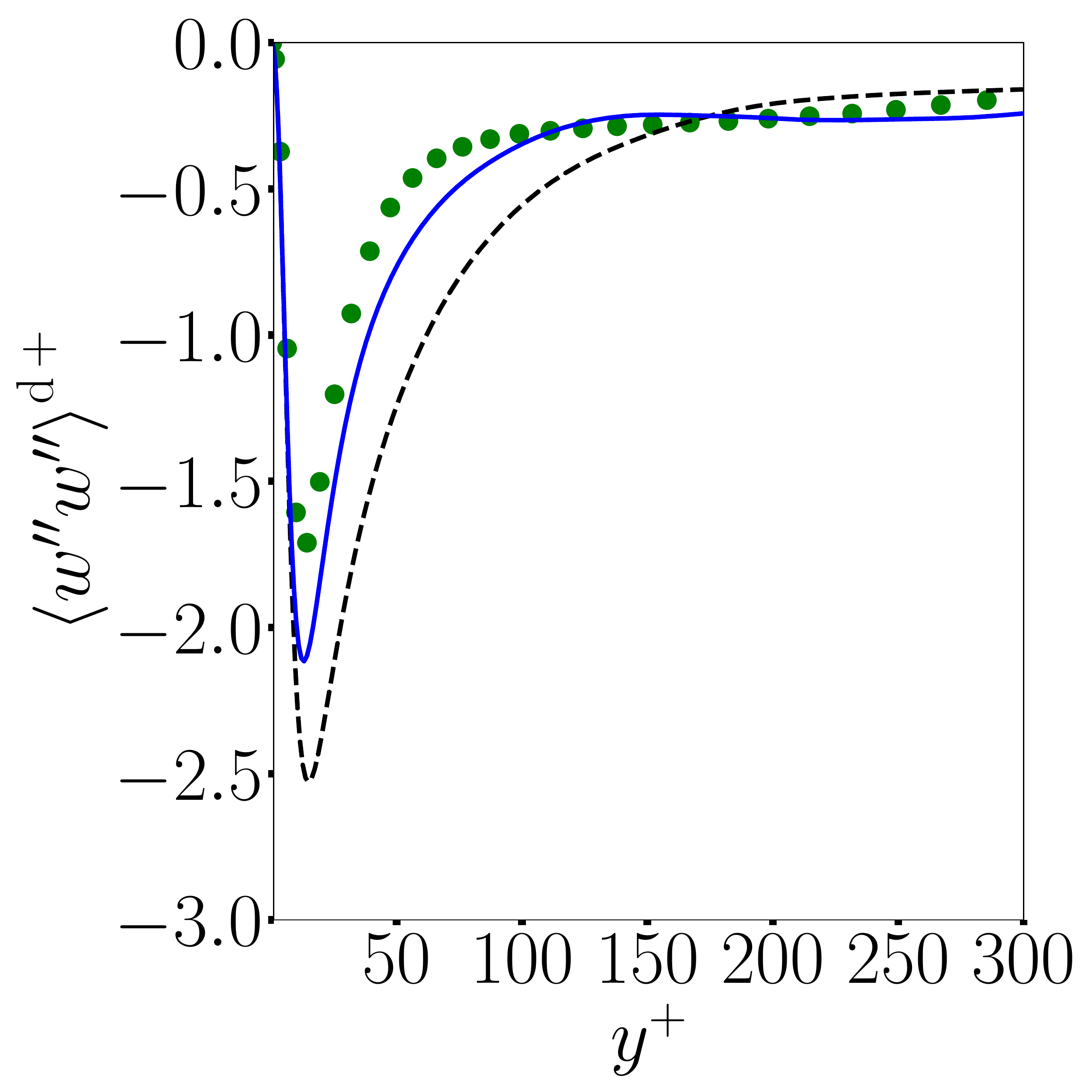}
\caption{\label{fig:Vincent_case4_R_dww}Deviatoric Reynolds stress $\left<w''w''\right>^{\mathrm{d}}$}
\end{subfigure}
\caption{\label{fig:Vincent_case4_turbulence}Second-order statistics of case 4 for DNS (\mycircle{black!50!green} \mycircle{black!40!green} \mycircle{black!40!green}), the Smagorinsky model (\dashed) and for final DA-LES (\full).}
\end{figure}

\begin{figure}
\centering
\begin{subfigure}{.28\linewidth}
\centering
\includegraphics[width=1.\textwidth]{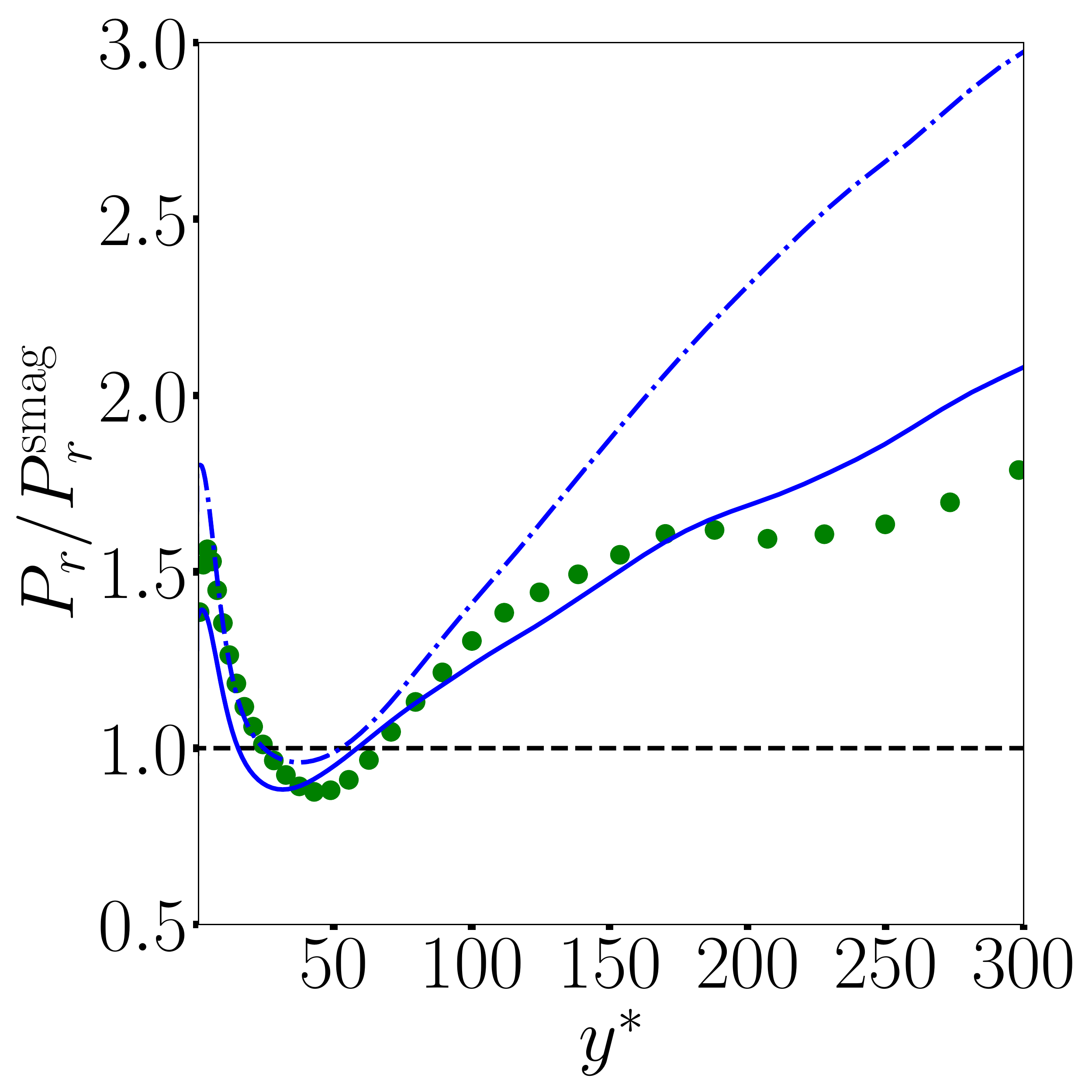}
\caption{\label{fig:Vincent_R_P_r}Production $P_r$}
\end{subfigure}%
\begin{subfigure}{.28\linewidth}
\centering
\includegraphics[width=1.\textwidth]{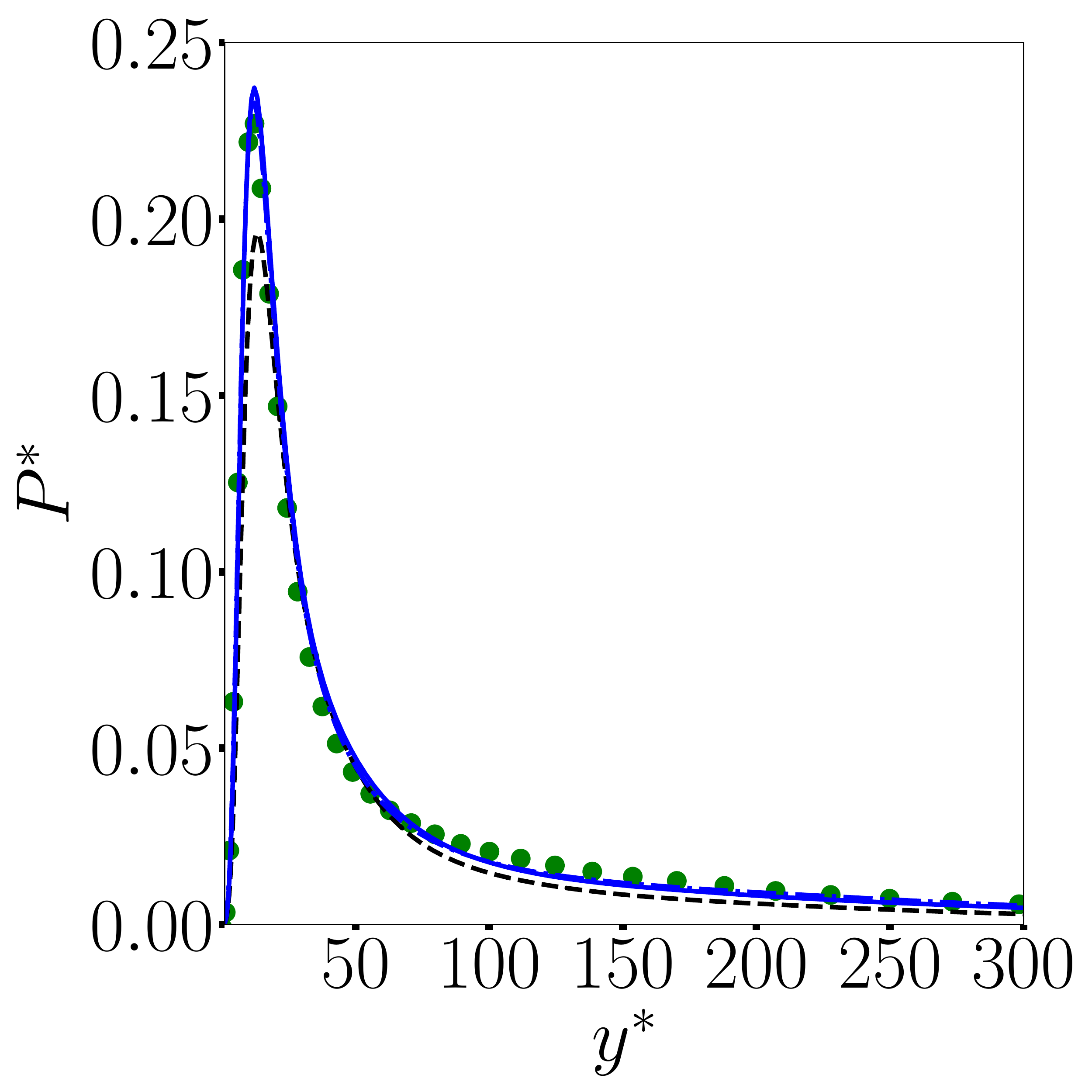}
\caption{\label{fig:Vincent_R_P}Production $P$}
\end{subfigure}
\caption{\label{fig:Vincent_production}Production of (a) resolved and (b) total turbulent kinetic energy for DNS (\mycircle{black!50!green} \mycircle{black!40!green} \mycircle{black!40!green}), the Smagorinsky model (\dashed) and for final DA-LES of cases 4 (\chain) and 5 (\full). Results in (a) are normalized by $P^{\mathrm{smag}}_r$, which refers to the production of resolved turbulent kinetic energy as predicted by the Smagorinsky model.}
\end{figure}

The data assimilation experiments in \S\ref{sec:results_DA_filtered} demonstrated that a good estimation of the subgrid stress tensor does not guarantee the fidelity of LES simulations. Therefore, here only statistical quantities of interest will be considered as observations $\boldsymbol{m}$. Data assimilation experiment 4 is examined in figures \ref{fig:Vincent_case4}-\ref{fig:Vincent_production}. The targeted friction Reynolds number is still $Re_{\tau}=590$, and the finest grid LES590f is employed. The reference mean flow from DNS, or more specifically its gradient $\mathrm{d}U/\mathrm{d}y$, is considered as sole observations $\boldsymbol{m}$. In order to foster the well-posedness of the data assimilation problem, and since variations in the coefficient $C_{\mathrm{s}}$ were shown to have only a mild effect on $U$, the control vector $\boldsymbol{\gamma}$ is formed by the forcing $\sigma$ only while $C_{\mathrm{s}}$ is kept as the profile (\ref{eq:smagorinsky_constant}) in this case. The final assimilated state is obtained with only 2 main iterations of the data assimilation procedure. The associated computational cost is thus equivalent to $N_{\mathrm{CFD}}=42$ LES calculations. The fact that only 2 main iterations are necessary for this and subsequent data assimilation experiments, and that the associated computational cost may even be lowered, will be discussed in \S\ref{sec:further_assessment}.

As illustrated in figure \ref{fig:Vincent_case4_R_U}, the reference mean flow is successfully recovered over the whole channel height, contrary to previous cases. This result has been achieved through the identification of an optimal profile for $\sigma$ with high positive values above $y^{*}=40$. This height roughly coincides with the region where the subgrid model does not have a large influence on $U$, and also where discrepancies between the true mean flow and its prediction by the standard Smagorinsky model are significant. Figure \ref{fig:Vincent_case4_R_tau12} confirms that this correction to the mean of the subgrid shear stress $\left<\tau_{12}\right>$ is almost negligible close to the wall, and starts to move away from the Smagorinsky profile and even from the reference one at the beginning of the log layer. The impact of the assimilated forcing $\sigma$ is also reflected in the resolved shear stress $\left<\overline{u}''\overline{v}''\right>$ which is reported in figure \ref{fig:Vincent_case4_R_uvr}, with a large overestimation of its magnitude in the same region. On the other hand, the total shear stress $\left<u''v''\right>$, which is more of interest here, is accurately predicted by DA-LES in accordance with the mean-flow equation (\ref{eq:meanflow_equation}), as illustrated in figure \ref{fig:Vincent_case4_R_uv}.

While the emphasis was previously placed on the correct estimation of the mean flow and other quantities that feature in its governing equation, we now turn to the Reynolds normal stresses from experiment 4 (figure \ref{fig:Vincent_case4_turbulence}). Contrary to previous results, non-dimensionalization is here based on the respective friction velocities of the various LES calculations, rather than the reference DNS, in order to facilitate comparisons and the distinction between results. Close to the wall, the data assimilation procedure is beneficial in correcting the initially overestimated component $\left<u''u''\right>$ and underestimated $\left<v''v''\right>$ and $\left<w''w''\right>$ (figures \ref{fig:Vincent_case4_R_uu}-\ref{fig:Vincent_case4_R_ww}). However, above $y^+ \sim 60$, the intensity of these components is significantly overestimated by DA-LES, which mirrors the elevated profile of the resolved shear stress in figure \ref{fig:Vincent_case4_R_uvr}. The productions of resolved and total kinetic kinetic energy are plotted in figure \ref{fig:Vincent_R_P_r}, where the dash-dotted blue line corresponds to present case 4. Due to the large over-estimation of $-\left<\overline{u}''\overline{v}''\right>$, the production $P_r=-\left<\overline{u}''\overline{v}''\right>\mathrm{d}U/\mathrm{d}y$ of resolved turbulent kinetic energy $\frac{1}{2}\langle \overline{u}_{i}''\overline{u}_{i}'' \rangle$ is significantly over-estimated, compared to the standard Smagorinsky model.
Nonetheless, the production $P=-\left<u''v''\right>\mathrm{d}U/\mathrm{d}y$ of total turbulent kinetic energy is satisfactorily predicted by DA-LES which significantly corrects the estimation by the Smagorinsky model.

It is however not necessarily fair to compare the reference full Reynolds stresses with those obtained from LES, as we here rely on the Smagorinsky model form (\ref{eq:smagorinsky_model}) which does not model the trace of the subgrid tensor $\tau_{ij}$ (see \S\ref{sec:Smagorinsky model}) and thus prevents an unambiguous estimation of the subgrid turbulent kinetic energy \citep{Winckelmans2002_pof}. As such, a comparison with the deviatoric part of Reynolds stress tensor, which is denoted by the superscript $\mathrm{d}$, is provided in figures \ref{fig:Vincent_case4_R_duu}-\ref{fig:Vincent_case4_R_dww}. Despite the perceived overestimation of the turbulent kinetic energy, DA-LES significantly improves the estimation of the deviatoric part of the Reynolds stress tensor compared to the Smagorinsky model over the whole channel height. The consideration of $dU/dy$ as the sole observed quantity in the data assimilation procedure thus appears sufficient to already enhance the predicted degree of anisotropy of the flow. However, it might still be desirable to prevent the overly large increase in the production $P_r$ of resolved turbulent kinetic energy reported in figure \ref{fig:Vincent_R_P_r}.

\begin{figure}
\centering
\begin{subfigure}{.28\linewidth}
\centering
\includegraphics[width=1.\textwidth]{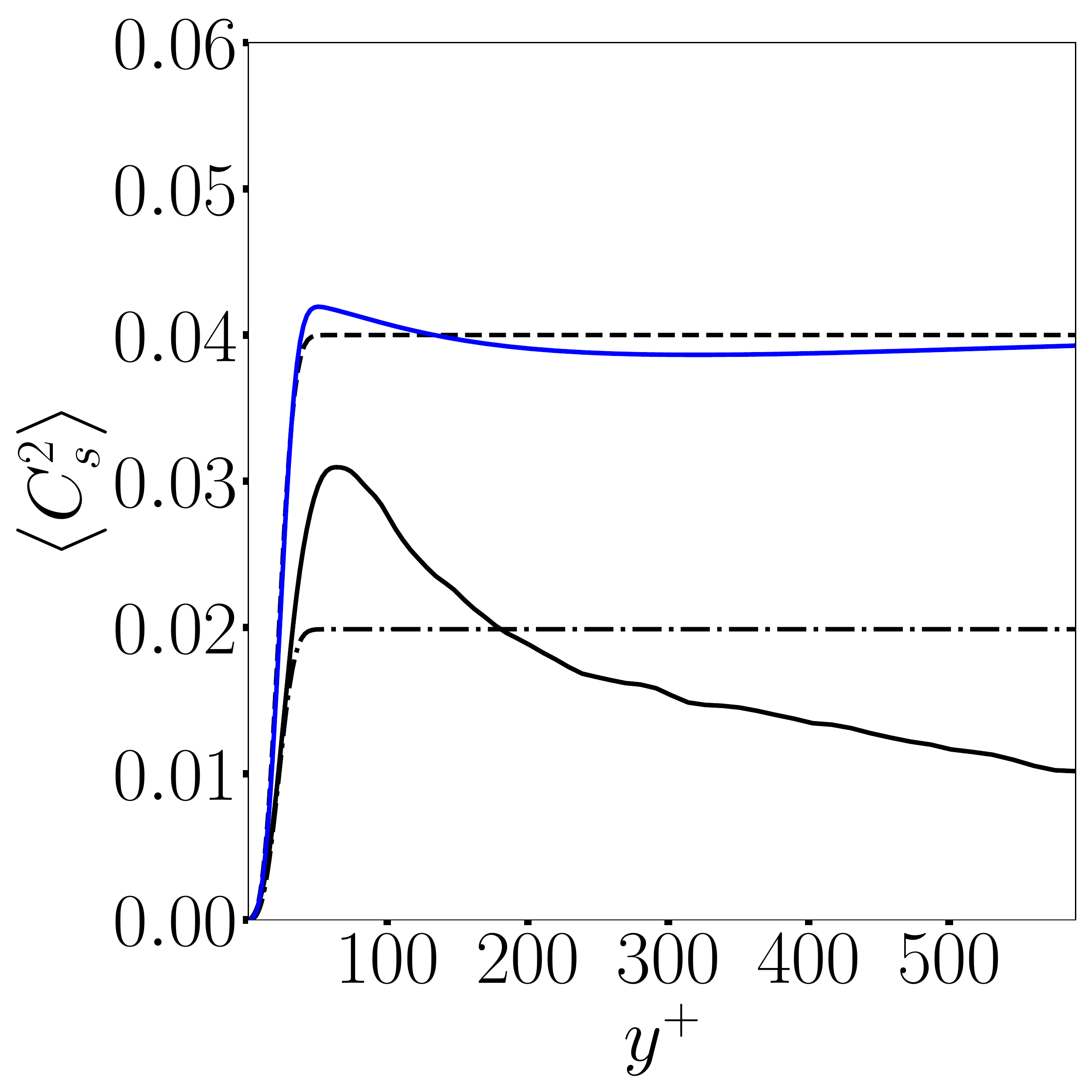}
\caption{\label{fig:Vincent_case5_C_C}Control vector $C_\mathrm{s}(y)$}
\end{subfigure}
\begin{subfigure}{.28\linewidth}
\centering
\includegraphics[width=1.\textwidth]{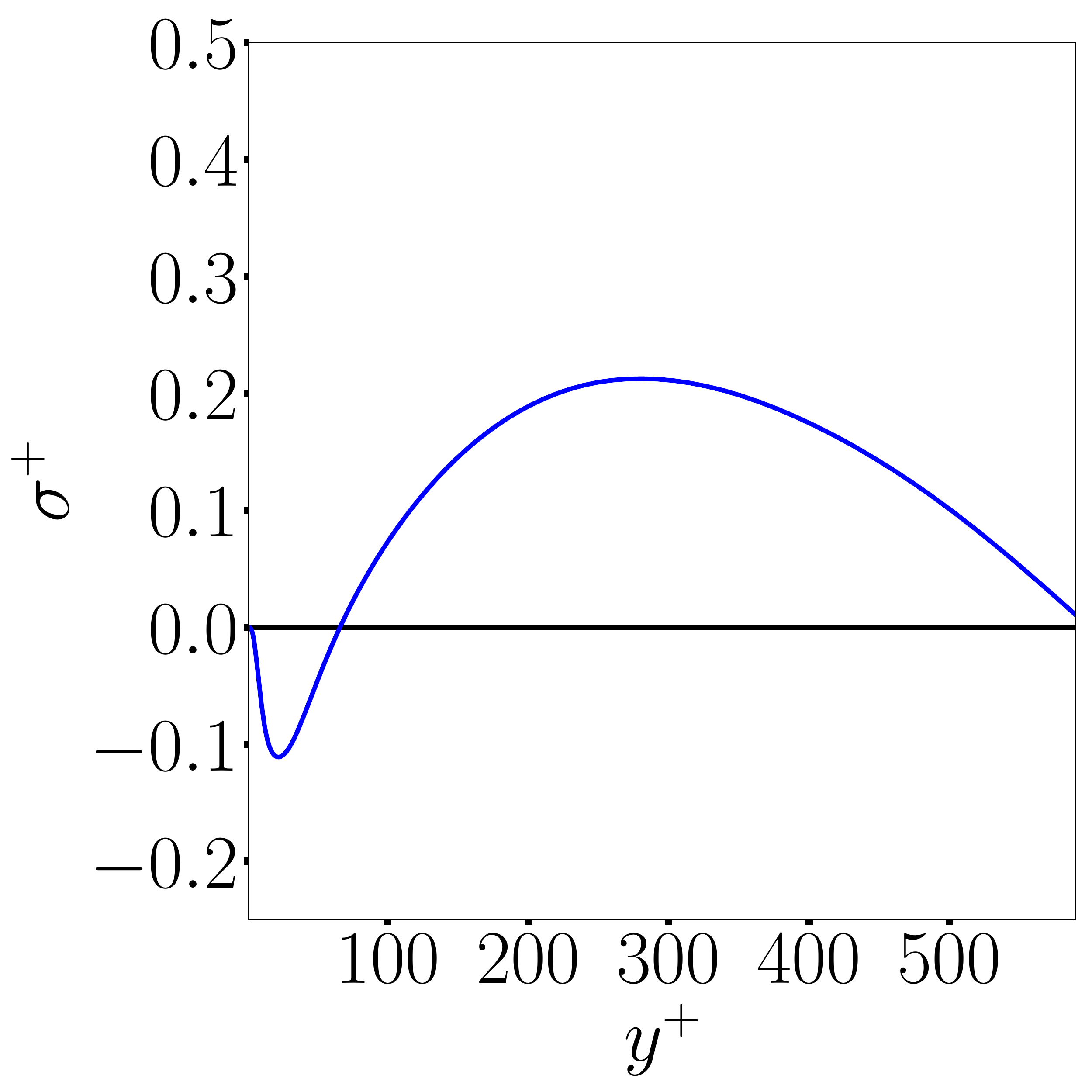}
\caption{\label{fig:Vincent_case5_C_sigma}Control vector $\sigma(y)$}
\end{subfigure}\\
\begin{subfigure}{.28\linewidth}
\centering
\includegraphics[width=1.\textwidth]{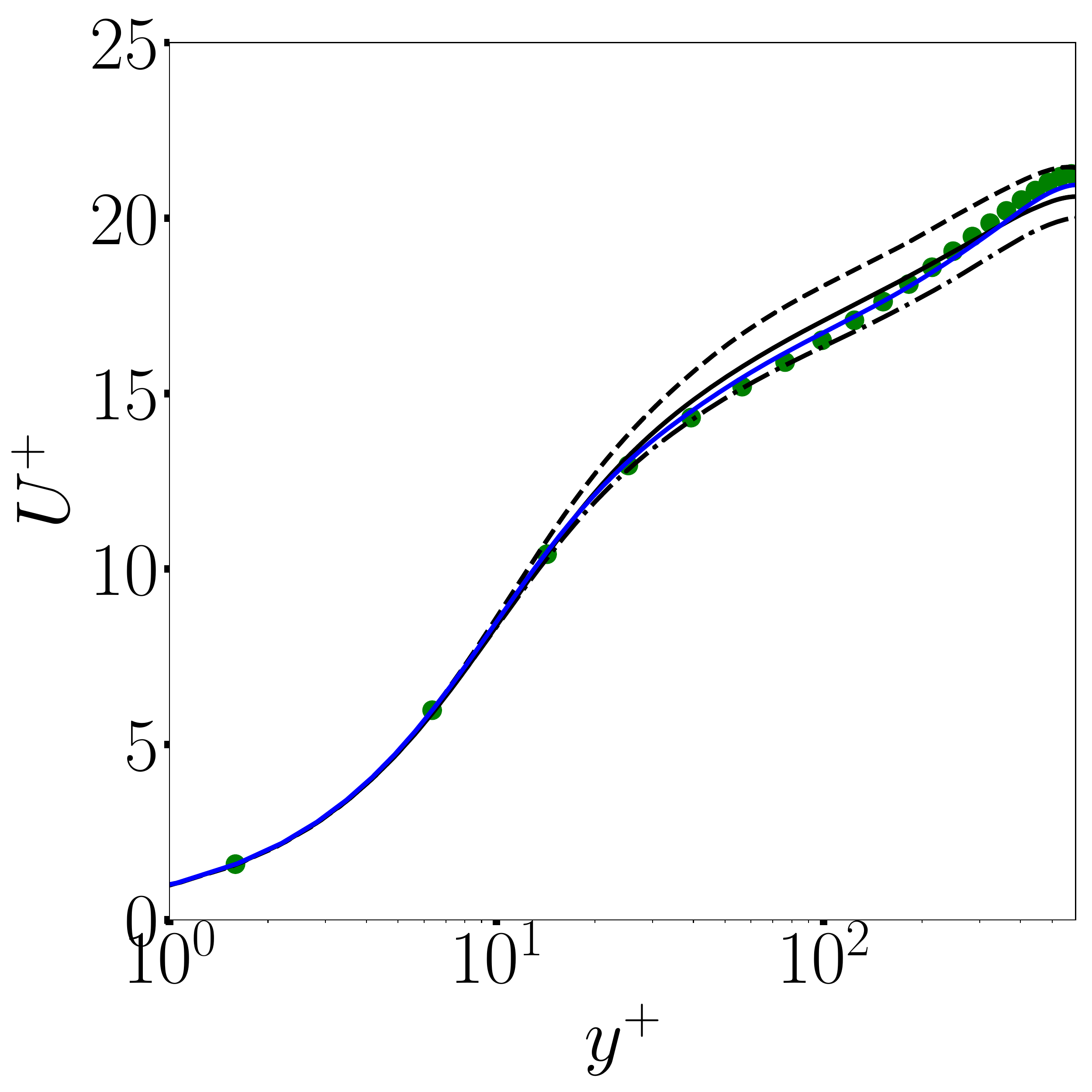}
\caption{\label{fig:Vincent_case5_C_U}Mean profile $U(y)$}
\end{subfigure}
\begin{subfigure}{.28\linewidth}
\centering
\includegraphics[width=1.\textwidth]{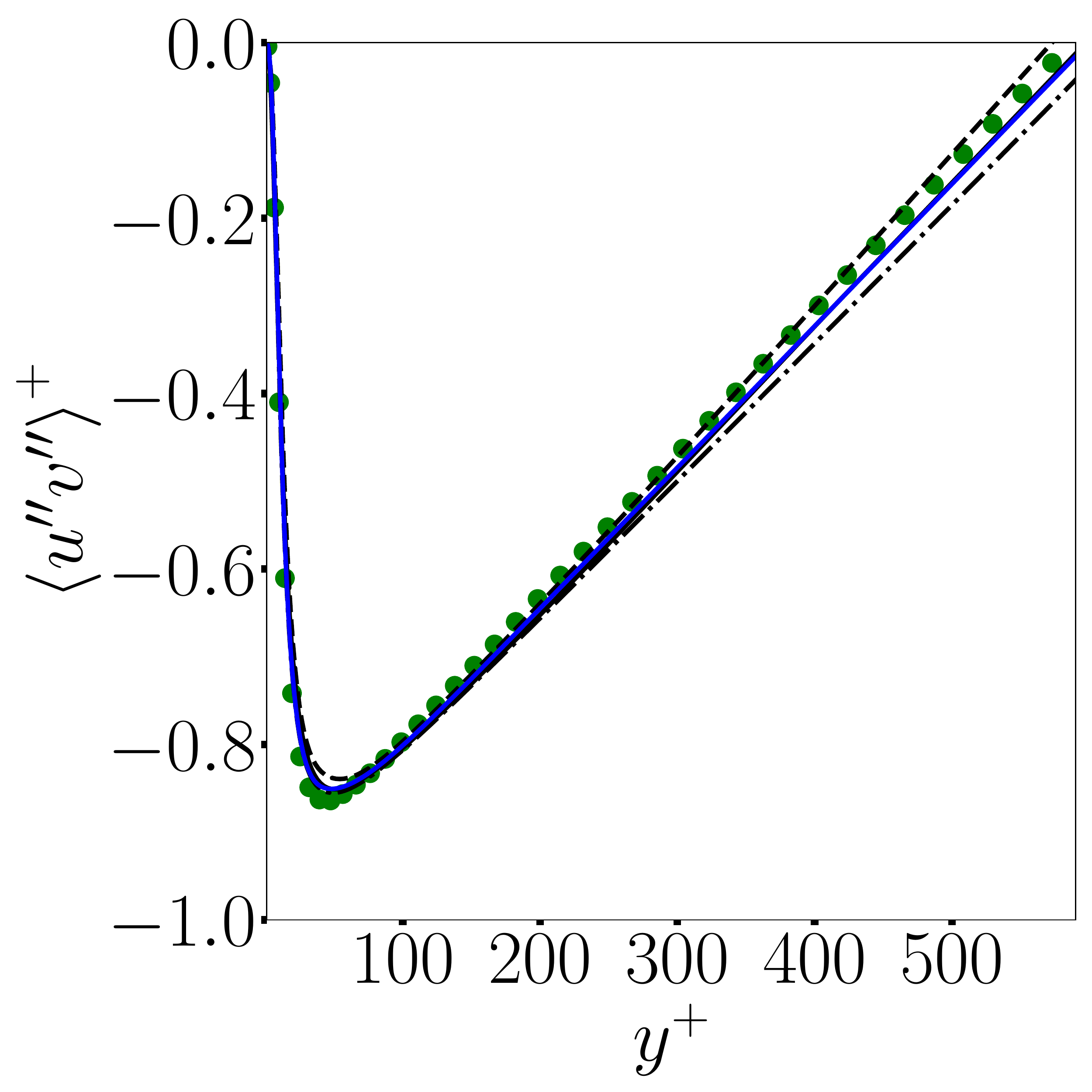}
\caption{\label{fig:Vincent_case5_C_uv}Total shear stress $\left<u''v''\right>$}
\end{subfigure}%
\caption{\label{fig:Vincent_case5}Control vector and flow statistics of case 5 for DNS (\mycircle{black!50!green} \mycircle{black!40!green} \mycircle{black!40!green}), the dynamic model (\fullblack), the mixed model (\chainblack), the Smagorinsky model (\dashed) and for final DA-LES (\full).}
\end{figure}

\subsection{Recovering reference statistics from mean-flow and Reynolds-stress observations (cases 5 and 5c)}\label{sec:results_DA_statistics_Re_590_cases5_5c}

\begin{figure}
\centering
\begin{subfigure}{.28\linewidth}
\centering
\includegraphics[width=1.\textwidth]{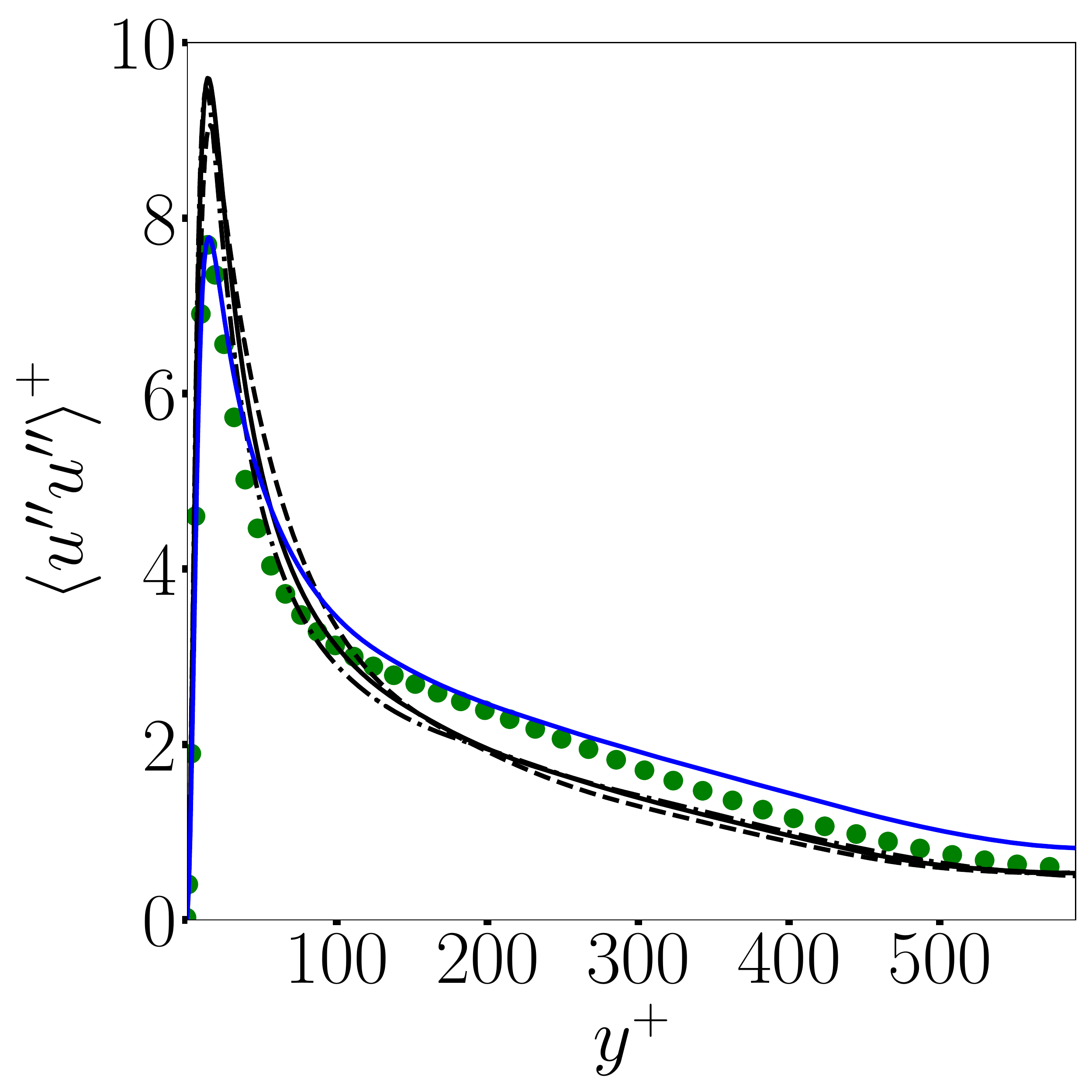}
\caption{\label{fig:Vincent_case5_C_uu}Reynolds stress $\left<u''u''\right>$}
\end{subfigure}%
\begin{subfigure}{.28\linewidth}
\centering
\includegraphics[width=1.\textwidth]{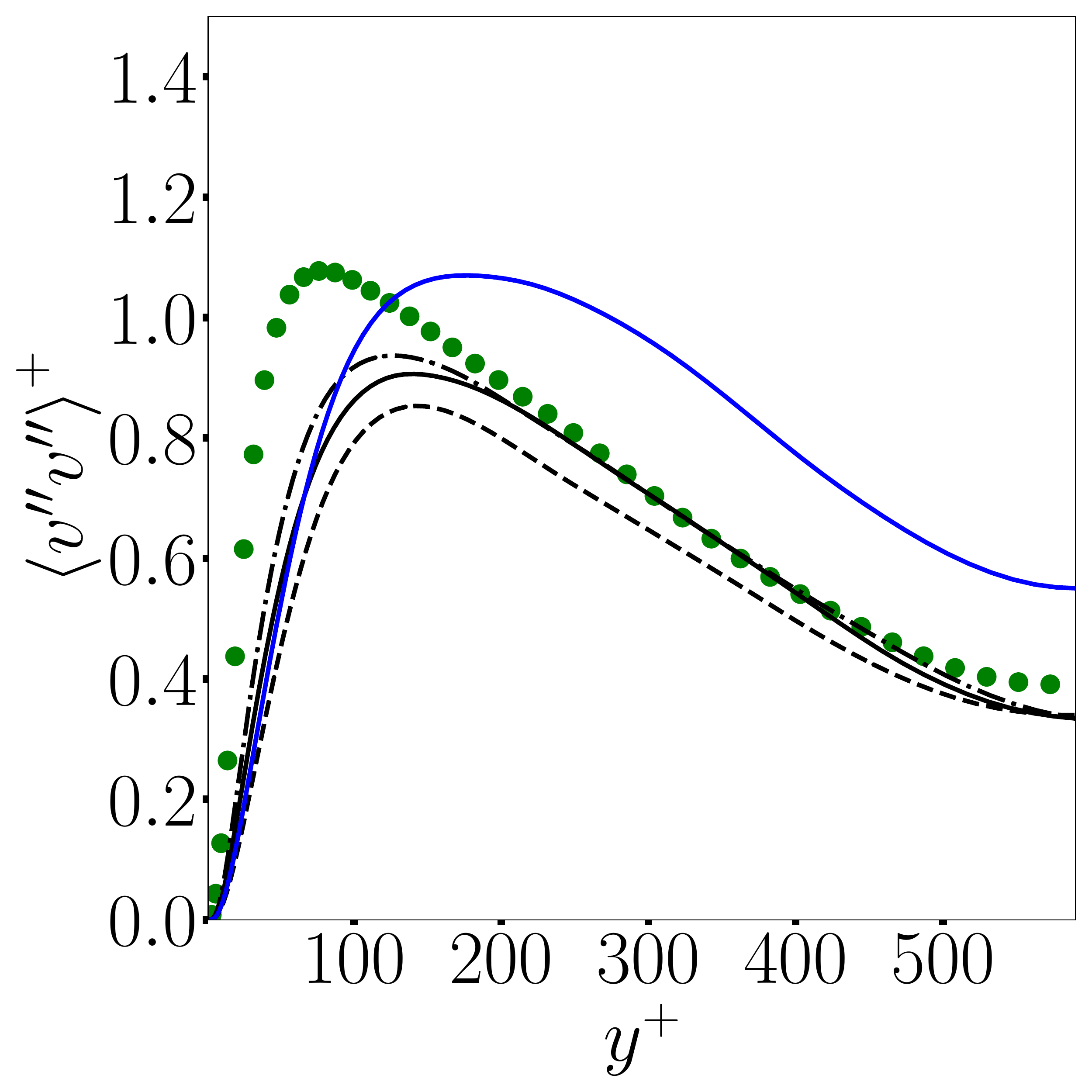}
\caption{\label{fig:Vincent_case5_C_vv}Reynolds stress $\left<v''v''\right>$}
\end{subfigure}
\begin{subfigure}{.28\linewidth}
\centering
\includegraphics[width=1.\textwidth]{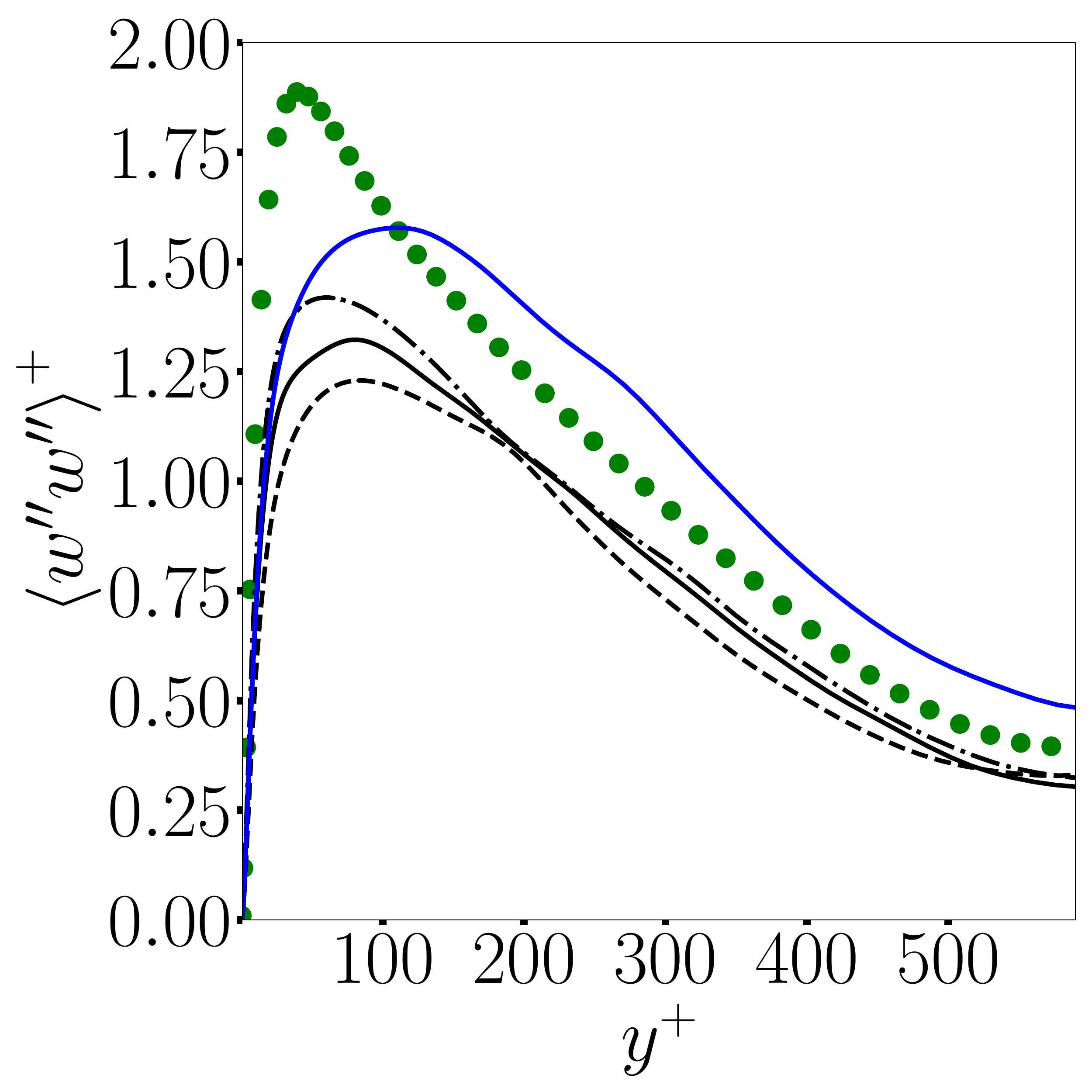}
\caption{\label{fig:Vincent_case5_C_ww}Reynolds stress $\left<w''w''\right>$}
\end{subfigure}
\begin{subfigure}{.28\linewidth}
\centering
\includegraphics[width=1.\textwidth]{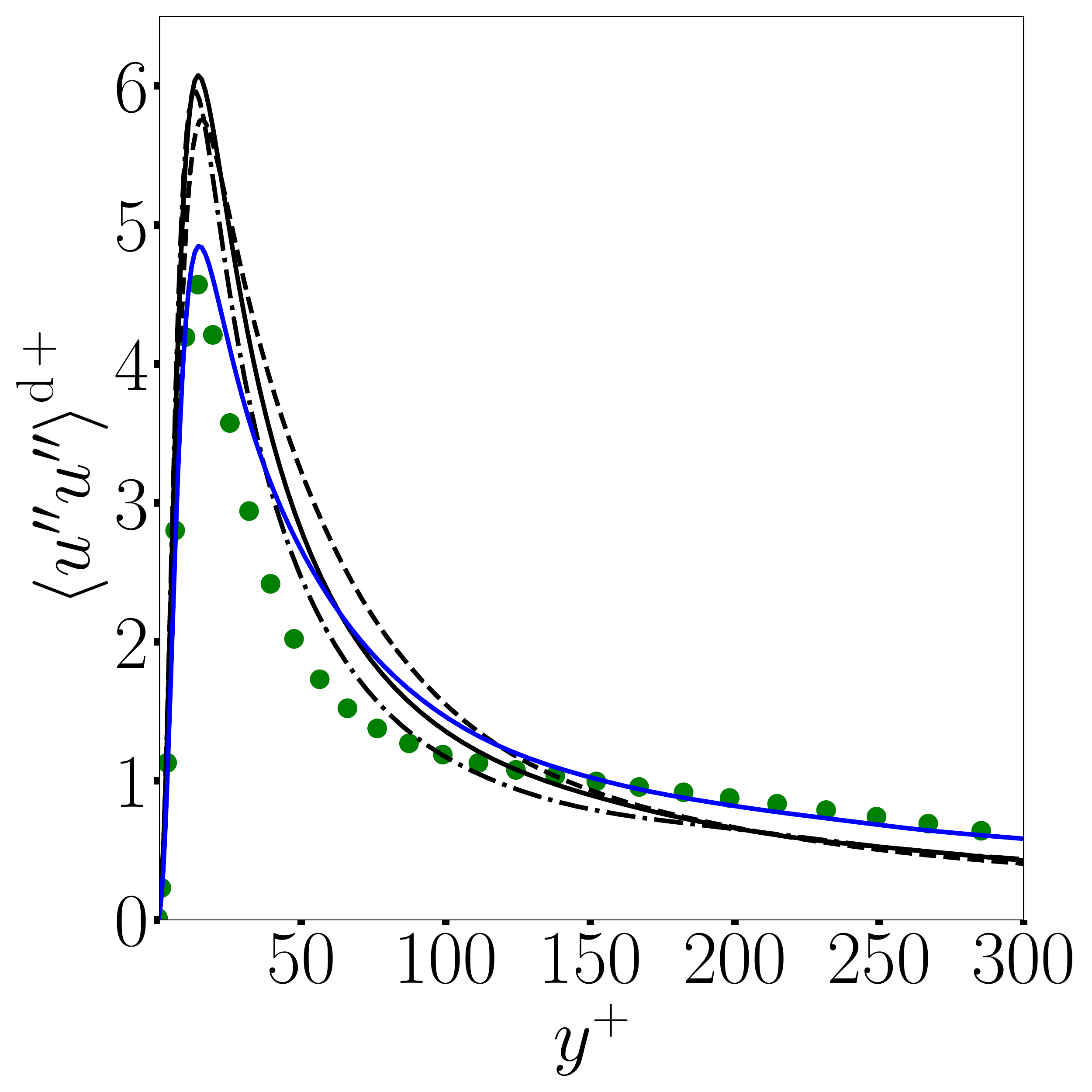}
\caption{\label{fig:Vincent_case5_C_duu}Deviatoric Reynolds stress $\left<u''u''\right>^{\mathrm{d}}$}
\end{subfigure}%
\begin{subfigure}{.28\linewidth}
\centering
\includegraphics[width=1.\textwidth]{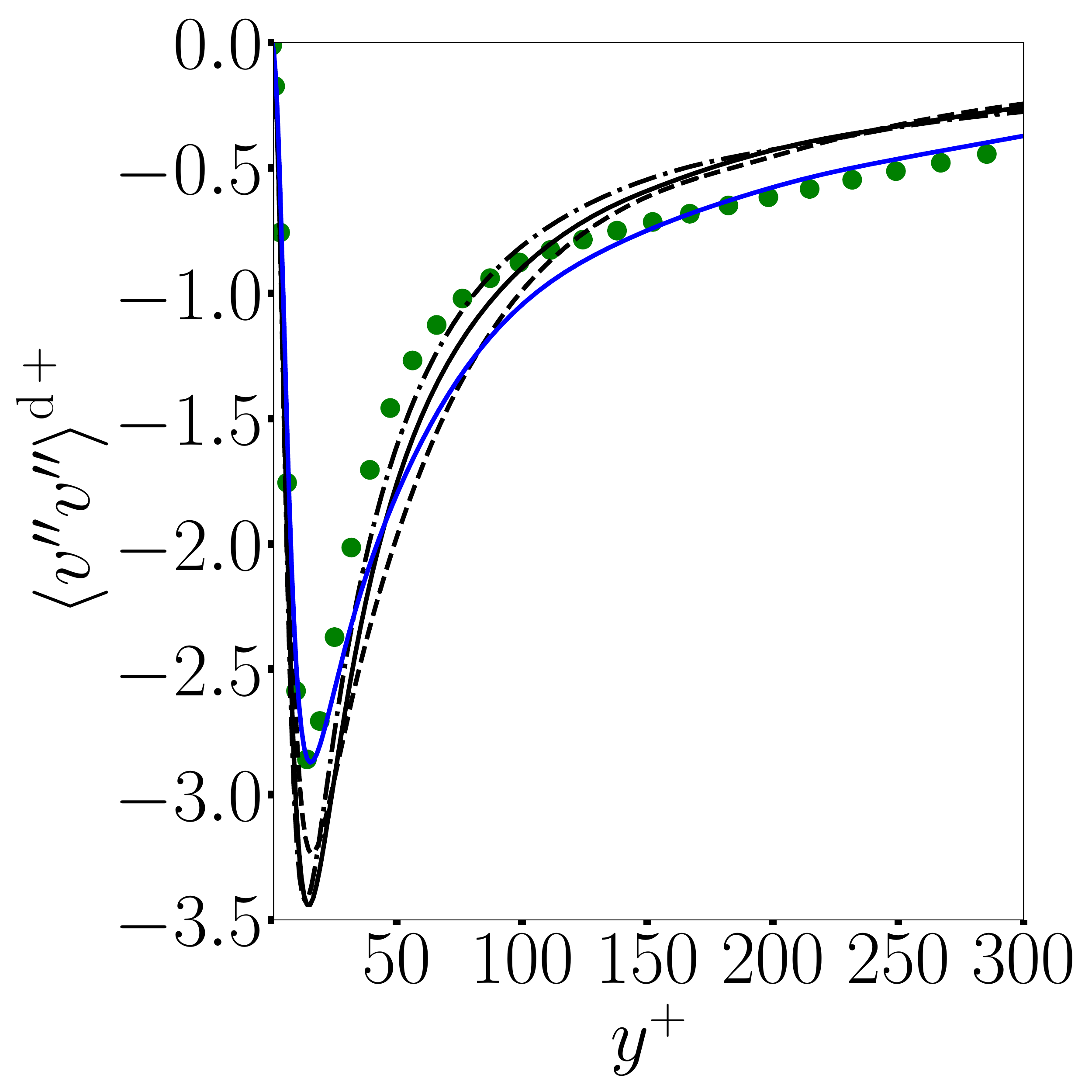}
\caption{\label{fig:Vincent_case5_C_dvv}Deviatoric Reynolds stress $\left<v''v''\right>^{\mathrm{d}}$}
\end{subfigure}
\begin{subfigure}{.28\linewidth}
\centering
\includegraphics[width=1.\textwidth]{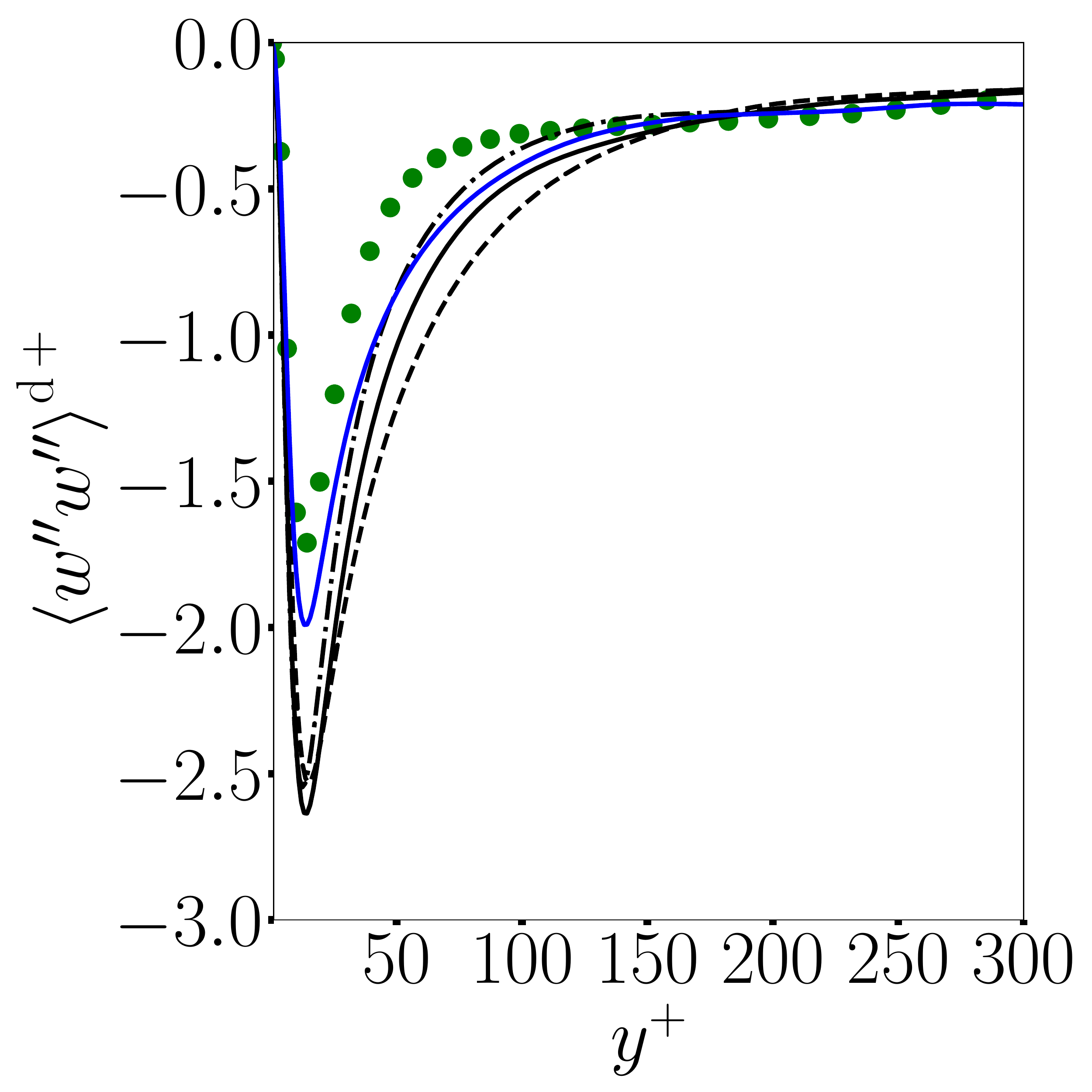}
\caption{\label{fig:Vincent_case5_C_dww}Deviatoric Reynolds stress $\left<w''w''\right>^{\mathrm{d}}$}
\end{subfigure}
\caption{\label{fig:Vincent_case5_turbulence}Second-order statistics of case 5 for DNS (\mycircle{black!50!green} \mycircle{black!40!green} \mycircle{black!40!green}), the dynamic model (\fullblack), the mixed model (\chainblack), the Smagorinsky model (\dashed) and for final DA-LES (\full).}
\end{figure}

\begin{figure}
\centering
\begin{subfigure}{.28\linewidth}
\centering
\includegraphics[width=1.\textwidth]{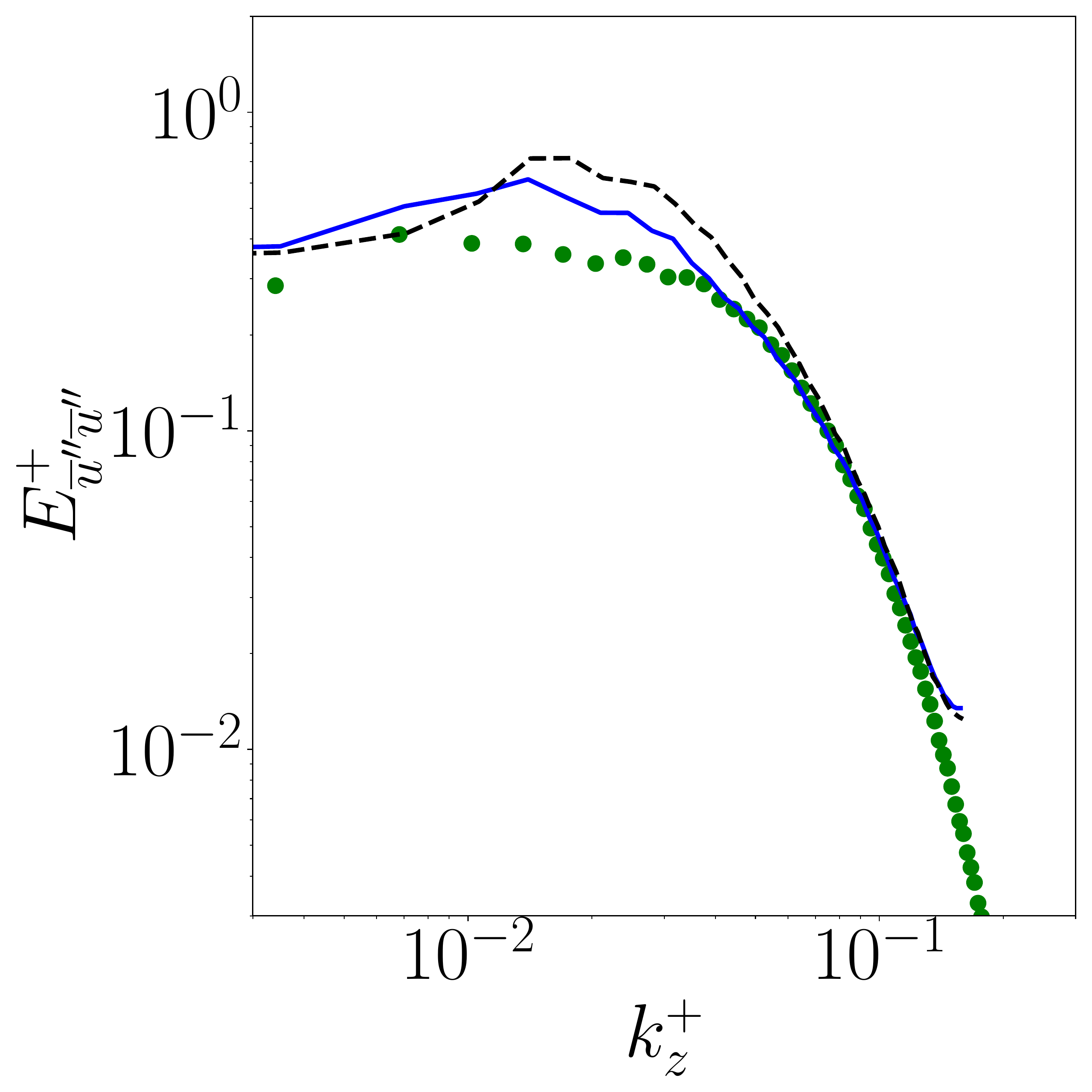}
\caption{Case 5, $y^{*}=19$}
\end{subfigure}
\begin{subfigure}{.28\linewidth}
\centering
\includegraphics[width=1.\textwidth]{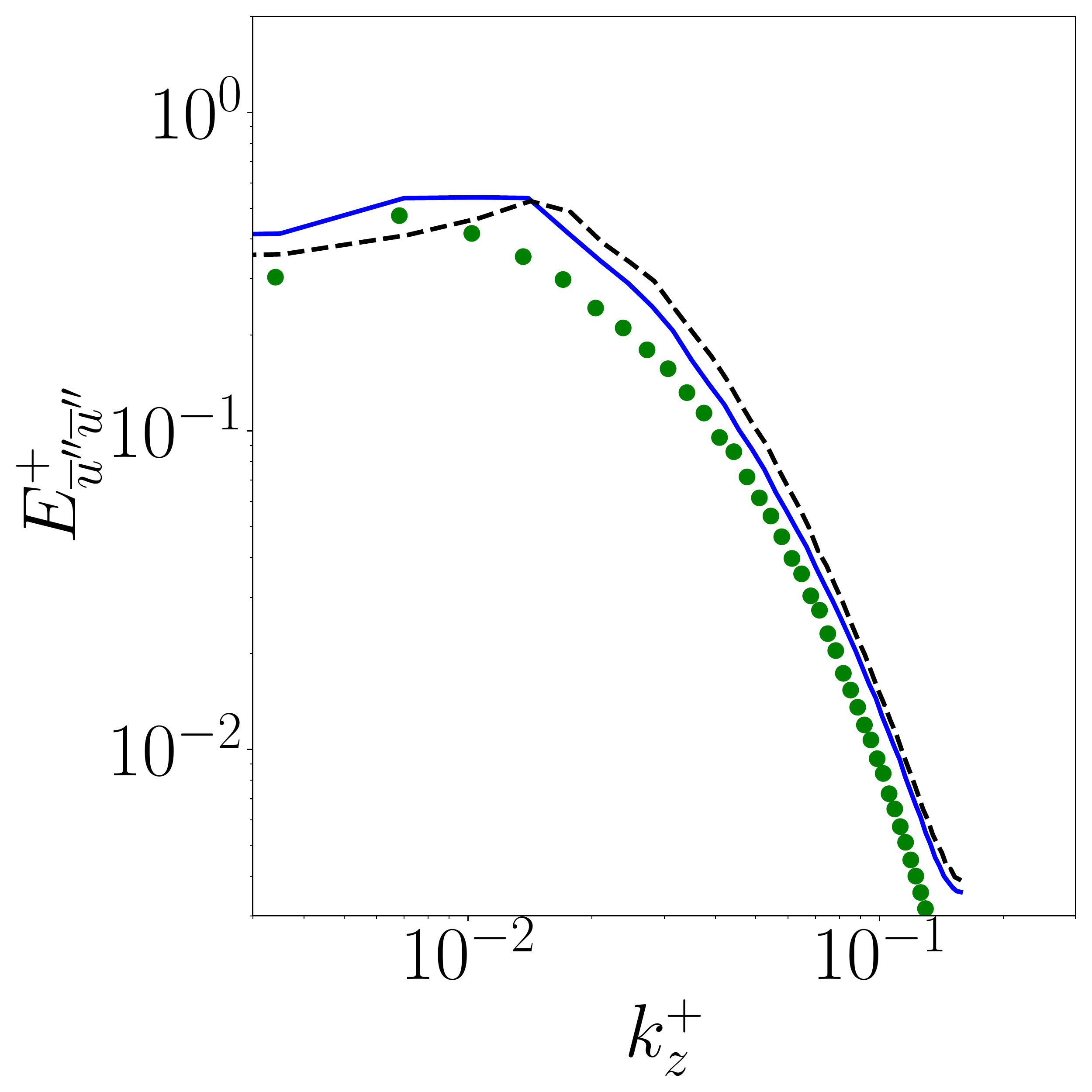}
\caption{Case 5, $y^{*}=59$}
\end{subfigure}\\
\begin{subfigure}{.28\linewidth}
\centering
\includegraphics[width=1.\textwidth]{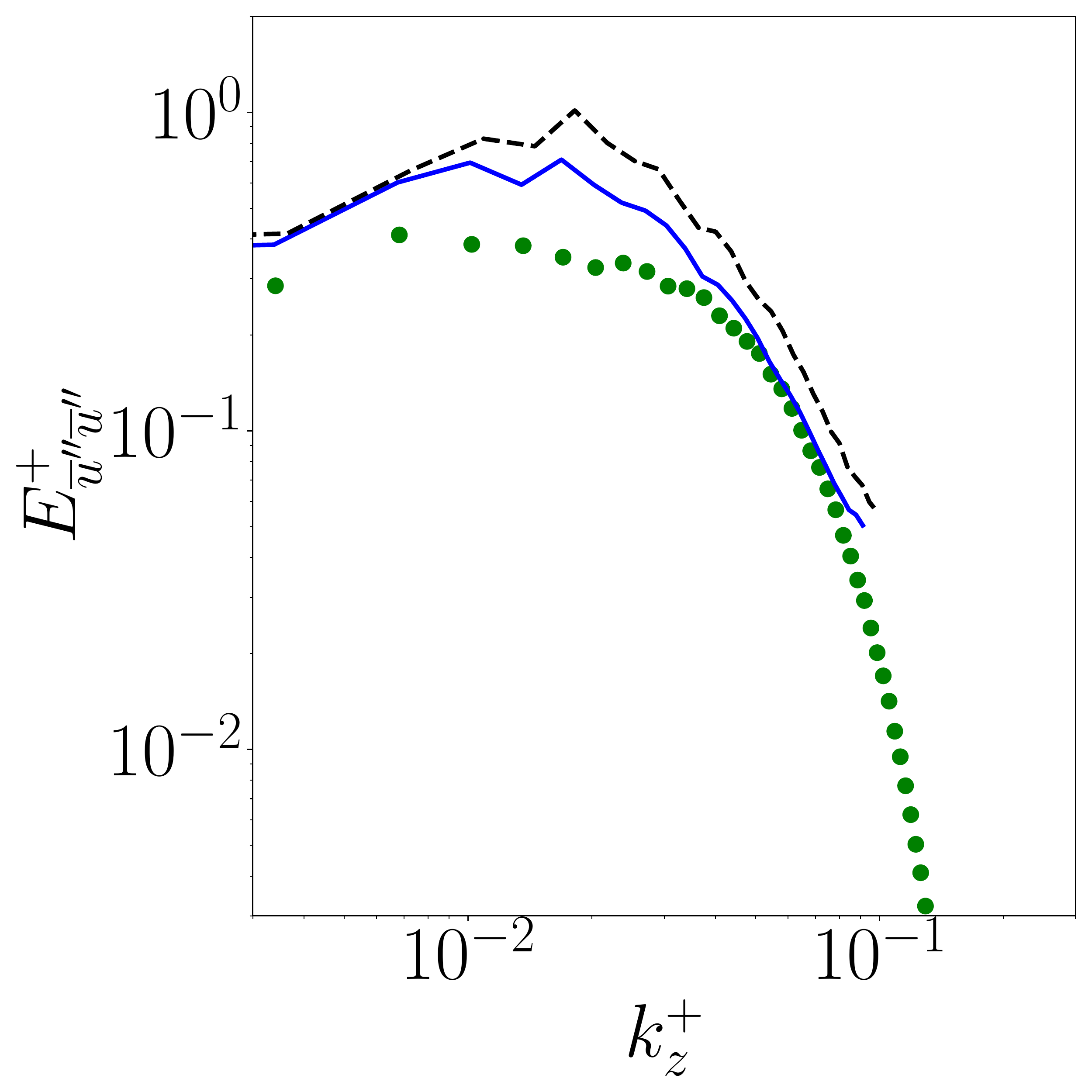}
\caption{\label{fig:case5c_spectrum_y_19}Case 5c, $y^{*}=19$}
\end{subfigure}
\begin{subfigure}{.28\linewidth}
\centering
\includegraphics[width=1.\textwidth]{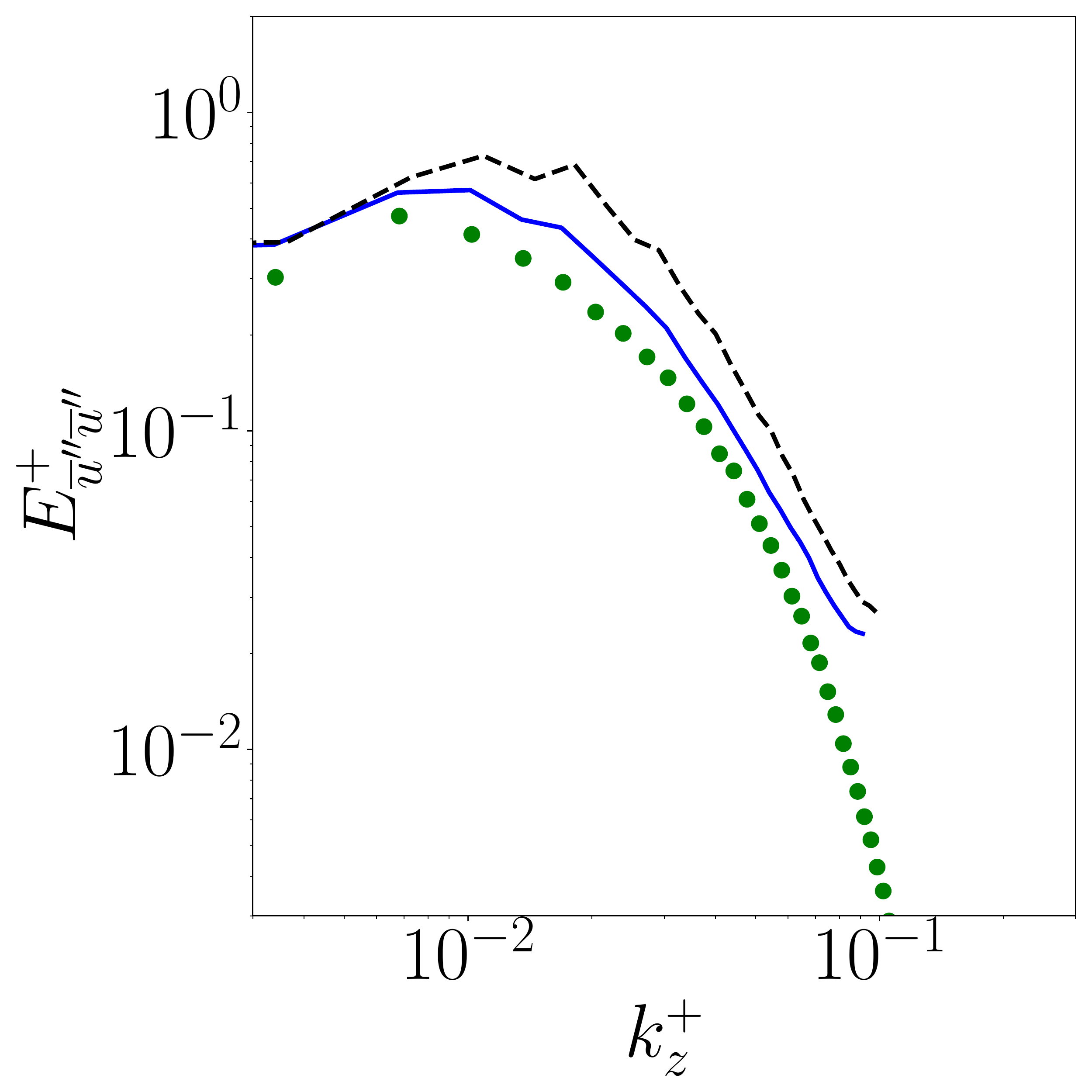}
\caption{\label{fig:case5c_spectrum_y_59}Case 5c, $y^{*}=59$}
\end{subfigure}
\caption{\label{fig:spectra_cases_5_5c}One-dimensional spanwise spectra of the resolved streamwise velocity fluctuations $E_{\overline{u}''\overline{u}''}(k_z)$ of cases 5 and 5c at $y^{*}=19$ and $y^{*}=59$. DNS (\mycircle{black!50!green} \mycircle{black!40!green} \mycircle{black!40!green}); Smagorinsky model (\dashed); DA-LES (\full).}
\end{figure}

We now consider the data assimilation experiment 5, which differs from the previous one in two respects: Firstly, the Reynolds stress $\left<u''u''\right>$ is included with the mean-flow gradient in the observations $\boldsymbol{m}$. This is mainly motivated by the above discussion and to better control the production $P_r$. Secondly, the control vector $\boldsymbol{\gamma}$ includes both the forcing $\sigma$ and the coefficient $C_{\mathrm{s}}$, the latter being reintroduced to better tackle the consideration of $\left<u''u''\right>$ as one of the observed quantities. The final assimilated state is obtained through two main iterations of the data assimilation procedure, and the corresponding results are reported in figures \ref{fig:Vincent_production}-\ref{fig:Vincent_case5_turbulence} and \ref{fig:spectra_cases_5_5c}. 
Figure \ref{fig:Vincent_R_P_r} shows that including the observation of the Reynolds stress component $\left<u''u''\right>$ in the data assimilation has indeed mitigated the increase in the production $P_r$ of resolved kinetic energy compared to case 4. The adjustment of $P_r$ with respect to the Smagorinsky model even allows DA-LES to roughly recover the reference DNS profile in the present case. In addition, as in case 4, the production $P$ of total turbulent kinetic energy is accurately predicted by DA-LES (figure \ref{fig:Vincent_R_P}).

In addition to comparisons with statistics from the reference DNS and the standard Smagorinsky model, the present DA-LES is also compared to results from the dynamic and mixed models in figures \ref{fig:Vincent_case5}-\ref{fig:Vincent_case5_turbulence}. The assimilated control vector $\boldsymbol{\gamma}$ is reported in figures \ref{fig:Vincent_case5_C_C}-\ref{fig:Vincent_case5_C_sigma}, and shows that the amplitude of the optimal forcing $\sigma$ is significantly reduced relative to case 4 (figure \ref{fig:Vincent_case4_R_sigma}). While the influence of $\sigma$ on the resolved shear stress $\left<\overline{u}''\overline{v}''\right>$ was previously noted, the present lower value of $\sigma$ at the start of the log layer leads to a better estimation of the production $P_r$ of resolved turbulent kinetic energy. The coefficient $C_{\mathrm{s}}$ is only mildly altered by the data assimilation procedure compared to the standard Smagorinsky model, in particular with a small increase around $y^+ =50$. Such localized peak is also qualitatively observed in the dynamic model at a similar location, although the corresponding values significantly differ. 

Figure \ref{fig:Vincent_case5_C_U} indicates that among all considered subgrid models, only DA-LES satisfactorily predicts the mean velocity $U$ over the whole channel height. The standard Smagorinsky, dynamic and mixed models predict an erroneous profile for $U$ above the start of the log layer. These mean-flow profiles would actually collapse close to the channel center if normalized using the same reference friction velocity. This result further underscores the lack of influence of the subgrid model in this region as discussed in \S\ref{sec:cases_1_2}, which is here observed even with the mixed model whose functional form significantly differs from the Smagorinsky and dynamic models.

The profiles of the diagonal Reynolds stress components are reported in figure \ref{fig:Vincent_case5_turbulence} for case 5. All baseline subgrid models significantly overestimate the intensity of $\left<u''u''\right>$ close to the wall. In contrast, the data assimilation procedure successfully recovers the observed reference profile apart from a slight overestimation close to the channel center, which also reflects in the unobserved components $\left<v''v''\right>$ and $\left<w''w''\right>$. The latter two still seem overall improved compared to the baseline models, in particular $\left<w''w''\right>$. This is confirmed by the examination of the deviatoric part of the Reynolds stress tensor as illustrated in figures \ref{fig:Vincent_case5_C_duu}-\ref{fig:Vincent_case5_C_dww}. DA-LES accurately predicts the degree of anisotropy of the flow over the whole channel height, even close to the channel center. These results demonstrate the efficacy of the present data assimilation procedure and confirm its validity in terms of choice of the control vector and observations. The present results also set a benchmark for the best achievable predictions by eddy-viscosity models in terms of prediction of the mean flow and the Reynolds stress tensor. Further quantitative assessment of the LES predictions will be provided in \S\ref{sec:further_assessment}.

\begin{figure}
\centering
\begin{subfigure}{.28\linewidth}
\centering
\includegraphics[width=1.\textwidth]{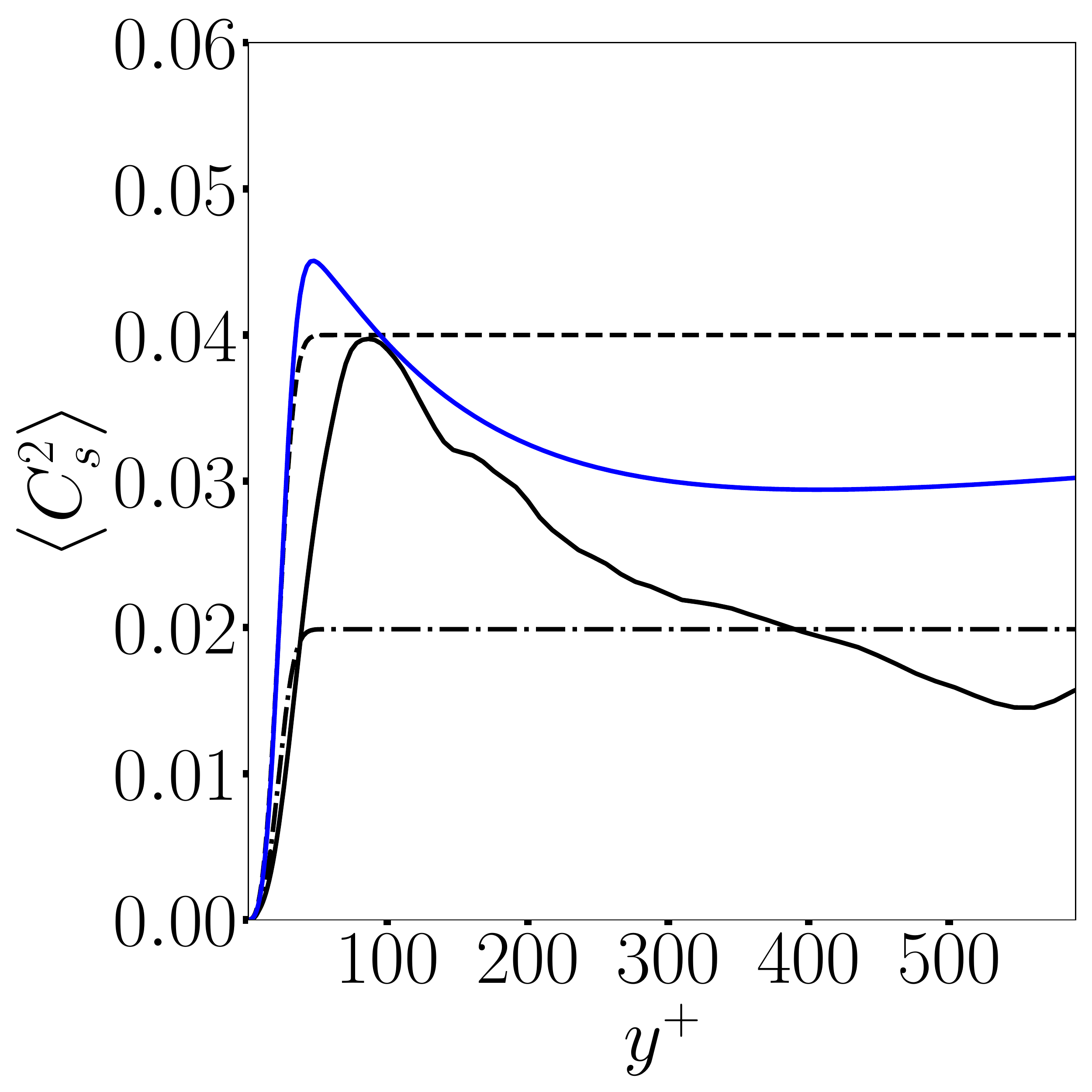}
\caption{\label{fig:Vincent_case5c_C_C}Control vector $C_\mathrm{s}(y)$}
\end{subfigure}%
\begin{subfigure}{.28\linewidth}
\centering
\includegraphics[width=1.\textwidth]{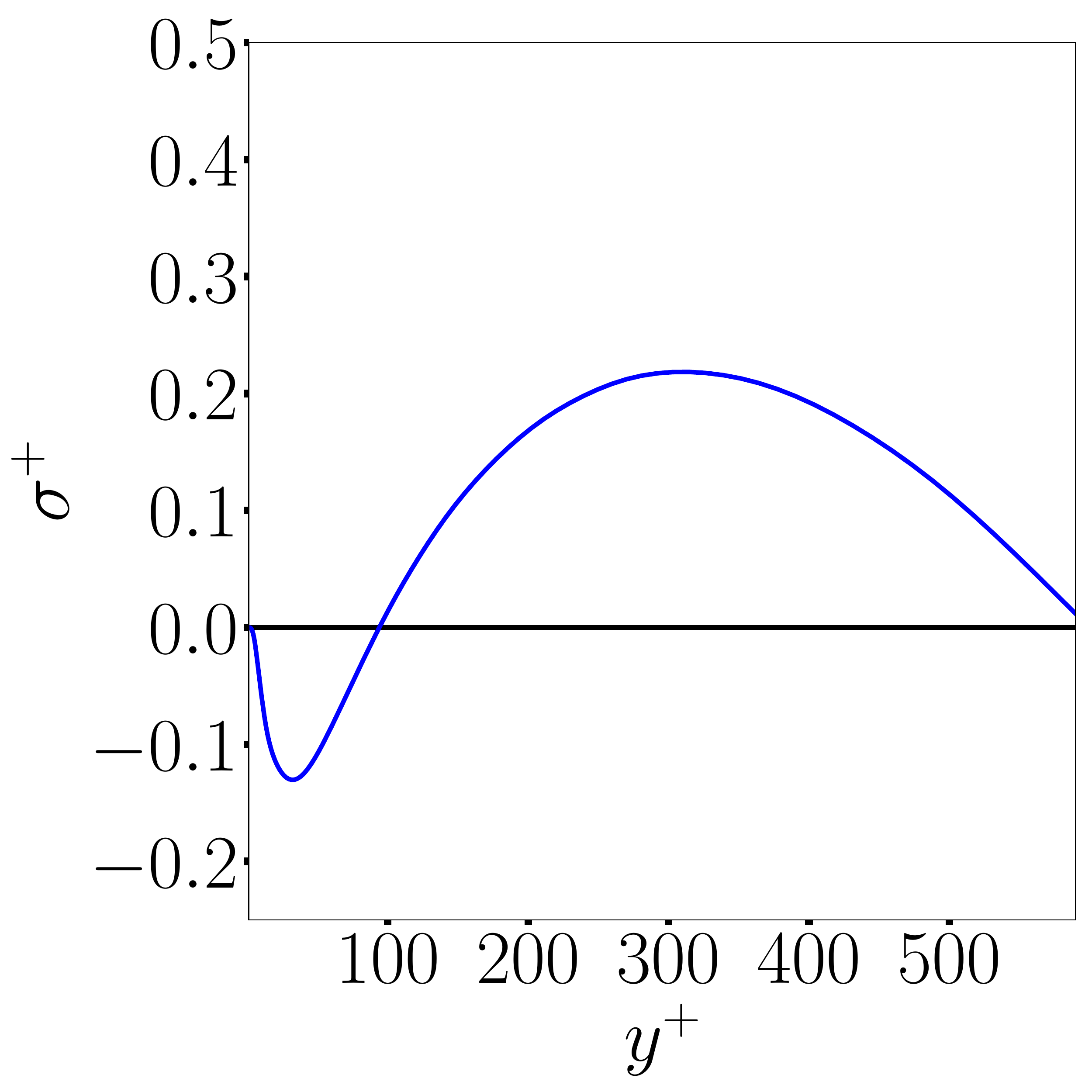}
\caption{\label{fig:Vincent_case5c_C_sigma}Control vector $\sigma(y)$}
\end{subfigure}
\begin{subfigure}{.28\linewidth}
\centering
\includegraphics[width=1.\textwidth]{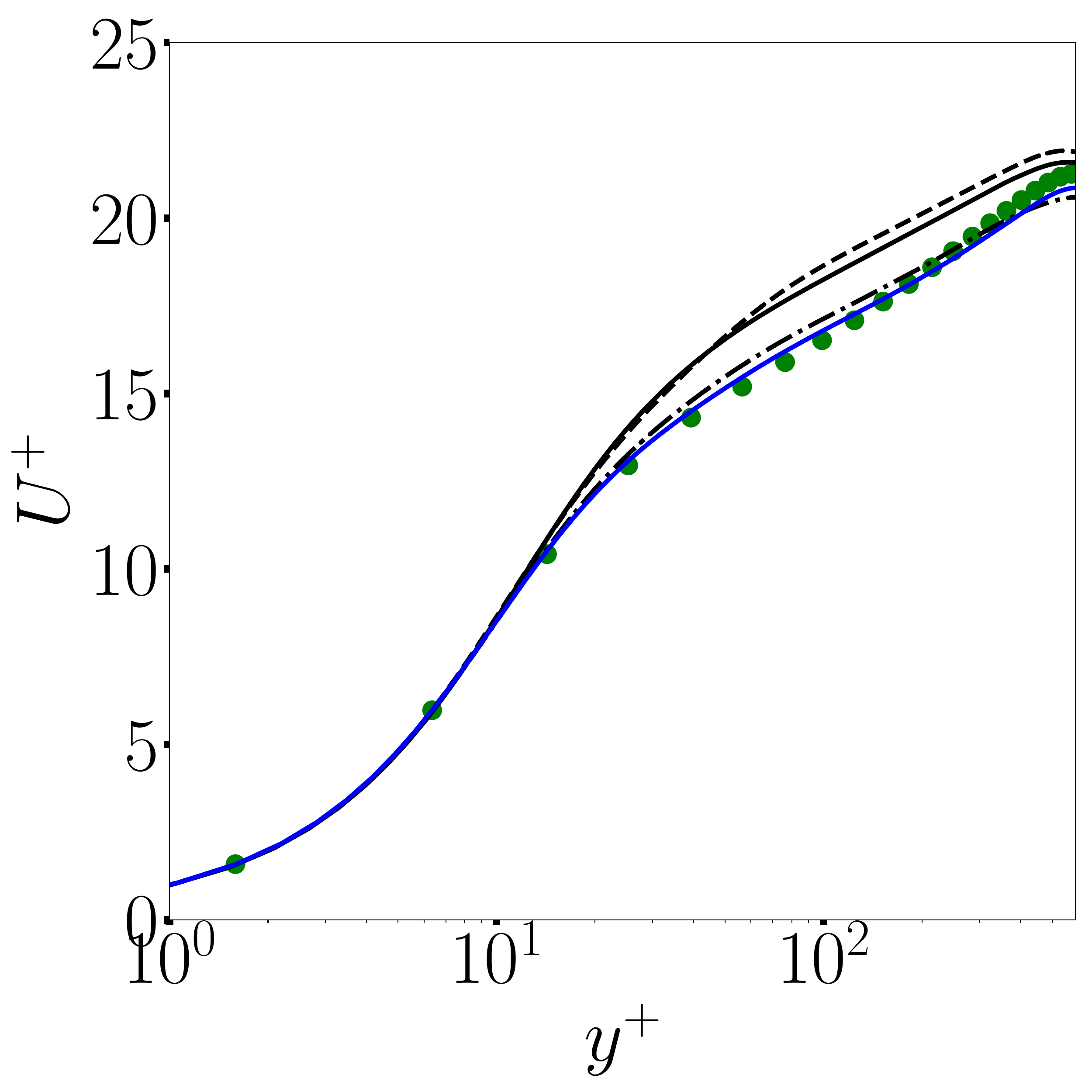}
\caption{\label{fig:Vincent_case5c_C_U}Mean profile $U(y)$}
\end{subfigure}
\begin{subfigure}{.28\linewidth}
\centering
\includegraphics[width=1.\textwidth]{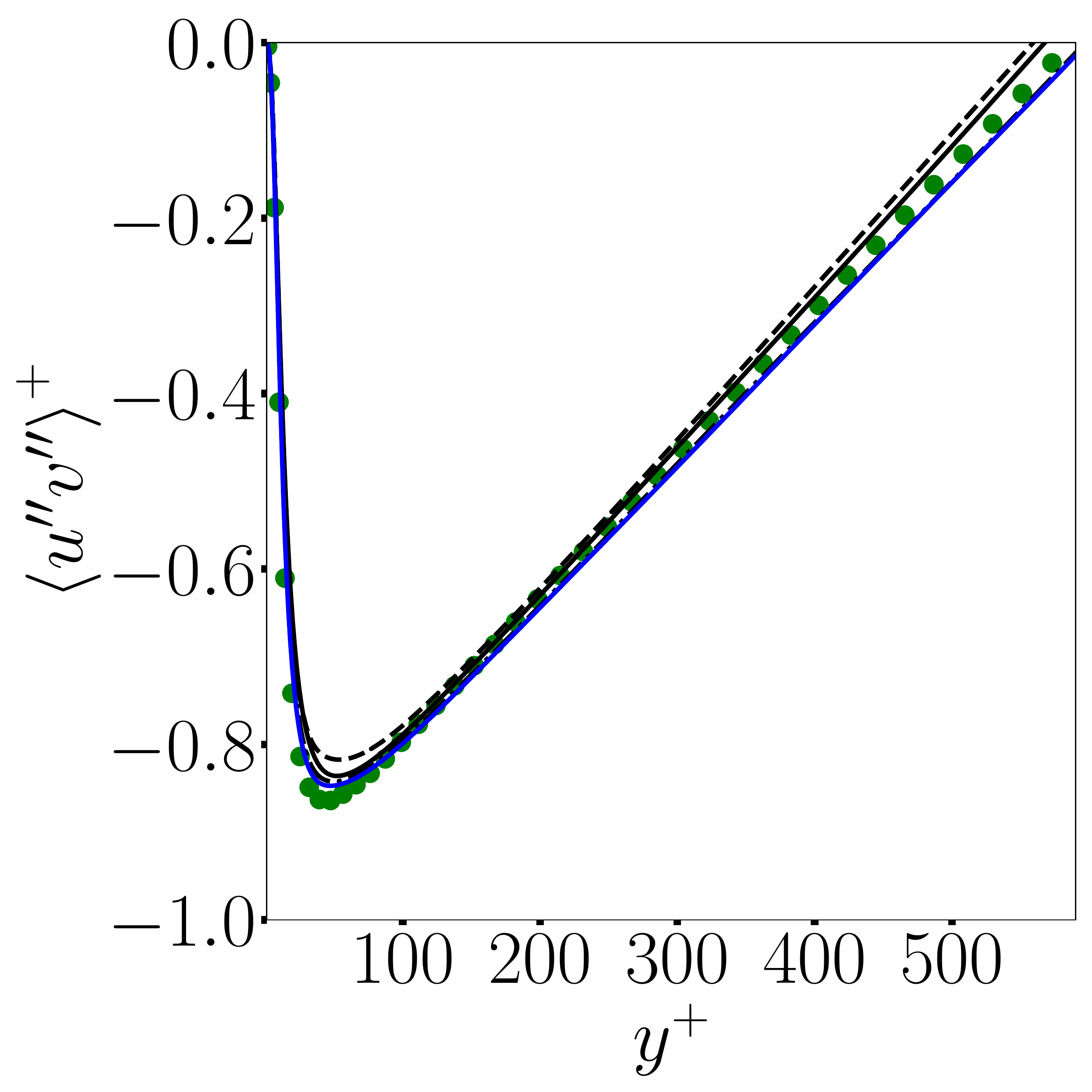}
\caption{\label{fig:Vincent_case5c_C_uv}Total shear stress $\left<u''v''\right>$}
\end{subfigure}
\begin{subfigure}{.28\linewidth}
\centering
\includegraphics[width=1.\textwidth]{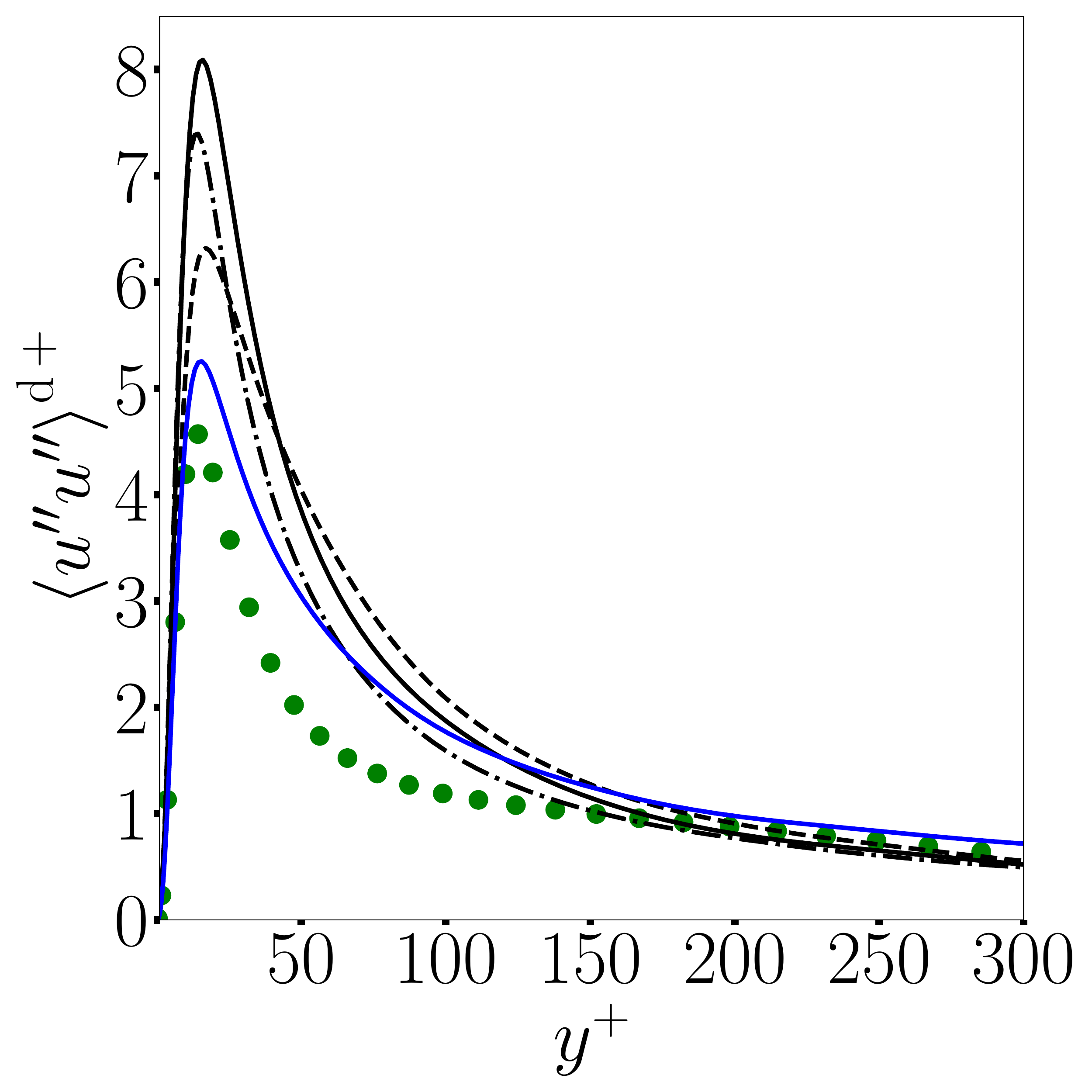}
\caption{\label{fig:Vincent_case5c_C_duu}Deviatoric Reynolds stress $\left<u''u''\right>^{\mathrm{d}}$}
\end{subfigure}%
\begin{subfigure}{.28\linewidth}
\centering
\includegraphics[width=1.\textwidth]{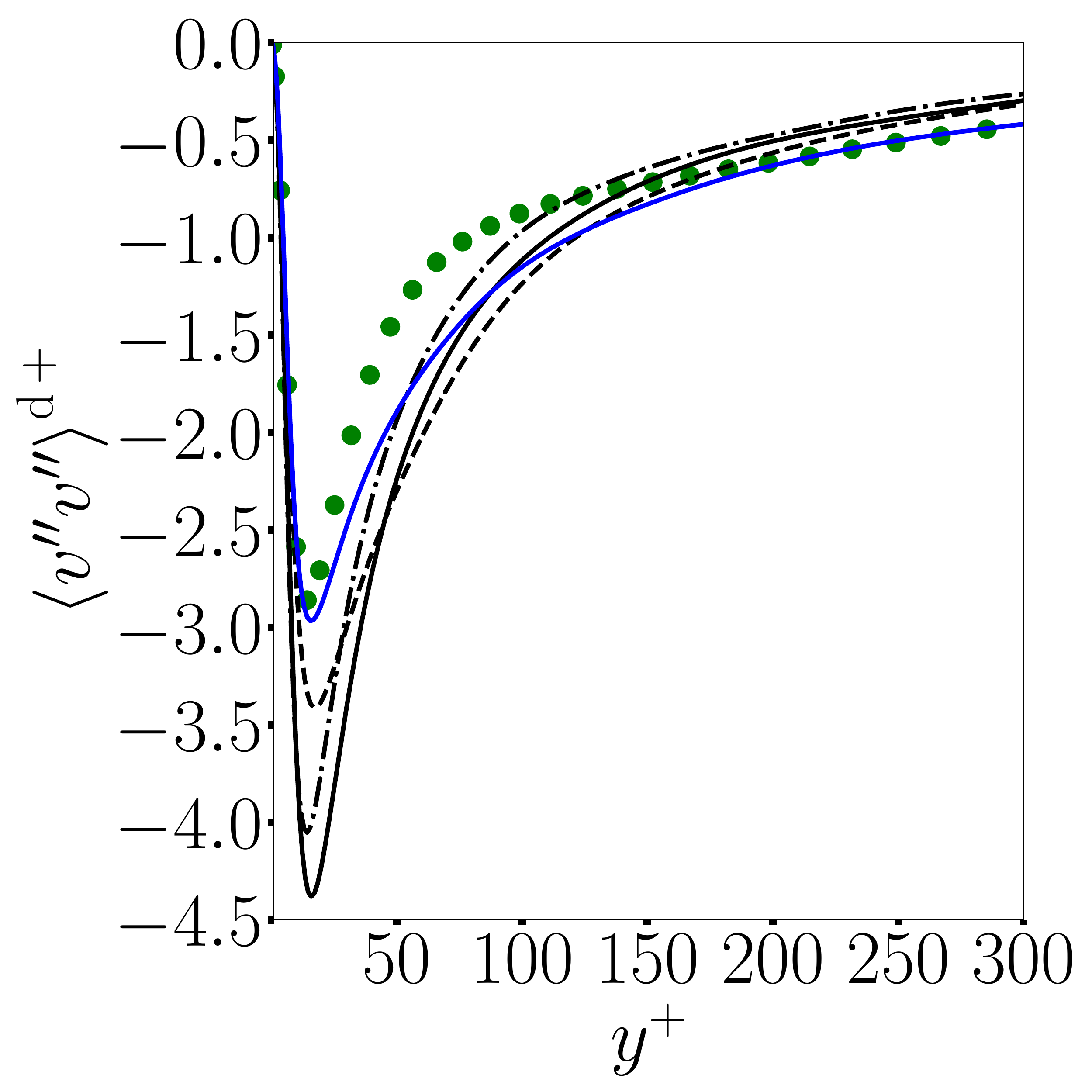}
\caption{\label{fig:Vincent_case5c_C_dvv}Deviatoric Reynolds stress $\left<v''v''\right>^{\mathrm{d}}$}
\end{subfigure}
\caption{\label{fig:Vincent_case5c}Control vector and flow statistics of case 5c for DNS (\mycircle{black!50!green} \mycircle{black!40!green} \mycircle{black!40!green}), the dynamic model (\fullblack), the mixed model (\chainblack), the Smagorinsky model (\dashed) and for final DA-LES (\full).}
\end{figure}

The robustness of the above findings is assessed in data assimilation experiment 5c, which adopts the coarser grid LES590c relative to the previous experiment. Again, only two main iterations of the data assimilation procedure are performed to obtain the state illustrated in figure \ref{fig:Vincent_case5c}. The profiles of the assimilated control vector in figures \ref{fig:Vincent_case5c_C_C}-\ref{fig:Vincent_case5c_C_sigma} are similar to those for case 5 (figures \ref{fig:Vincent_case5_C_C}-\ref{fig:Vincent_case5_C_sigma}), with the exception of the value of the coefficient $C_{\mathrm{s}}$ close to the channel center which is overall lower in the present case. This point will be revisited in \S\ref{sec:results_DA_statistics_uncertainties} to assess its significance in the context of uncertainty of the assimilated state.
As in the previous case 5, the mean flow in case 5c is best predicted by DA-LES. Concerning the Reynolds stress tensor, only its deviatoric part and more specifically the components $\left<u''v''\right>^{\mathrm{d}}=\left<u''v''\right>$, $\left<u''u''\right>^{\mathrm{d}}$ and $\left<v''v''\right>^{\mathrm{d}}$ are reported in figures \ref{fig:Vincent_case5c_C_duu}-\ref{fig:Vincent_case5c_C_dvv} for the sake of conciseness, keeping in mind that $\left<w''w''\right>^{\mathrm{d}}=-\left<u''u''\right>^{\mathrm{d}}-\left<v''v''\right>^{\mathrm{d}}$. The discrepancies between the baseline subgrid models and the reference profiles for $\left<u''u''\right>^{\mathrm{d}}$ and $\left<v''v''\right>^{\mathrm{d}}$ appear larger than in case 5 (figures \ref{fig:Vincent_case5_C_duu}-\ref{fig:Vincent_case5_C_dvv}) due to the use of a coarser grid. DA-LES provides an appreciable improvement in the prediction of these quantities and thus significantly outperforms the three considered baseline models.

The improvement in the estimation of single-point statistics using DA-LES will be examined further in \S\ref{sec:further_assessment}.  Here, we turn to the spectral content of the predictions using the Smagorinsky model and DA-LES in cases 5 and 5c.  Figure \ref{fig:spectra_cases_5_5c} shows the one-dimensional spanwise spectra of the resolved streamwise fluctuations $E_{\overline{u}''\overline{u}''}(k_z)$ at two wall-normal heights: at $y^{*}=19$ which is in the buffer layer and relatively close to where $\left<u''u''\right>$ reaches its maximum value ($y^{*}\simeq 14$), and at $y^{*}=59$ which lies in the log layer. The results in figure \ref{fig:spectra_cases_5_5c} confirm that, for both cases 5 and 5c, the improvement in the prediction of the (observed) Reynolds stress $\left<u''u''\right>$ between the first-guess Smagorinsky model and DA-LES also translates in a better scale-by-scale estimation of the flow. In order to facilitate comparisons, the relative discrepancies between DNS and LES for estimation of $E_{\overline{u}''\overline{u}''}(k_z)$ are evaluated as
\begin{equation}\label{eq:spectral_error}
    e_{E_{\overline{u}''\overline{u}''}}=\frac{\int\left| E^{+}_{\overline{u}''\overline{u}''\,\mathrm{DNS}}(k_z^{+})-E^{+}_{\overline{u}''\overline{u}''\,\mathrm{LES}}(k_z^{+})  \right|dk_z^{+}}{\int E^{+}_{\overline{u}''\overline{u}'' \,\mathrm{DNS}}(k_z^{+}) dk_z^{+}},
\end{equation}
where the integrals are performed over the range of spanwise wavenumbers that are resolved by LES. The value of $e_{E_{\overline{u}''\overline{u}''}}$ for the different LES calculations from figure \ref{fig:spectra_cases_5_5c} are reported in table \ref{tab:errors_spectra}. 
From a comparison of cases 5 and 5c (fine and coarse grids, LES590f and LES590c in table \ref{tab:grid_resolutions}), the deterioration in the estimation of the spectra $E_{\overline{u}''\overline{u}''}(k_z)$ is significant for the Smagorinsky model, in particular with a further overestimation of the energy at relatively large scales ($k_z^{+} \leq 10^{-1.5}$), and discrepancies become overall of the same order magnitude as the energy density itself for case 5c ($e_{E_{\overline{u}''\overline{u}''}}\geq 0.8$). While improvements in case 5 through the data assimilation procedure are already appreciable, they are especially significant in case 5c. The spectra appear improved at all wavenumbers for DA-LES (see figures \ref{fig:case5c_spectrum_y_19}-\ref{fig:case5c_spectrum_y_59}), although remaining discrepancies are still noticeable. Furthermore, the errors $e_{E_{\overline{u}''\overline{u}''}}$ are decreased by half compared to the first-guess Smagorinsky model in case 5c.

\begin{table}
\begin{center}
\begin{tabular}{ ccccc} 
 \hline
 \multirow{2}{*}{Model} & \multicolumn{2}{c}{case 5 } & \multicolumn{2}{c}{case 5c}  \\
       & $y^{*}=19$  & $y^{*}=59$ & $y^{*}=19$ & $y^{*}=59$\\
       \hline
       Smagorinsky & $0.39$ & $0.42$ & $0.81$ & $0.84$\\
       DA-LES & $0.24$ & $0.34$ & $0.46$ & $0.42$\\
\hline
\end{tabular}
\caption{\label{tab:errors_spectra}Error on the spanwise spectrum $e_{E_{\overline{u}''\overline{u}''}}$ in (\ref{eq:spectral_error}) for the Smagorinsky and final DA-LES calculations of cases 5 and 5c.}
\end{center}
\end{table}

All the above results confirm the ability of the present data assimilation procedure to significantly enhance LES predictions, firstly in terms of single-point statistics and also in terms of spectral content, even with grid resolutions that are away from usual recommendations. In particular, as detailed in \S\ref{sec:numerical_method} and table \ref{tab:grid_resolutions}, the grid LES590c for case 5c corresponds to a spanwise resolution of $\Delta z^{*}=35$, while usual recommendations are $\Delta z^{*}\sim 20$ (as in case 5 with grid LES590f). The streamwise resolution of grid LES590c ($\Delta x^{*}=70$) is also lower than the guidelines ($\Delta x^{*}=50$, \cite{Sagaut2006_springer,Meyers2007_pof}). The good performance of the data assimilation procedure in case 5c may motivate the consideration of even coarser grids. However, as confirmed by figure \ref{fig:case5c_spectrum_y_19}, if the spanwise resolution is further decreased (i.e.~further reducing the maximum resolved $k_z$), energetic scales would be eliminated from the LES which can not longer be considered wall-resolved. In such case, the observation of resolved Reynolds stresses (e.g.~$\left<\overline{u}''\overline{u}''\right>$) may become more appropriate than total ones (e.g.~$\left<u''u''\right>$), as the difference between the two becomes appreciable.

\subsection{Recovering reference statistics at $Re_{\tau}=1{,}000$ (case 6)}\label{sec:results_DA_statistics_Re_1000}

\begin{figure}
\centering
\begin{subfigure}{.28\linewidth}
\centering
\includegraphics[width=1.\textwidth]{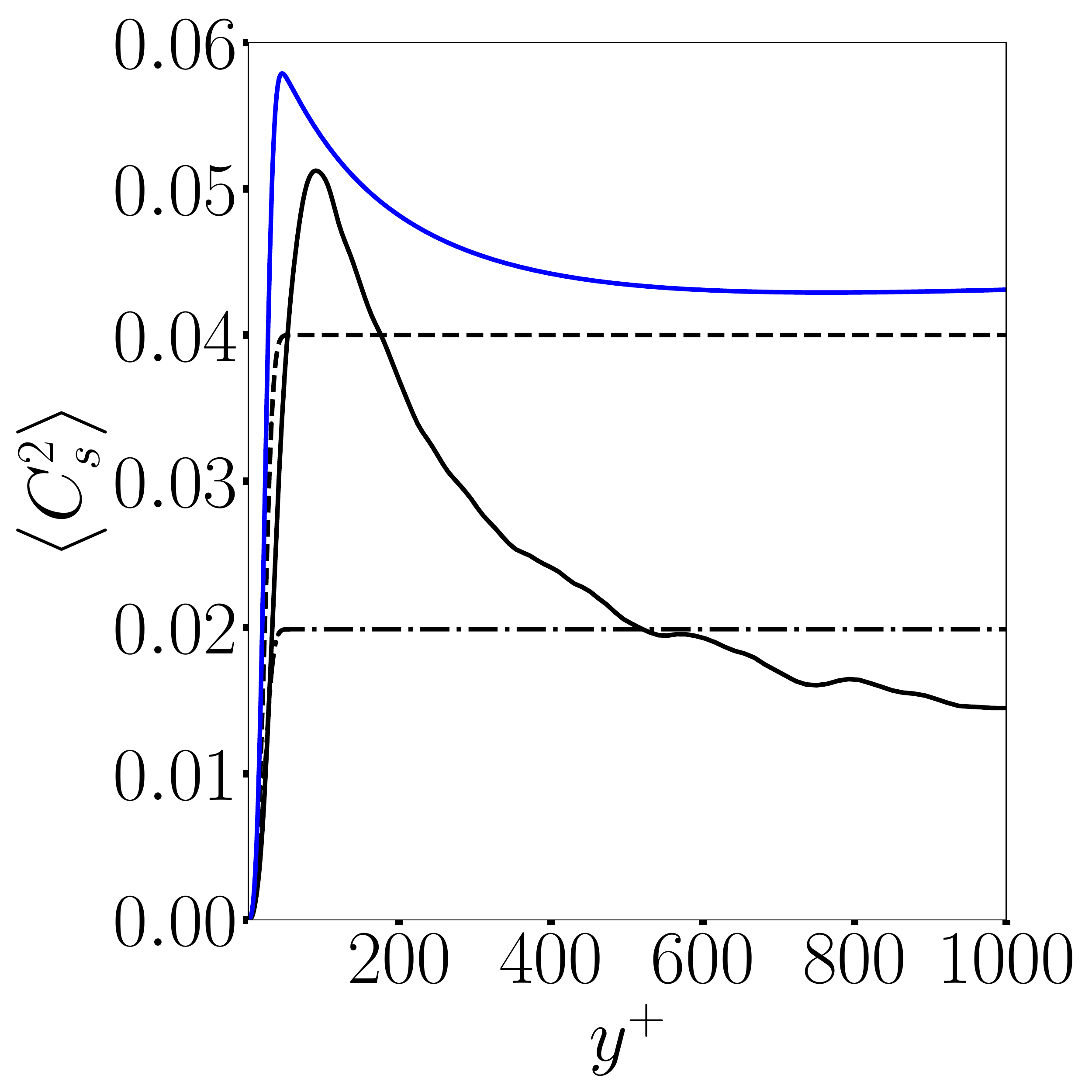}
\caption{\label{fig:Vincent_case6_C_C}Control vector $C_\mathrm{s}(y)$}
\end{subfigure}%
\begin{subfigure}{.28\linewidth}
\centering
\includegraphics[width=1.\textwidth]{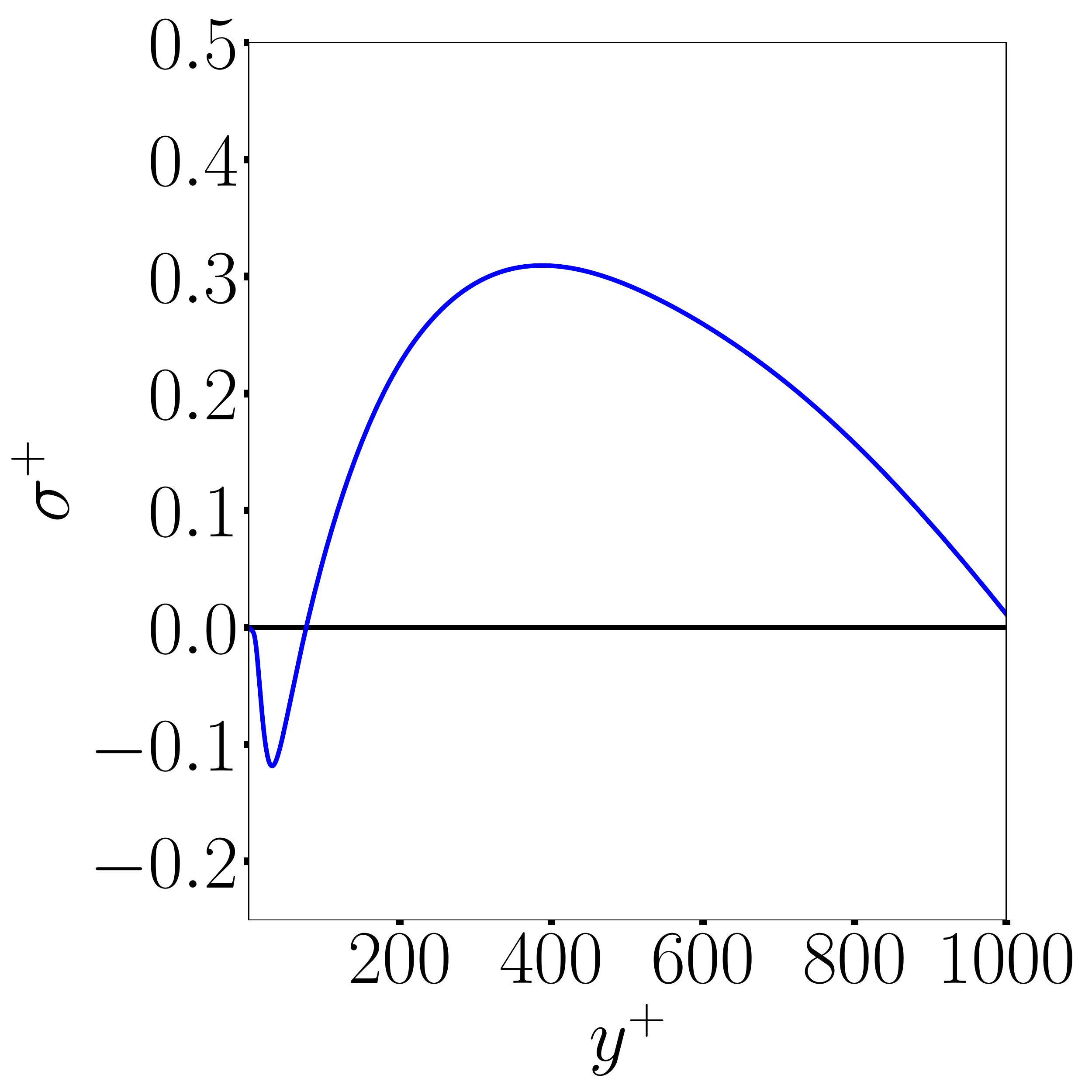}
\caption{\label{fig:Vincent_case6_C_sigma}Control vector $\sigma(y)$}
\end{subfigure}
\begin{subfigure}{.28\linewidth}
\centering
\includegraphics[width=1.\textwidth]{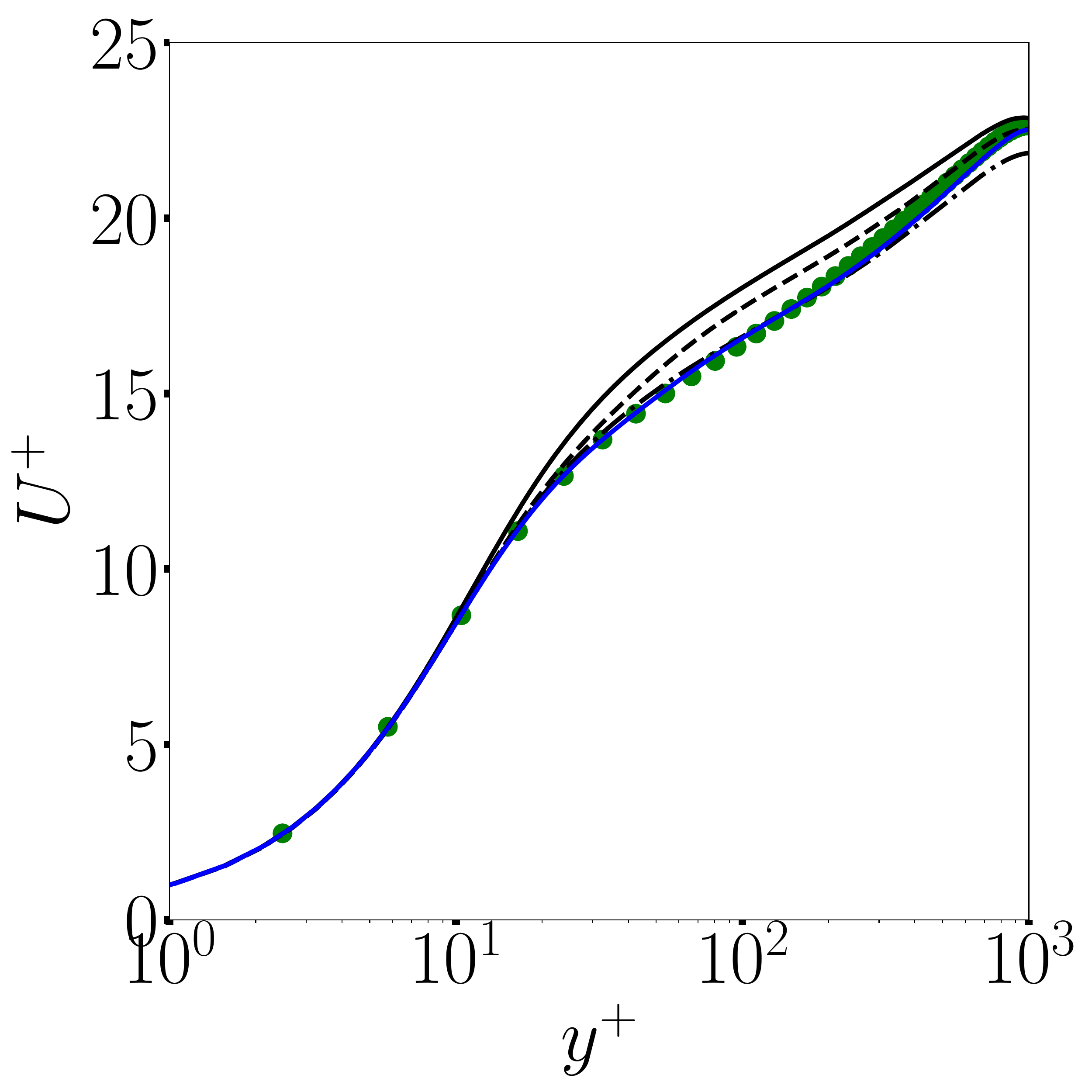}
\caption{\label{fig:Vincent_case6_C_U}Mean profile $U(y)$}
\end{subfigure}
\begin{subfigure}{.28\linewidth}
\centering
\includegraphics[width=1.\textwidth]{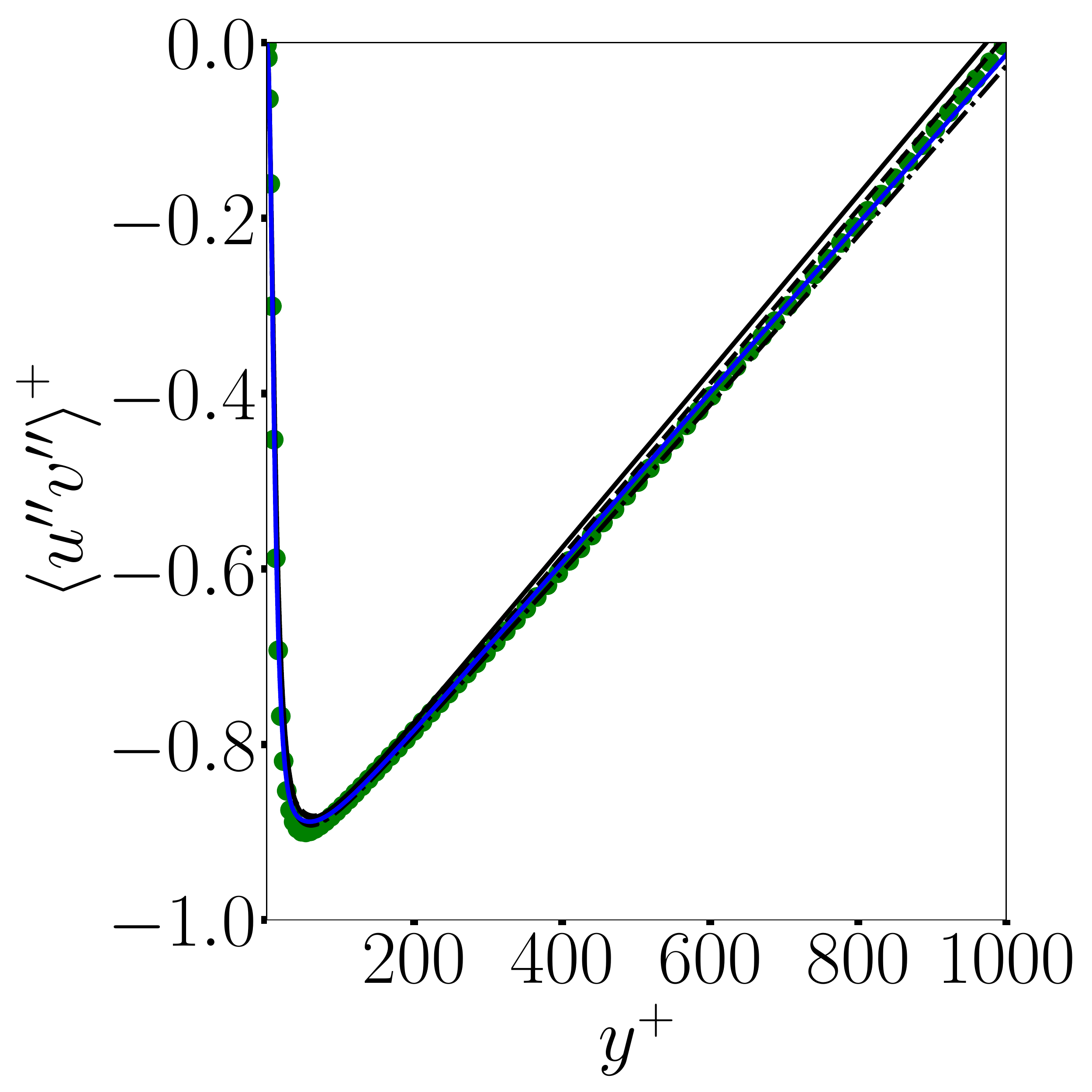}
\caption{\label{fig:Vincent_case6_C_uv}Total shear stress $\left<u''v''\right>$}
\end{subfigure}
\begin{subfigure}{.28\linewidth}
\centering
\includegraphics[width=1.\textwidth]{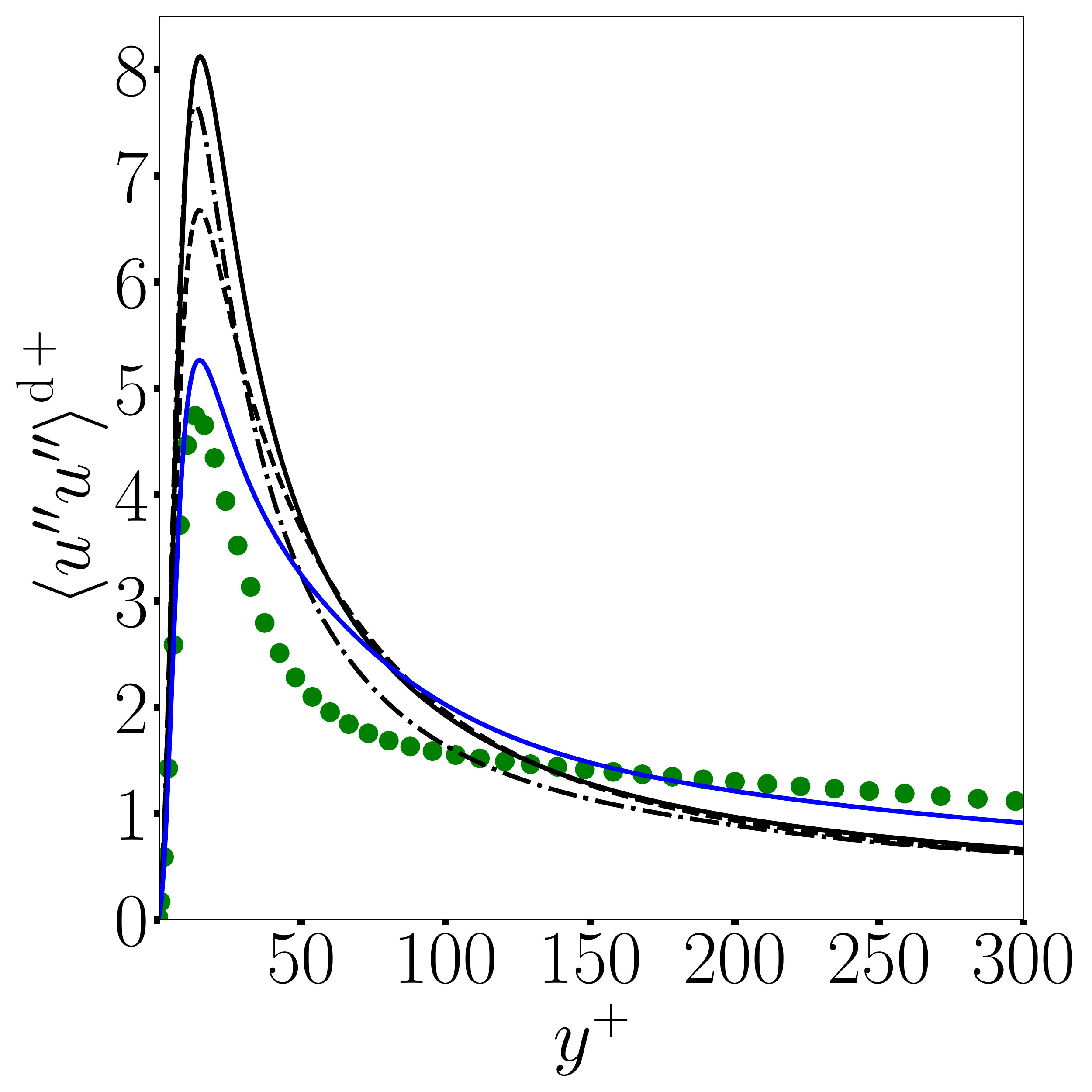}
\caption{\label{fig:Vincent_case6_C_duu}Deviatoric Reynolds stress $\left<u''u''\right>^{\mathrm{d}}$}
\end{subfigure}%
\begin{subfigure}{.28\linewidth}
\centering
\includegraphics[width=1.\textwidth]{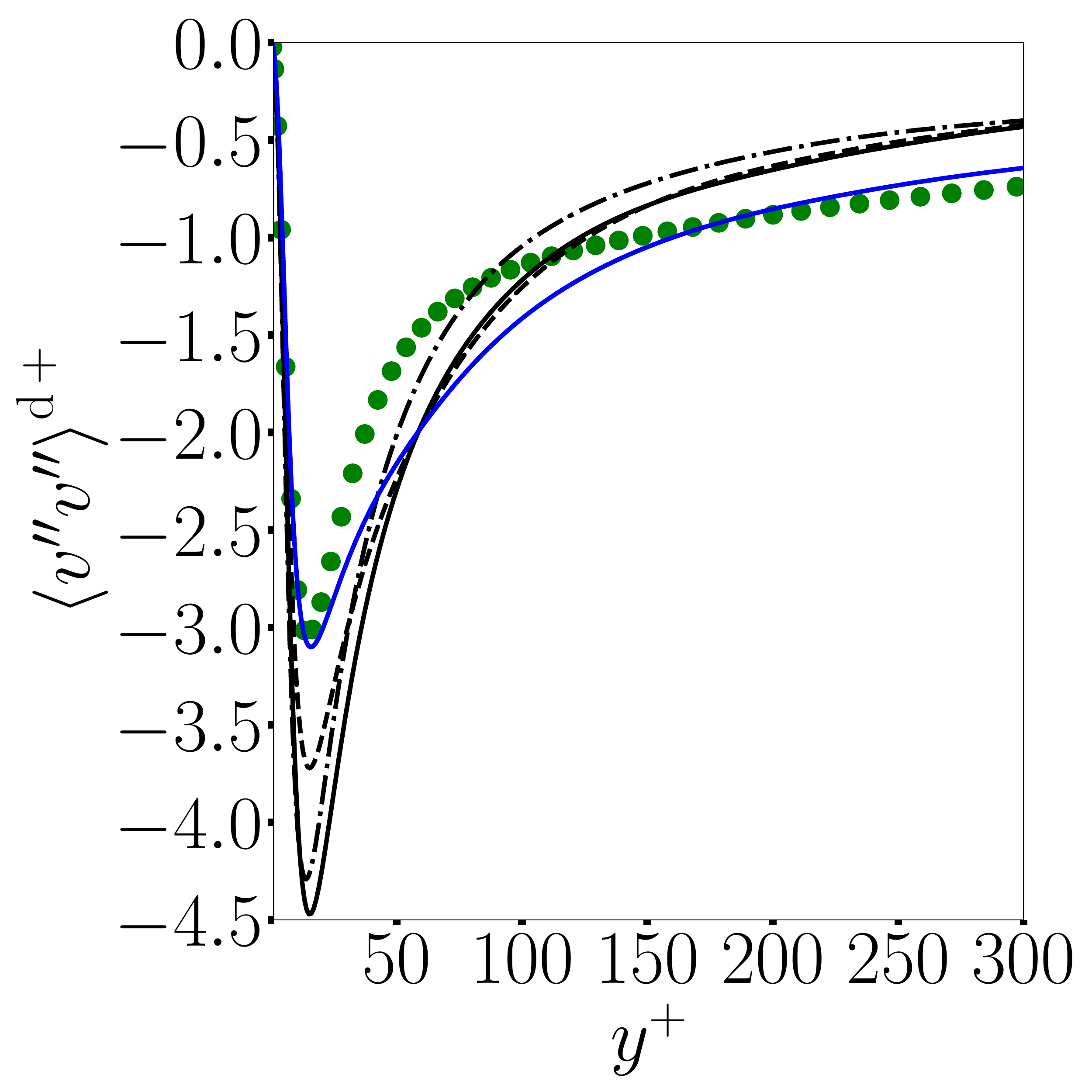}
\caption{\label{fig:Vincent_case6_C_dvv}Deviatoric Reynolds stress $\left<v''v''\right>^{\mathrm{d}}$}
\end{subfigure}
\caption{\label{fig:Vincent_case6}Control vector and flow statistics of case 6 for DNS (\mycircle{black!50!green} \mycircle{black!40!green} \mycircle{black!40!green}), the dynamic model (\fullblack), the mixed model (\chainblack), the Smagorinsky model (\dashed) and for final DA-LES (\full).}
\end{figure}

The final data assimilation experiment is case 6, where the target friction Reynolds number is $Re_{\tau}=1{,}000$. The setup is otherwise similar to case 5c: the employed grid (LES1000) has the same relatively low resolution in wall units, the control vector $\boldsymbol{\gamma}$ is formed by both $C_{\mathrm{s}}(y)$ and $\sigma(y)$, and the observations include the mean flow gradient $\mathrm{d}U/\mathrm{d}y$ and the Reynolds stress $\left<u''u''\right>$. These reference statistics are extracted from the Johns Hopkins Turbulence Databases \citep{Graham2016_jot}. Figure \ref{fig:Vincent_case6} reports results for the final assimilated state, which has been obtained in two main iterations of the data assimilation procedure, along with the predictions obtained with the considered baseline subgrid models. It appears from figure \ref{fig:Vincent_case6_C_C} that the value of the coefficient $C_{\mathrm{s}}$ has been notably increased compared with the first-guess Smagorinsky profile (\ref{eq:smagorinsky_constant}) around $y^+=50$ through the assimilation of the observations. This increase in $C_{\mathrm{s}}$, also compared with case 5c (figure \ref{fig:Vincent_case5c_C_C}), is also observed in the profile that is predicted by the dynamic model. As in previous cases in this section, DA-LES accurately recovers the reference mean flow in figure \ref{fig:Vincent_case6_C_U}. In addition, as reported in figures \ref{fig:Vincent_case6_C_duu}-\ref{fig:Vincent_case6_C_dvv}, the diagonal components of the deviatoric part of the Reynolds stress tensor are poorly predicted by all baseline models, while DA-LES provides a large improvement compared with these profiles over the whole channel height. These results demonstrate the robustness of the present data assimilation methodology for the considered $Re_{\tau}$. Since $Re_{\tau}=1{,}000$ may be considered as a lower bound above which low-Reynolds-number effects in turbulent channel flows are mitigated, and as most features of the considered statistical quantities should exhibit mild ($\propto \log(Re_{\tau})$) variations with increasing $Re_{\tau}$ \citep{Schultz2013_pof,Lee2015_jfm}, the data assimilation procedure is expected to perform satisfactorily at higher Reynolds numbers than considered herein.

\subsection{Further assessment of LES results and computational cost (cases 5, 5c and 6)}\label{sec:further_assessment}

For the three last cases (5, 5c and 6), the control vector $\boldsymbol{\gamma}$ was formed by the coefficient $C_{\mathrm{s}}$ and the forcing $\sigma$, and the observations $\boldsymbol{m}$ corresponded to the mean-flow gradient $\mathrm{d}U/\mathrm{d}y$ and the Reynolds stress $\left<u''u''\right>$. In all three, 
the data assimilation procedure has significantly enhanced LES predictions compared to the use of baseline models.  In this section, this improvement is further quantified and we discuss the computational costs for the baseline and DA-LES calculations.

Following \cite{Toosi2017_cf,Sun2018_jcp}, table \ref{tab:errors} reports error indicators that quantify the discrepancies between LES and DNS in the estimation of the mean flow $U$ and the components of the deviatoric part of the Reynolds stress tensor $\langle u_{i}^{''}u_{j}^{''}\rangle^{\mathrm{d}}$, according to
\begin{equation}\label{eq:errors_wr_DNS_1}
    e_U=\frac{\int \left| U^{+}_{\mathrm{LES}}(y^{+})-U^{+}_{\mathrm{DNS}}(y^{+})  \right|dy^{+}}{\int U^{+}_{\mathrm{DNS}}(y^{+}) dy^{+}}, \quad e_{\left\langle u_{i}''u_{j}''\right\rangle^{\mathrm{d}}}=\frac{\int \left| \left\langle u_{i}''u_{j}''\right\rangle^{\mathrm{d}\,+}_{\mathrm{LES}}(y^{+})-\left\langle u_{i}''u_{j}''\right\rangle^{\mathrm{d}\,+}_{\mathrm{DNS}}(y^{+}) \right|dy^{+}}{\int \frac{1}{2}\left\langle u_{i}''u_{i}''\right\rangle^{+}_{\mathrm{DNS}}(y^{+}) dy^{+}},
\end{equation}
where the integrals are performed over the entire channel height. Note that the errors $e_{\langle u_{i}^{''}u_{j}^{''}\rangle^{\mathrm{d}}}$ are normalized by the reference total turbulent kinetic energy. An averaged error indicator $e_{\mathrm{avg}}$ is also considered in table \ref{tab:errors}, which is based on the error in the mean flow $e_U$ and those in the non-zero components of $\langle u_{i}^{''}u_{j}^{''}\rangle^{\mathrm{d}}$, according to  
\begin{equation}\label{eq:errors_wr_DNS_2}
    e_{\mathrm{avg}}=\frac{1}{5}\left(e_U+ e_{\left\langle u''v''\right\rangle} + e_{\left\langle u''u''\right\rangle^{\mathrm{d}}}+e_{\left\langle v''v''\right\rangle^{\mathrm{d}}}+e_{\left\langle w''w''\right\rangle^{\mathrm{d}}}\right).
\end{equation}

\begin{figure}
\centering
\begin{subfigure}{.28\linewidth}
\centering
\includegraphics[width=1.\textwidth]{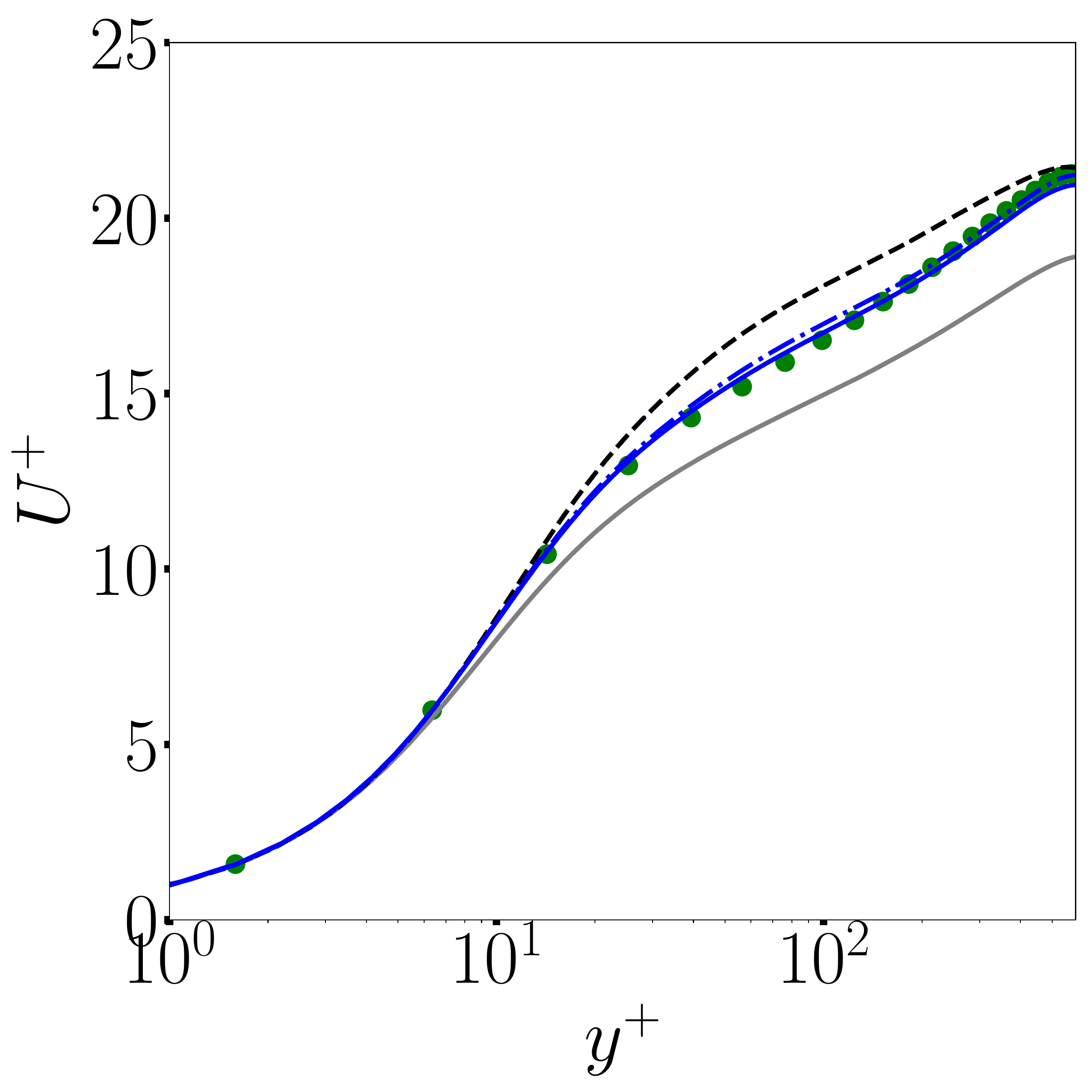}
\caption{Case 5, Mean profile $U(y)$}
\end{subfigure}%
\begin{subfigure}{.28\linewidth}
\centering
\includegraphics[width=1.\textwidth]{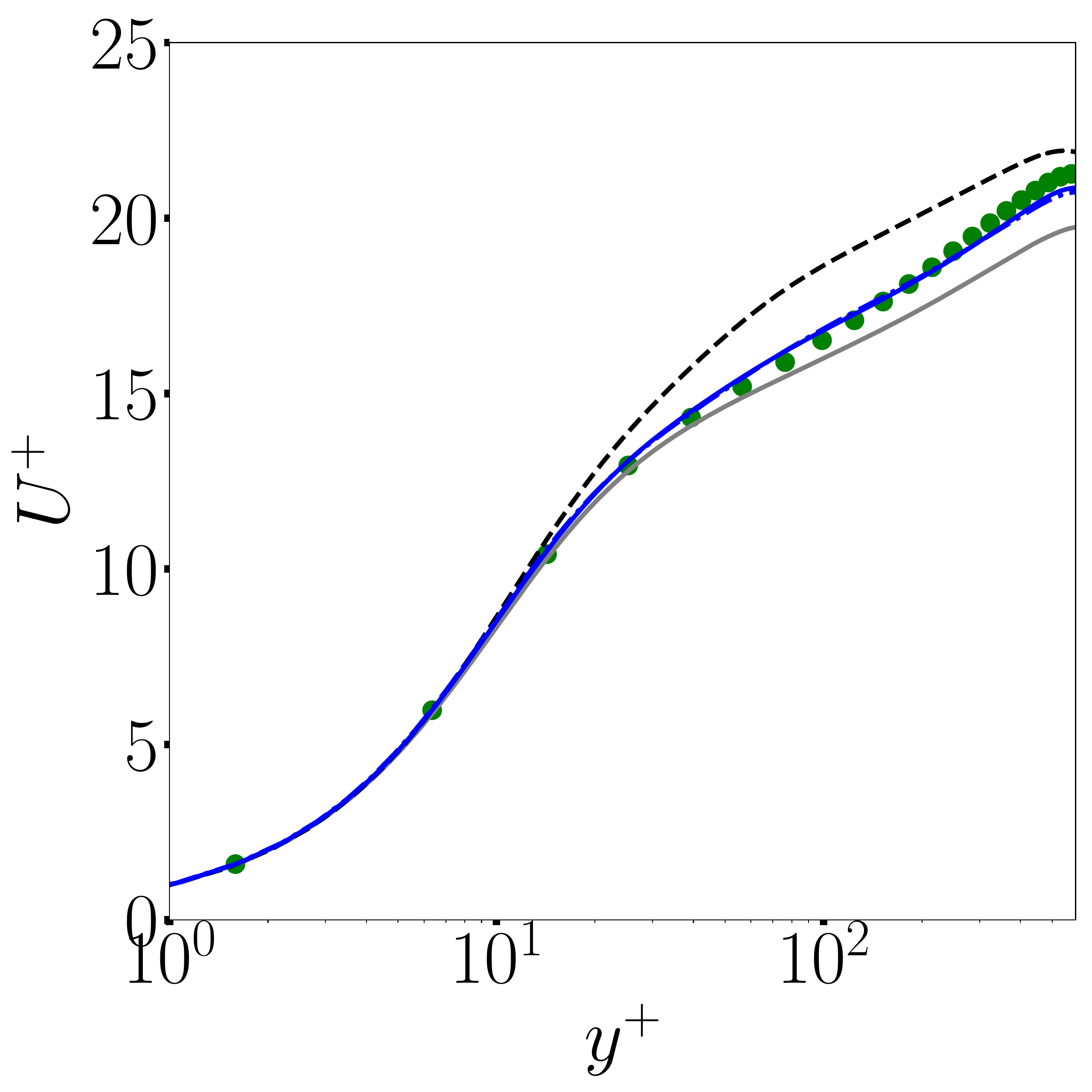}
\caption{Case 5c, Mean profile $U(y)$}
\end{subfigure}
\begin{subfigure}{.28\linewidth}
\centering
\includegraphics[width=1.\textwidth]{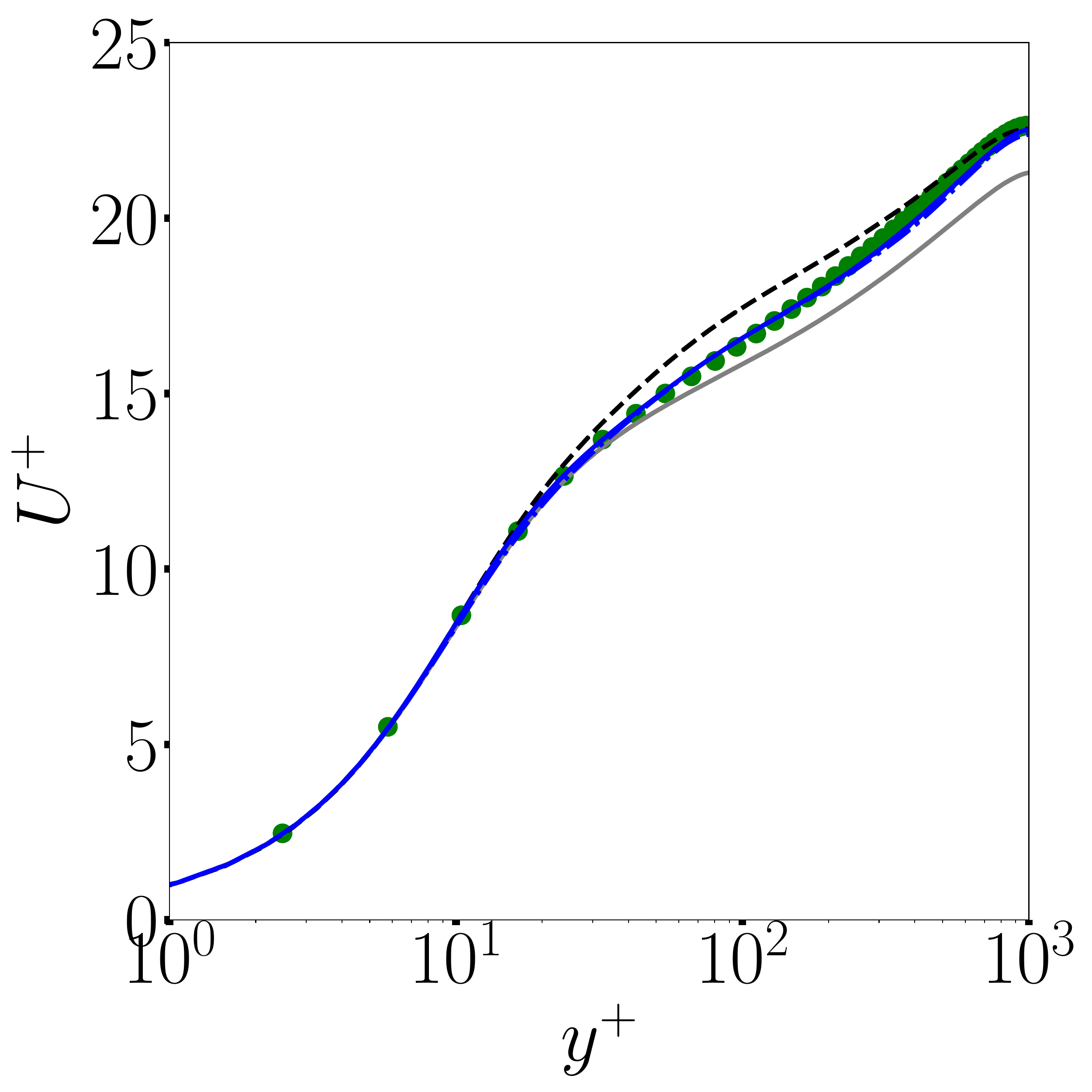}
\caption{Case 6, Mean profile $U(y)$}
\end{subfigure}
\begin{subfigure}{.28\linewidth}
\centering
\includegraphics[width=1.\textwidth]{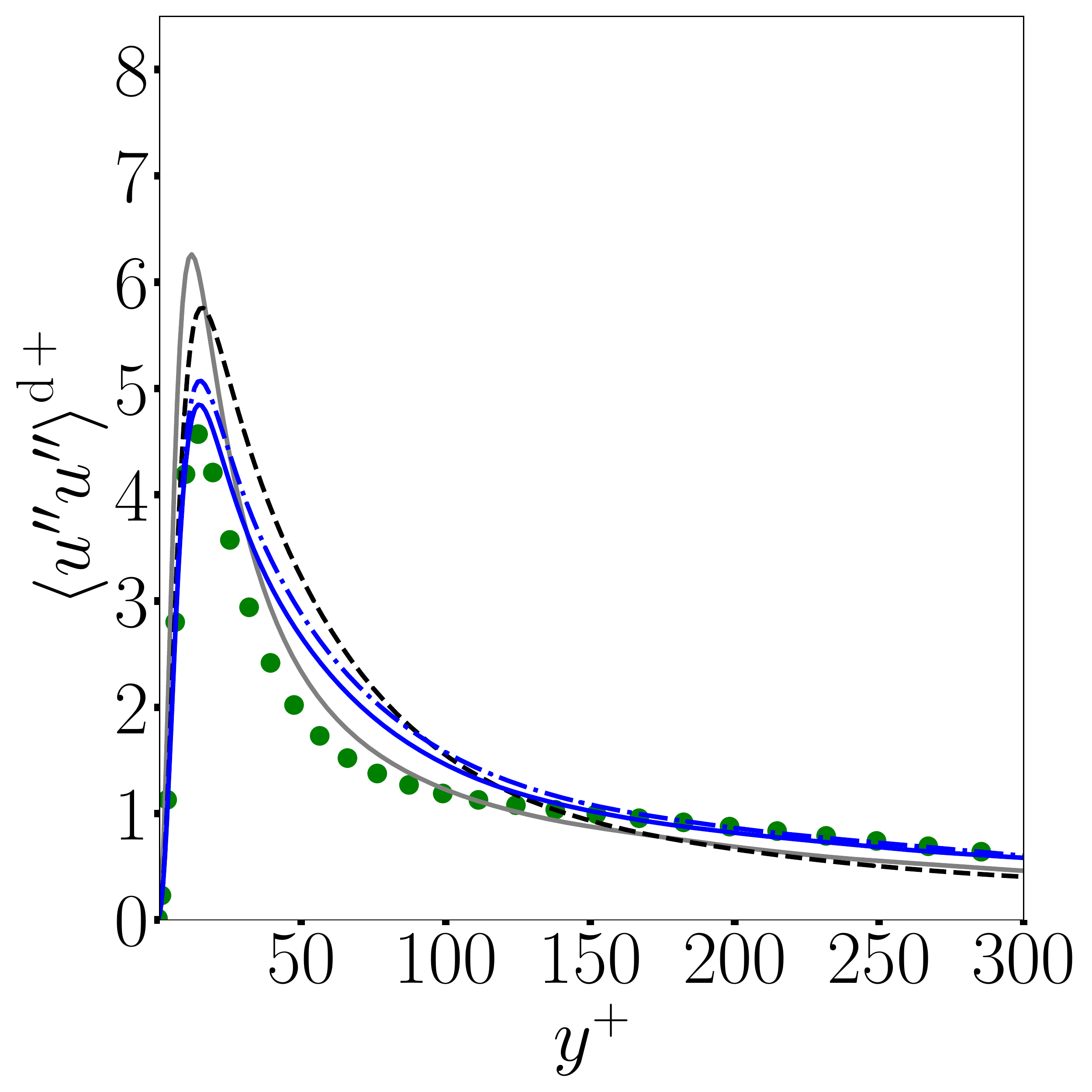}
\caption{Case 5, Deviatoric Reynolds stress $\left<u''u''\right>^{\mathrm{d}}$}
\end{subfigure}
\begin{subfigure}{.28\linewidth}
\centering
\includegraphics[width=1.\textwidth]{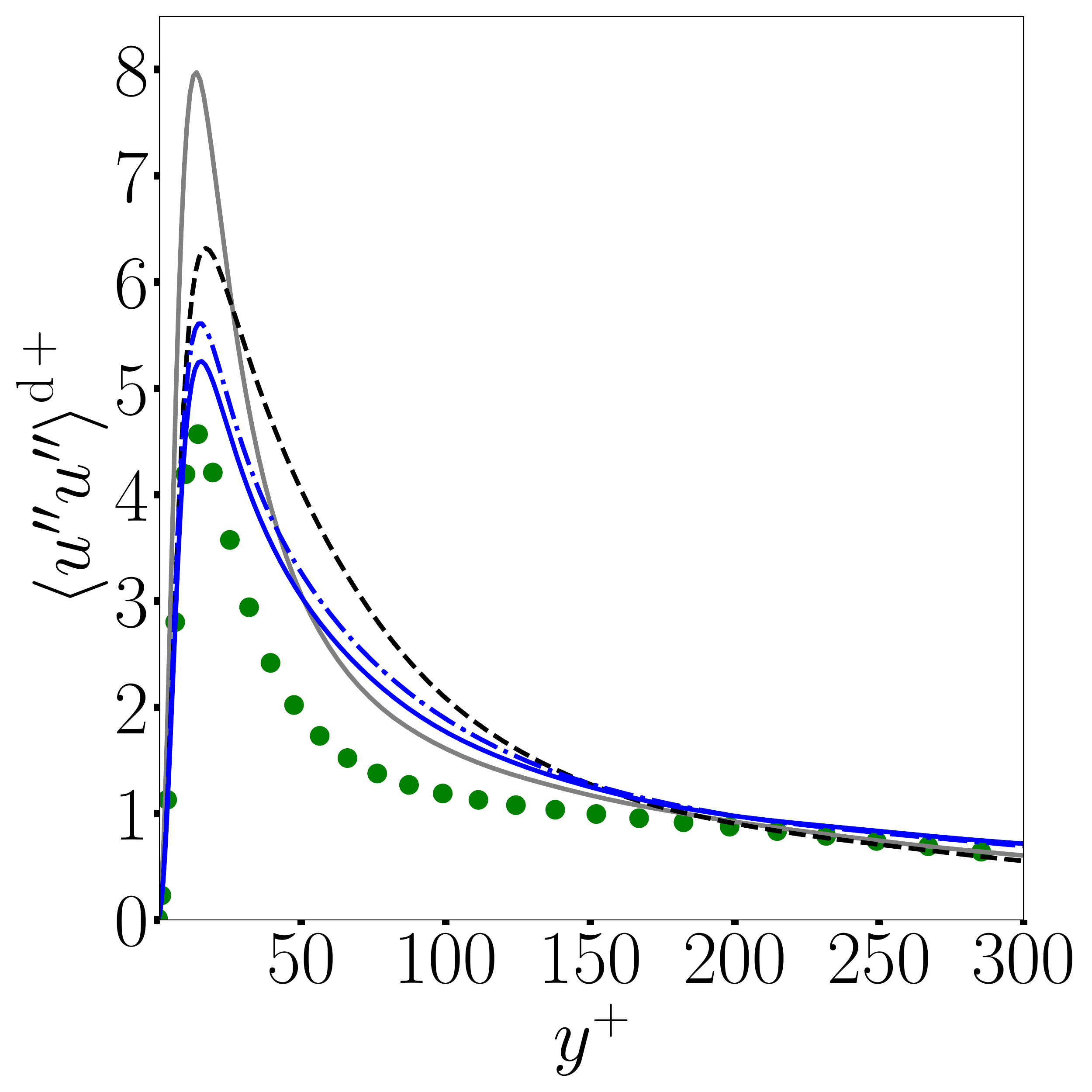}
\caption{\label{fig:nomodel_10it_duu_5c}Case 5c, Deviatoric Reynolds stress $\left<u''u''\right>^{\mathrm{d}}$}
\end{subfigure}%
\begin{subfigure}{.28\linewidth}
\centering
\includegraphics[width=1.\textwidth]{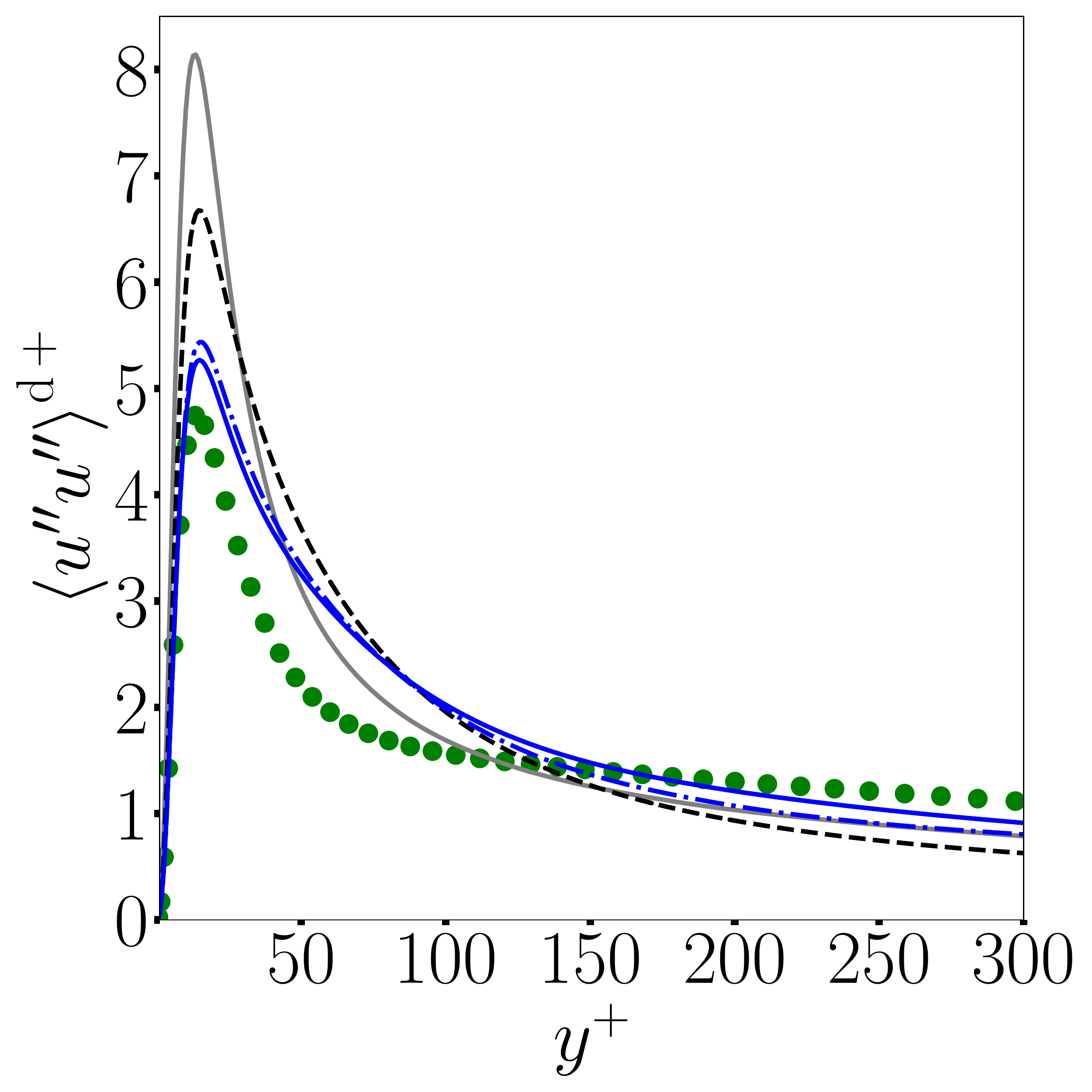}
\caption{Case 6, Deviatoric Reynolds stress $\left<u''u''\right>^{\mathrm{d}}$}
\end{subfigure}
\caption{\label{fig:nomodel_10it}Flow statistics of cases 5, 5c and 6 for DNS (\mycircle{black!50!green} \mycircle{black!40!green} \mycircle{black!40!green}), no-model calculations (\fullgrey), the Smagorinsky model (\dashed) and DA-LES results obtained through $N_{\mathrm{CFD}}=11$ (\chain) or $N_{\mathrm{CFD}}=42$ (\full).}
\end{figure}

In addition to results from LES that have already been discussed in the previous sections, table \ref{tab:errors} reports error indicators for two new sets of LES, which are also illustrated in figure \ref{fig:nomodel_10it} for cases 5, 5c and 6. The first set is referred to as no-model simulations, namely LES without a subgrid model, or implicit LES. These calculations are computationally cheap and indirectly help to demonstrate the influence of the subgrid model in LES (see \S\ref{sec:cases_1_2}). The other new results in table \ref{tab:errors} and figure \ref{fig:nomodel_10it} are from DA-LES, and examine the effect of reducing the number of ensemble members and iterations of the data assimilation Algorithm \ref{tab:EnVar_algo}. Specifically, we perform DA-LES with $N_{\mathrm{ens}}=10$ ensemble members and only $N_{\mathrm{it}}=1$ iteration of Algorithm \ref{tab:EnVar_algo}, and therefore $N_{\mathrm{CFD}}=N_{\mathrm{it}}\times (N_{\mathrm{ens}}+1)=11$ LES calculations. For comparison, the previous results for cases 5, 5c and 6 (\S\ref{sec:results_DA_statistics_Re_590_cases5_5c}-\ref{sec:results_DA_statistics_Re_1000}) were obtained using $N_{\mathrm{ens}}=20$ and $N_{\mathrm{it}}=2$, and therefore $N_{\mathrm{CFD}}=42$ LES computations. Those results along with the first-guess Smagorinsky curves are also reproduced in figure \ref{fig:nomodel_10it}.

\begin{table}
\begin{center}
\begin{tabular}{ ccccccccc } 
 \hline
Case & model & $e_U$ & $e_{\left\langle u''v''\right\rangle}$  & $e_{\left\langle u''u''\right\rangle^{\mathrm{d}}}$ & $e_{\left\langle v''v''\right\rangle^{\mathrm{d}}}$ & $e_{\left\langle w''w''\right\rangle^{\mathrm{d}}}$  & $e_{\mathrm{avg}}$ & $T_{\mathrm{CPU}}(h)$\\ 
 \hline
\multirow{5}{*}{5} 
& no-model & 9.6  & 0.2  & 10.5 & 8.5 & 3.2 & 6.4 & 89\\
& mixed & 3.7  & 0.2  & 11.1 & 7.5 & 3.8  & 5.3 & 110\\
& dynamic & 1.9  & 0.4  & 12.2 & 7.7 & 5.1  & 5.5 & 109 \\
\rowcolor{SkyBlue!40} \cellcolor{white} 
&  Smagorinsky & 3.8  & 0.7  & 13.6 & 7.6 & 6.6  & 6.5 & 101\\
\rowcolor{Tan!40} \cellcolor{white} 
&  DA-LES ($N_{\mathrm{CFD}}=11$) & 0.7  & 0.5 & 6.5 & 3.8 & 4.3  & 3.2 & 1111 \\
\rowcolor{Tan!40} \cellcolor{white} 
& DA-LES ($N_{\mathrm{CFD}}=42$) & 1.2  & 0.2  & 5.8 & 3.4 & 3.8  & 2.9 & 4242 \\
\hline
\multirow{5}{*}{5c} 
& no-model & 5.2  & 0.2  & 13.1 & 8.2 & 9.3  & 7.2 & 49 \\
& mixed & 2.0  & 0.2  & 14.9 & 8.5 & 9.1  & 7.0 & 62 \\
& dynamic & 4.8  & 0.6  & 20.7 & 10.4 & 12.8 & 9.8 & 60 \\
\rowcolor{SkyBlue!40} \cellcolor{white} 
& Smagorinsky & 6.4  & 0.6  & 19.8 & 9.2 & 13.1  & 9.8 & 55\\
\rowcolor{Tan!40} \cellcolor{white} 
& DA-LES ($N_{\mathrm{CFD}}=11$) & 1.6  & 0.3  & 12.5 & 4.7 & 9.2  & 5.7 & 605 \\
\rowcolor{Tan!40} \cellcolor{white} 
& DA-LES ($N_{\mathrm{CFD}}=42$) & 1.3  & 0.2  & 12.2 & 4.1 & 8.5  & 5.3 & 2310 \\
\hline
\multirow{5}{*}{6} 
& no-model & 5.1  & 0.4 & 16.1 & 10.7 & 6.1  & 7.7 & 196\\
& mixed & 2.3  & 0.3  & 19.8 & 12.3 & 7.9  & 8.5 & 232 \\
& dynamic & 3.5 & 0.6  & 20.5 & 11.7 & 9.0  & 9.1 & 232\\
\rowcolor{SkyBlue!40} \cellcolor{white} 
& Smagorinsky & 1.7  & 0.5  & 19.6 & 11.0 & 8.8  & 8.3 & 219\\
\rowcolor{Tan!40} \cellcolor{white} 
& DA-LES ($N_{\mathrm{CFD}}=11$) & 1.3  & 0.5  & 11.1 & 6.1 & 5.0  & 4.8 & 2409 \\
\rowcolor{Tan!40} \cellcolor{white} 
& DA-LES ($N_{\mathrm{CFD}}=42$) & 0.8  & 0.5  & 7.9 & 4.6 & 3.3  & 3.4 & 9198 \\
\hline
\end{tabular}
\caption{\label{tab:errors}Errors in (\ref{eq:errors_wr_DNS_1})-(\ref{eq:errors_wr_DNS_2}) and CPU time for LES calculations in cases 5 ($Re_{\tau}=590$, fine grid LES590f), 5c ($Re_{\tau}=590$, coarse grid LES590c) and 6 ($Re_{\tau}=1{,}000$, coarse grid LES1000).}
\end{center}
\end{table}

Table \ref{tab:errors} does not suggest a clear hierarchy among the LES using baseline subgrid models.  For case 5, which corresponds to $Re_{\mathrm{\tau}}=590$ and the use of the fine grid LES590f, the dynamic and mixed models provide overall better results (among these two models, the mean flow is better estimated with the former while turbulence anisotropy is more satisfactorily recovered by the latter). When employing the coarser grid LES590c (case 5c), the mixed model again provides more satisfactory results than other baseline models, in particular in terms of mean flow estimation. 
These relatively good performances are followed by the no-model calculation.
It should be emphasized, however, that the errors in (\ref{eq:errors_wr_DNS_1})-(\ref{eq:errors_wr_DNS_2}) are integrated over the whole channel height and that more nuanced conclusions may be drawn when considering local discrepancies. For example, figure \ref{fig:nomodel_10it_duu_5c} reports the deviatoric stress $\left<u''u''\right>^{\mathrm{d}}$ for case 5c, and demonstrates that the standard Smagorinsky model, due to its associated subgrid dissipation, provides a better prediction of the flow anisotropy around the peak turbulent kinetic energy than the no-model calculation. Subgrid dissipation thus still seems desirable to improve the estimation of the flow in the near-wall region. Further away from the wall, no-model results become better than Smagorinsky ones, which explains the respective errors of $e_{\left\langle u''u''\right\rangle^{\mathrm{d}}}$ reported in the table. 
For case 6, the Reynolds number is $Re_{\tau}=1{,}000$ and a relatively coarse grid was employed, with the same resolution in wall units as in case 5c.  The DA-LES results aside, the no-model calculation may appear to provide the best prediction based on the average integrated errors $e_{\mathrm{avg}}$. We must emphasize, however, that this measure masks the local inaccuracy in important dynamical regions. In addition, the no-model computation yields the most inaccurate prediction of the mean flow.  Taken all together, the results of the baseline LES computations confirm the difficulty in identifying one model that consistently outperforms the others.

Table \ref{tab:errors} also confirms the superiority of DA-LES, which systematically outperforms the other models in the estimation of both the mean flow and turbulence anisotropy. For cases 5, 5c and 6, using the baseline models yields errors in the mean flow, $e_U$, at least $1.5$ folds, and often $4$ or more folds, those from DA-LES.
Concerning the diagonal components of the deviatoric part of the Reynolds stress tensor, the assimilation procedure overall halves the errors $e_{\left\langle u_{i}''u_{j}''\right\rangle^{\mathrm{d}}}$, in particular compared to the standard Smagorinsky model which is the first guess of the DA-LES procedure. This applies to both the assimilated states that are obtained through $11$ and $42$ LES calculations. While in \S\ref{sec:results_DA_statistics_Re_590_cases5_5c}-\ref{sec:results_DA_statistics_Re_1000} results were obtained using $42$ LES calculations to ensure the convergence of the data assimilation procedure, it appears that approximately $10$ computations are sufficient in order to achieve large improvements with respect to the baseline model. Increasing the number of ensemble members and/or of iterations of the data assimilation procedure brings only marginal further improvements. This is well illustrated by figure \ref{fig:nomodel_10it}, where the results for the assimilated states obtained through $11$ or $42$ calculations are very similar.

The computational cost in terms of CPU time $T_{\mathrm{CPU}}$ is also reported in table \ref{tab:errors}. No-model calculations set the lowest computational cost for a given configuration. Simulations based on the Smagorinsky model are associated with an increase by $\sim 10\%$ in computational cost for the present implementation, while the dynamic and mixed models correspond to an increase by $\sim 20\%$ due to the explicit filtering operations (see \S\ref{sec:subgrid_models}). The cost of DA-LES is equivalent to that for $N_{\mathrm{CFD}}$ Smagorinsky calculations (here $N_{\mathrm{CFD}}=11$ or $N_{\mathrm{CFD}}=42$). The variations in the computational cost between cases 5, 5c and 6 reflect the scaling $T_{\mathrm{CPU}} \propto N_tN_xN_yN_z$ and the changes in the total number of grid points $N_xN_yN_z$ between cases, while the number of time steps to collect converged statistical results $N_t$ was kept constant for all calculations.

The DA-LES predictions using $N_{\mathrm{CFD}}\sim O(10)$ outperform baseline models. The associated computational cost should be viewed in context. Firstly, this cost is less than that associated with other data-assimilation techniques including adjoint- and ensemble-methods, as demonstrated in previous studies from ocean and atmospheric sciences \citep{Lewis2006_cambridge,Evensen2009_springer} and aerodynamic applications \citep{Colburn2011_jfm,Gronskis2013_jcp,Mons2016_jcp}. Furthermore, performing on the order of $10$ LES remains significantly more affordable than a single DNS. 
In the latter, the total number of grid points $N_xN_yN_z$ is typically between $10^{{3}/{2}}$ and $10^{2}$ larger than for LES, or more \citep{hartel1994_pof,Yang2021_pof}. Assuming that the number of required time steps scales as $N_t \propto (N_xN_yN_z)^{{1}/{3}}$ \citep{Piomelli2002_arfm}, this results in a computational cost that is $10^{2}$ to $10^{{8}/{3}}$ larger for DNS than for LES, which is verified with the present code. 
Assuming that $O(10)$ computations are performed for DA-LES, the computational cost remains at least $10$ or $10^{{5}/{3}}$ times smaller than that required for a single DNS. 

The gap between the computational cost of DNS and DA-LES becomes larger with Reynolds number. Common estimations of the required number of grid points for DNS and LES are $(N_xN_yN_z)_{\mathrm{DNS}} \propto Re_{L}^{{9}/{4}}$ and $(N_xN_yN_z)_{\mathrm{LES}} \propto Re_{L}^{1.8}$ \citep{Piomelli2002_arfm}, respectively, where $Re_{L}$ is an integral scale-based Reynolds number. These requirements lead to $T_{\mathrm{CPU}\,\mathrm{DNS}} \propto Re_{L}^{3}$ and $T_{\mathrm{CPU}\,\mathrm{LES}} \propto Re_{L}^{2.4}$. No matter the resolution estimates adopted for LES \citep{Choi2012_pof,Yang2021_pof}, the ratio of DNS to LES computational cost remains  $T_{\mathrm{CPU}\,\mathrm{DNS}}/T_{\mathrm{CPU}\,\mathrm{LES}} \propto Re_{L}^{a}$ with $a>0$; for the herein quoted $Re_L$ dependence, $a=0.6$.  As a result, the present data assimilation approach remains significantly less expensive than DNS at higher Reynolds numbers, by one or more orders of magnitude.

\subsection{Uncertainties in the assimilated control vector $\boldsymbol{\gamma}$ (cases 5, 5c and 6)}\label{sec:results_DA_statistics_uncertainties}

\begin{figure}
\centering
\begin{subfigure}{.28\linewidth}
\centering
\includegraphics[width=1.\textwidth]{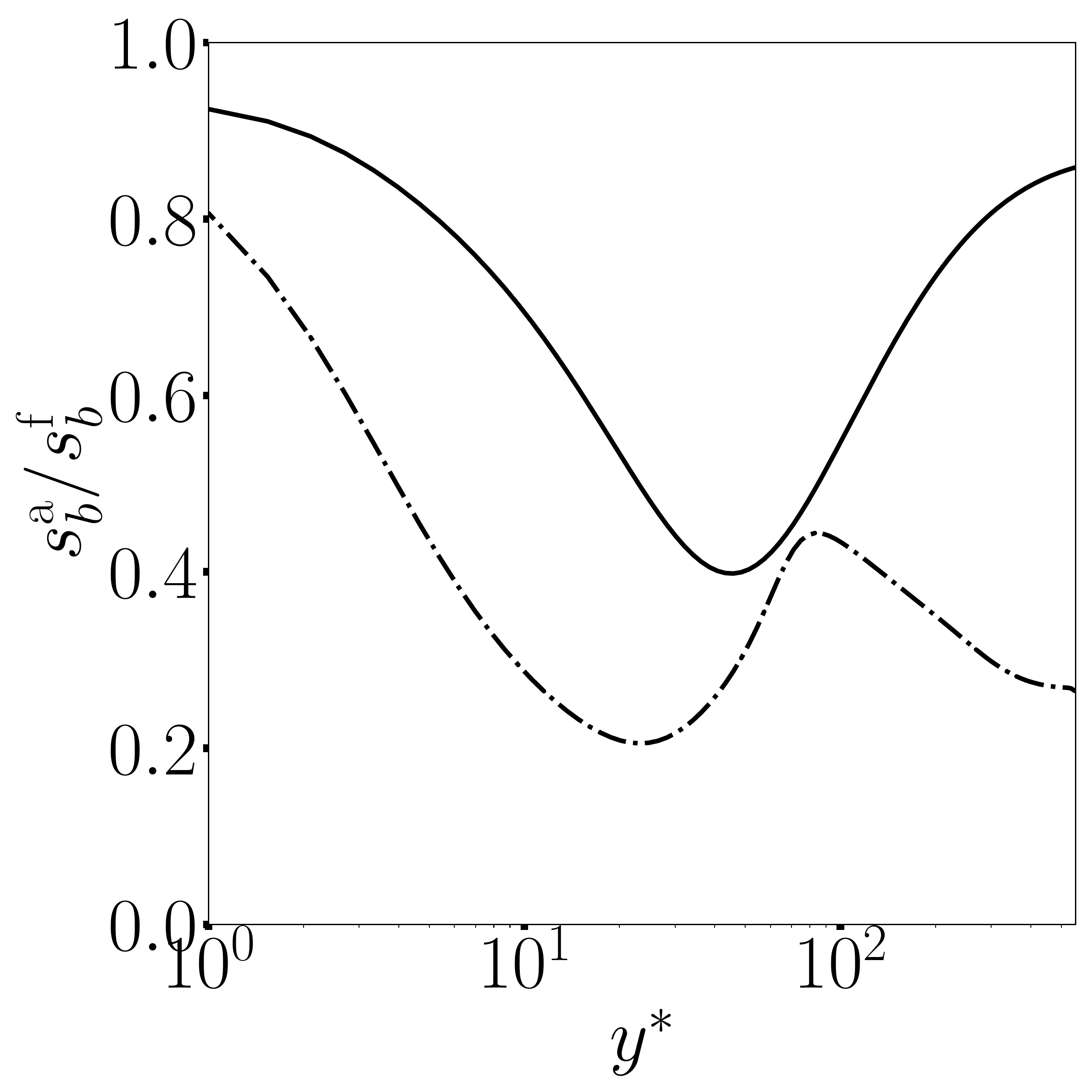}
\caption{Case 5}
\end{subfigure}
\begin{subfigure}{.28\linewidth}
\centering
\includegraphics[width=1.\textwidth]{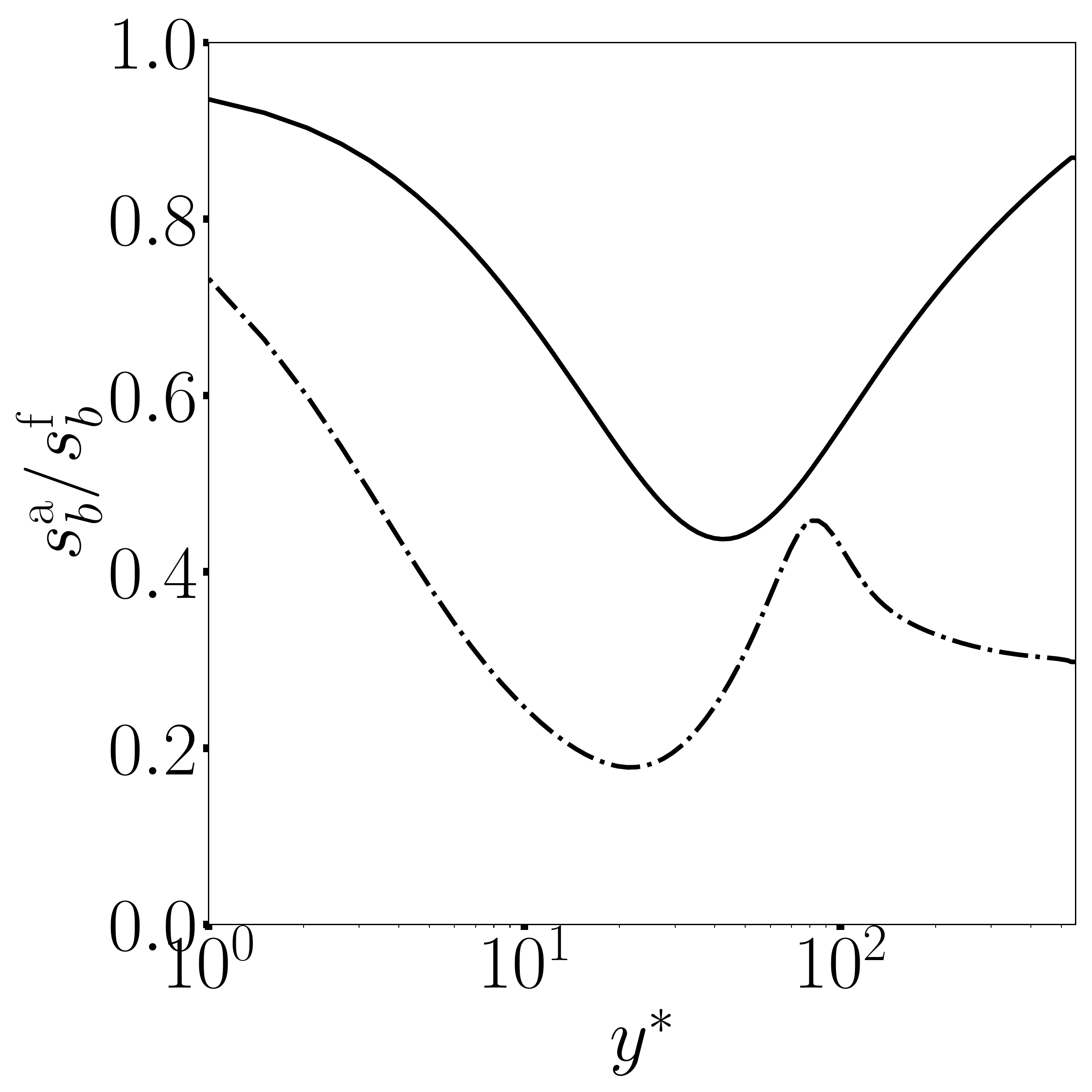}
\caption{Case 5c}
\end{subfigure}
\begin{subfigure}{.28\linewidth}
\centering
\includegraphics[width=1.\textwidth]{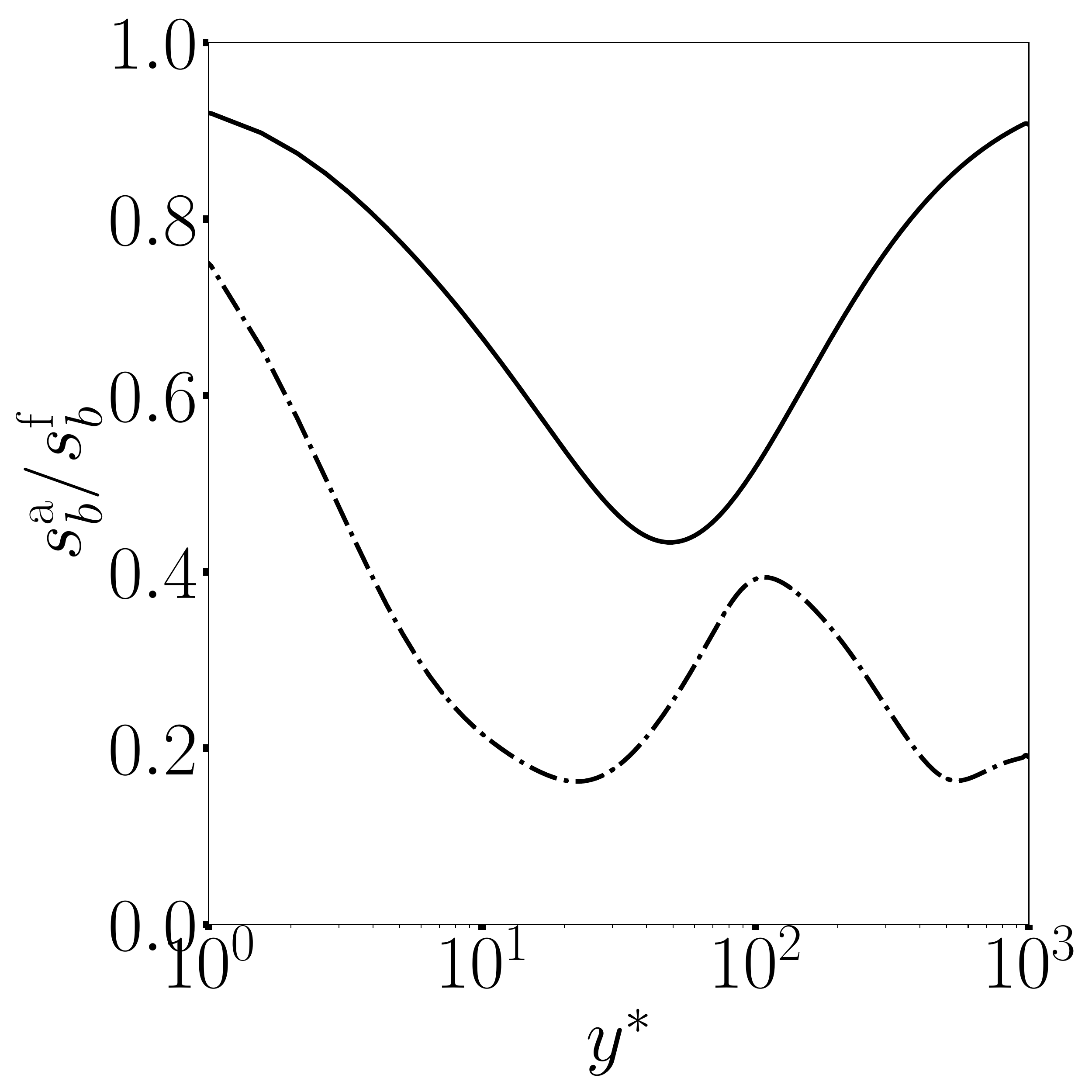}
\caption{Case 6}
\end{subfigure}
\caption{\label{fig:post_stat}Ratio between the standard deviation at the end of the data assimilation procedure $s^{\mathrm{a}}_b$ and at its beginning $s^{\mathrm{f}}_b$ for a quantity $b$ that refers either to the coefficient $C^2_{\mathrm{s}}$ (\fullblack) or the forcing $\sigma$ (\chainblack) for cases 5 (a), 5c (b) and 6 (c).}
\end{figure}

As emphasized in \S\ref{sec:EnVar}, the present data assimilation methodology not only provides an assimilated state, but also the associated uncertainties, or posterior statistics, which are quantified through the covariance matrix $\boldsymbol{\mathrm{P}}^{\mathrm{a}}$ in (\ref{eq:posterior_statistics}). While these posterior statistics have been previously exploited to design new ensembles in the iterative procedure summarized in Algorithm \ref{tab:EnVar_algo}, they are here considered to assess the quality and accuracy in the final assimilated control vectors that have been obtained in data assimilation experiments 5, 5c and 6 ($N_{\mathrm{CFD}}=42$). In these three cases, $\boldsymbol{\gamma}$ is formed by the coefficient $C_{\mathrm{s}}$ and the forcing $\sigma$, while the observations are the mean-flow gradient $\mathrm{d}U/\mathrm{d}y$ and the Reynolds stress $\left<u''u''\right>$. The confidence in the retrieved control parameters can be quantified by the associated standard deviations $s^{\mathrm{a}}_b(y)$ evaluated at the end of the data assimilation procedure, where $b$ refers to either $C^2_{\mathrm{s}}$ or $\sigma$. However, as the absolute values of these posterior statistics are inherently dependent on the choice of the prior statistics, which are quantified by the covariance matrix  $\boldsymbol{\mathrm{B}}$ in (\ref{eq:cov_init}), it is more informative to consider the ratio $s^{\mathrm{a}}_b/s^{\mathrm{f}}_b$ instead, where $s^{\mathrm{f}}_b$ refers to the prior standard deviation of $b$. This ratio thus captures the reduction in the uncertainties in the control vector through the data assimilation procedure, and is reported in figure \ref{fig:post_stat} for cases 5, 5c and 6.  In all three experiments, the uncertainty in the coefficient $C_{\mathrm{s}}$ is significantly decreased only in relatively narrow region around $y^*=50$. This location coincides with the region where variations in $C_{\mathrm{s}}$ have the largest impact on the flow statistics as discussed in \S\ref{sec:cases_1_2}. In contrast, elsewhere the ratio $s^{\mathrm{a}}_b/s^{\mathrm{f}}_b$ remains close to unity in particular at the channel center. This suggests that the alterations in $C_{\mathrm{s}}$ by the data assimilation procedure should be interpreted cautiously in this region, in the sense that the assimilated profiles $C_{\mathrm{s}}(y)$ remains uncertain. In particular, the significant decrease in $C_{\mathrm{s}}$ beyond $y^*=200$ in case 5c (figure \ref{fig:Vincent_case5c_C_C}) might not be meaningful, in accordance with the fact that the sensitivity analysis of \S\ref{sec:cases_1_2} suggests that variations in $C_{\mathrm{s}}$ at the channel center barely have an influence on the flow. Only the increase in $C_{\mathrm{s}}$ around $y^*=50$ by the data assimilation procedure, which is observed in the three cases, should be considered as a robust feature. It should also be noted that this adjustment is somewhat in agreement with the prediction of $C_{\mathrm{s}}(y)$ by the dynamic model, in particular in cases 5c (figure \ref{fig:Vincent_case5c_C_C}) and 6 (figure \ref{fig:Vincent_case6_C_C}).

In comparison, the assimilated forcing $\sigma(y)$ is generally evaluated with significantly lower uncertainty. This reflects the ability of $\sigma$ to directly correct the mean flow. The adjustment of $C_{\mathrm{s}}$ remains useful to independently tune the subgrid dissipation $\varepsilon^{\mathrm{fs}}$ of resolved turbulent kinetic energy and to improve the prediction of second-order statistics.

\section{Conclusions}\label{sec:conclusions}

An algorithm was developed to infer corrections to LES subgrid models through data assimilation (DA-LES) of reference statistical quantities. The approach was examined in the context of turbulent channel flow, with the Smagorinsky model as the baseline subgrid model. The fluctuations and mean of the subgrid tensor, and ultimately the subgrid dissipation of resolved mean and turbulent kinetic energies, were independently adjusted through a steady forcing term $\sigma(y)$ in the momentum equations and the Smagorinsky coefficient $C_{\mathrm{s}}(y)$, respectively. The data assimilation algorithm adopts an ensemble-variational (EnVar) approach, which merges the relative strengths of standard variational and stochastic techniques, namely the robustness of the former and the non-intrusiveness and ease of implementation of the latter. As in any stochastic data assimilation approach, the EnVar methodology not only provides an assimilated state, but also the associated uncertainties, which helps in assessing the quality/robustness of the retrieved model corrections in the present case.

Preliminary data assimilation experiments confirmed the limitations of only optimizing the profile $C_{\mathrm{s}}(y)$, as recovering first- and second-order statistics yields competitive objectives in this case. The experiments also confirmed the lack of influence of the subgrid model on the flow above $y^+=30$, which is also the region where dicretization errors become dominant \cite{Majander2002_ijnmf}. In order to disentangle the adjustment of mean and turbulent subgrid dissipations, and to enable a more profound alteration of the Smagorinsky structure, a second set of data assimilation experiments simultaneously considered the coefficient $C_{\mathrm{s}}(y)$ and the forcing $\sigma(y)$ as control vectors in conjunction with the use of statistical quantities of interest as observations. These tests confirmed the efficacy of the present data assimilation procedure in accurately reproducing both first- and second-order single-point statistics, namely the mean flow and the Reynolds stresses, while also providing some improvement in terms of spectral content. For these data assimilation experiments, DA-LES systematically outperformed more sophisticated models, namely the dynamic model and a particular mixed model variant. DA-LES predictions were robust with respect to the Reynolds number. Interestingly, the data assimilation procedure was also able to satisfactorily correct the Smagorinsky model when adopting relatively coarse grids according to the standards for wall-resolved LES. This suggests that the present methodology can enable, in the future, use of grids for which standard models start to become deficient, possibly using different observations that are more suitable in the absence of some of the energetic scales, thus contributing to further alleviating the computational cost of LES. 

An important extension of this study is to apply the developed data assimilation procedure to more complex flow configurations. While turbulent channel flows exhibit only one direction of statistical inhomogeneity, which here leads to the introduction of one-dimensional adjustable quantities, the consideration of more elaborate settings would require the definition of two- or three-dimensional correction forms. While this would lead to a significant increase in the size of the control vector in the data assimilation procedure, previous studies \citep{Mons2016_jcp} have illustrated the robustness of the EnVar approach. On the other hand,
the present results do not particularly encourage the consideration of other baseline subgrid models in the data assimilation procedure. Instead, a more promising alternative that should facilitate both the use of coarser grids and the consideration of higher Reynolds number flows is the extension of the present methodology to wall modeling \citep{Piomelli2008_pas,Larsson2016_mer}.

\section*{Acknowledgements}
The authors acknowledge financial support from the Office of Naval Research (N00014-20-1-2715,  N00014-21-1-2375). Computational resources were provided by the Maryland Advanced Research Computing Center (MARCC).

\bibliography{bibli_rev}

\end{document}